\documentclass[fleqn,10pt]{wlscirep}
\usepackage[utf8]{inputenc}
\usepackage[T1]{fontenc}
\usepackage{algorithm}
\usepackage{algpseudocode}
\usepackage{lineno,hyperref}
\usepackage{amsmath}
\usepackage{amssymb}
\usepackage{float}
\usepackage{tikz}
\usetikzlibrary{patterns}

\usepackage{multirow}
\usepackage{mathtools}
\usepackage{graphicx}
\usepackage{textcomp}
\usepackage{xcolor}
\usepackage{array}
\usepackage[FIGTOPCAP]{subfigure}

\usepackage{mathtools}
\usepackage{xcolor}
\usepackage{lineno}
\DeclarePairedDelimiter\floor{\lfloor}{\rfloor}

\newcolumntype{C}[1]{>{\centering\arraybackslash}m{#1}}
\newcolumntype{L}[1]{>{\flushleft\arraybackslash}m{#1}}

\title{Forecasting Financial Market Structure from Network Features using Machine Learning}

\author[1,5,*,+]{Douglas Castilho}
\author[2,+]{Th\'{a}rsis T. P. Souza}
\author[3]{Soong Moon Kang}
\author[4]{Jo\~{a}o Gama}
\author[1]{Andr\'{e} C. P. L. F. de Carvalho}
\affil[1]{Institute of Mathematics and Computer Sciences (ICMC), University of S\~{a}o Paulo (USP), S\~{a}o Carlos, Brazil}
\affil[2]{Department of Computer Science, University College London, Gower Street, London, WC1E 6BT, UK}
\affil[3]{School of Management, University College London, Gower Street, London, WC1E 6BT, UK}
\affil[4]{Institute for Systems and Computer Engineering, Technology and Science, University of Porto (UP), Porto, Portugal}
\affil[5]{Laboratory of Technology and Innovation (LATIN), 
Federal Institute of South of Minas Gerais (IFSULDEMINAS), Po\c{c}os de Caldas, Brazil}

\affil[*]{douglas.braz@ifsuldeminas.edu.br}

\affil[+]{these authors contributed equally to this work}


\begin{abstract}
We propose a model that forecasts market correlation structure from link- and node-based financial network features using machine learning. For such, market structure is modeled as a dynamic asset network by quantifying time-dependent co-movement of asset price returns across company constituents of major global market indices. We provide empirical evidence using three different network filtering methods to estimate market structure, namely Dynamic Asset Graph (DAG), Dynamic Minimal Spanning Tree (DMST) and Dynamic Threshold Networks (DTN). Experimental results show that the proposed model can forecast market structure with high predictive performance with up to $40\%$ improvement over a time-invariant correlation-based benchmark. Non pair-wise correlation features showed to be important compared to traditionally used pair-wise correlation measures for all markets studied, particularly in the long-term forecasting of stock market structure. Evidence is provided for stock constituents of the DAX30, EUROSTOXX50, FTSE100, HANGSENG50, NASDAQ100 and NIFTY50 market indices. Findings can be useful to improve portfolio selection and risk management methods, which commonly rely on a backward-looking covariance matrix to estimate portfolio risk.  

\end{abstract}
\begin{document}

\flushbottom
\maketitle
\thispagestyle{empty}

% ----------------------------------------------------------
% Introduction
% ----------------------------------------------------------
\section{Introduction}

% Covariance matrix is heavily used in finance. However, static analysis leads to issues.
Multi-asset financial analyses, particularly optimal portfolio selection and portfolio risk management, traditionally rely 
on the usage of a covariance matrix representative of market structure, which is commonly assumed to be time invariant. 
Under this assumption, however, non-stationarity~\cite{1742-5468-2012-07-P07025,Morales20136470} and long range memory~\cite{Cont2005} can lead to misleading conclusions and spoil the ability to explain future market structure dynamics. 

% Financial Networks have been used to analyze dynamic changes in market structure
Empirical analyses of networks in finance have been used successfully to study market structure dynamics, particularly to explain market interconnectedness from high-dimensional data~\cite{Mantegna1999,Tumminello10421,IORI2018637,marti2021review}. Under this approach, market structure is modeled as a network whose nodes represent different financial assets and edges represent one or many types of relevant relationships among those assets. There is a vast literature applying financial networks to descriptive analysis of market and portfolio dynamics, including market stability~\cite{morales2012dynamical}, information extraction~\cite{song2008analysis}, asset allocation~\cite{pozzi2013spread,Mineo2018} and dependency structure~\cite{Mantegna1999,Tumminello201040,musmeci2014clustering,song2012hierarchical,musmeci2017multiplex}. However, there is little research on the application of financial networks in market structure forecasting. Recent research on market structure inference makes use of information filtering networks to produce a robust estimate of the global sparse inverse covariance matrix~\cite{PhysRevE.94.062306}, achieving computationally efficient results. In a later study~\cite{SOUZA2019122343}, the authors forecast market structure based on a model that uses a principle of link formation by triadic closure in stock market networks. Spelta~\cite{spelta2017financial} proposed a method to predict abrupt market changes, inferring the future dynamics of stock prices by predicting future distances between them, using a tensor decomposition technique. Musmeci et al.~\cite{Musmeci2016} proposed a new tool to predict future market volatility using correlation-based stock networks, meta-correlation and logistic regression. Park et al.~\cite{park2020link} analyzed the evolution of Granger causality network of global currencies and proposed a link prediction method incorporating the squared eta of the causality directions of two nodes as the weight of future edges. To build the causality network, they used the effective exchange rate of $61$ countries and showed that the predictive capacity of their model outperforms other static methods for predicting links. Other related work~\cite{castilho2019weighted} proposed a model for predicting links in weighted financial networks, used to define input variables for the portfolio management problem, increasing the financial return of the investment.

In this article, financial market structure forecasting is formulated as a link prediction problem where we estimate the probability of adding or removing links in future networks. To tackle this problem, we developed a machine learning-based model that uses node- and link-specific financial network features to forecast stock to stock links based on past market structure. Applying machine learning algorithms in the decision-making process on stock markets is not a recent task~\cite{trippi1992neural}. An increasing number of applications have been created using machine learning-based models to predict the behavior of price time series~\cite{long2019deep}, volatility forecasting~\cite{liu2019novel}, sentiment analysis for investment~\cite{pagolu2016sentiment} and automatic trading rules~\cite{potvin2004generating}. This paper provides a set of empirical experiments designed to address the following research questions:

% ----------------------------------------------------------
% Research Questions
% ----------------------------------------------------------
\begin{enumerate}
    \item To what extent can dynamic financial networks help forecast stock market correlation structure? 
    \item How do financial network topology features perform relative to traditionally used pair-wise correlation data to forecast stock market structure?
    \item How does the predictability of market structure vary across multiple financial markets for the proposed models? 
    %\item How does the algorithm to model the market structure affects forecast results? \\
\end{enumerate}

%Despite new works focusing on predict stock networks, correlation and co-variance matrices, 
%To the best of our knowledge, this is the first work that uses financial network topological features and machine learning to forecast stock market structure. 
Findings can be particularly useful to improve portfolio selection and risk management, which commonly rely on a backward-looking correlation matrix to estimate portfolio risk. To the best of our knowledge, this is the first study that combines financial network features and machine learning to forecast stock market structure. %To clarify how the predictive performance is influenced by financial topological features
% Link prediction in the Granger causality network of the global currency market
% Interpretable Machine Learning for Diversified Portfolio Construction
% Matrix Evolutions: Synthetic Correlations and Explainable Machine Learning for Constructing Robust Investment Portfolios
The remainder of this paper is organized as follows: Section~\ref{sec:material-and-methods} describes the Materials and Methods used to provide the experiments; Section~\ref{sec:results-and-discussion}, which is the Results and Discussion, presents a descriptive analysis of the temporal stock networks and predictive analysis of market structure forecasting, and Section~\ref{sec:conclusion} draws the Conclusions.

% ----------------------------------------------------------
% Related Works
% ----------------------------------------------------------
%\input{literature}

% ----------------------------------------------------------
% Methodology and Data Set
% ----------------------------------------------------------
\section{Materials and Methods}
\label{sec:material-and-methods}

%Next, we briefly describe the main alternatives used in the experiments carried out to assess the predictive performance of the proposed method.

%\textbf{O TÍTULO DESTA SEÇÃO DEVE SER MÉTHODS OU PROPOSED METHOD?}
%Methods

\begin{figure*}[ht!]
	\centering
	\includegraphics[trim=0.7cm 16.4cm 0.7cm 2cm, clip=true, width=0.8\textwidth]{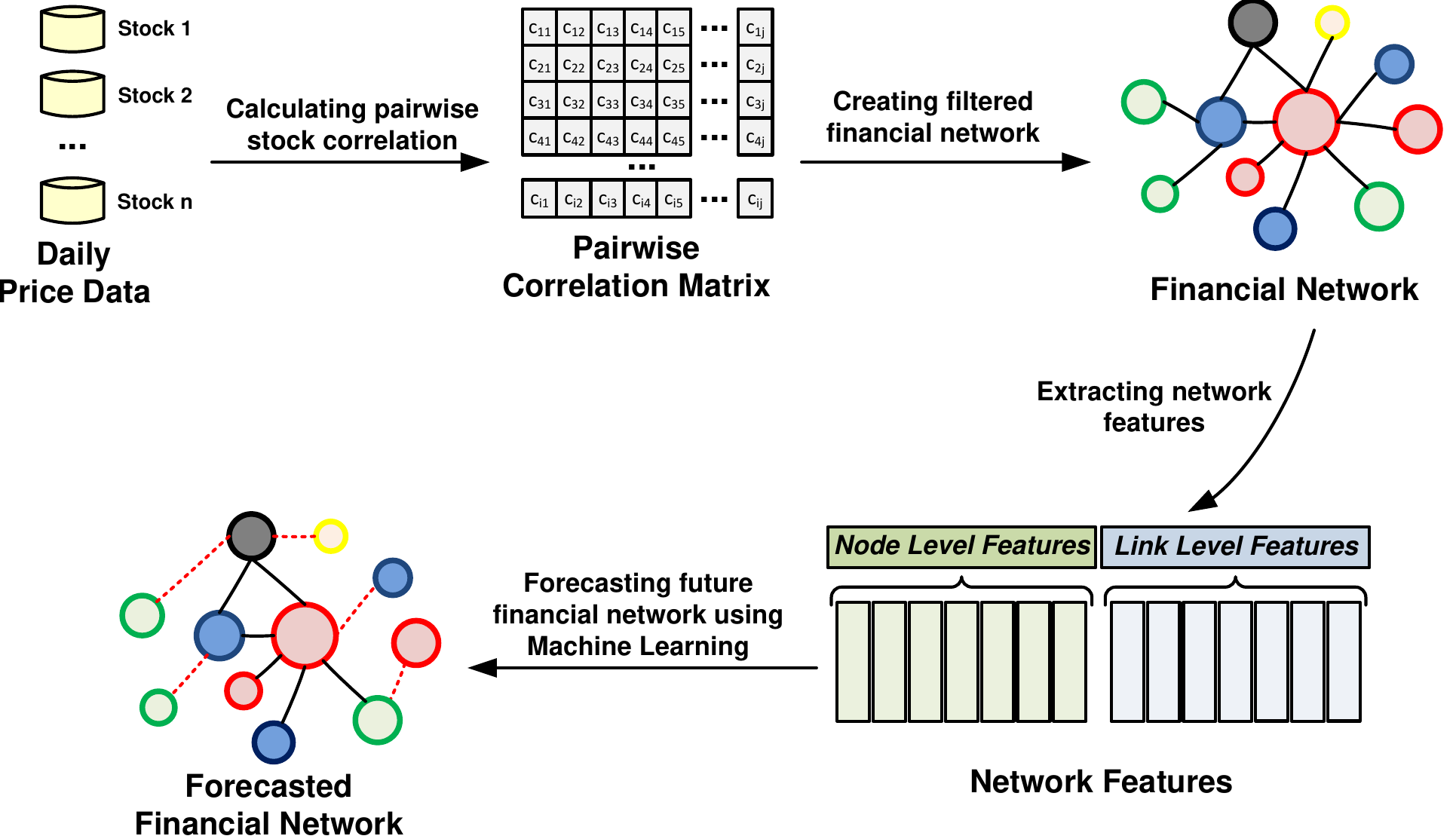}
	\caption{\textbf{Main steps of the methodology used in this work.} Based on daily asset closing prices of stocks constituents of a target stock market index, we calculate a pairwise correlation matrix and create a filtered financial network using three different network filtering algorithms. Given the financial network, we create a graph embedding by extracting network derived features at node and link levels. These features are used as input for a machine learning algorithm to forecast future financial networks.}
	\label{fig:methodology}
\end{figure*}

In this section, we describe the main steps of the proposed method to forecast market structure from financial network features using machine learning. Figure~\ref{fig:methodology} presents the methodology. 

Initially, we calculate the pairwise correlation matrix based on daily closing price series of assets. Given the correlation matrix, the market structure is modeled as a financial network by calculating the assets' distance matrix and applying a network filtering method. In this article, we evaluated three different network filtering methods to model financial market structure, described in Section~\ref{sec:dynamic-financial-networks}. We then extract a set of network features, used as input attributes for the machine learning model, by calculating node- and link-level network features, as described in Section~\ref{sec:network-based-features}. Finally, we applied a machine learning model, described in Section~\ref{sec:machine-learning-based-approach}, to forecast financial networks using network information itself as input. 

%We then create an embedding of this network by extracting node and link level features, described in Section~\ref{sec:network-based-features}. Finally, these features are used as input to a machine learning algorithm in order to forecast future financial network, described in Section~\ref{sec:machine-learning-based-approach}.

\subsection{Dynamic Financial Networks}
\label{sec:dynamic-financial-networks}

There are many methods in the literature to model financial market structure. Some of the most commonly used methods include correlation based networks and network filtering methods~\cite{marti2021review}. Network filtering methods allow prompt and temporal analysis of the market structure by exploring market data snapshots to model financial networks that represent the topology and the structure of the market. Using a rolling window approach, we can take snapshots in each time window of arbitrary length, allowing to explore temporal analysis of the market evolution~\cite{musmeci2014clustering}, also called as dynamic or temporal networks. Some examples of the most common methods include Minimal Spanning Tree approach~\cite{Mantegna1999}, the Planar Maximally Filtered Graph~\cite{Tumminello26072005}, the Directed Bubble Hierarchical Tree~\cite{song2012hierarchical}, asset graphs~\cite{onnela2003} and other approaches based on the threshold networks~\cite{onnela2004clustering}. 

In this study, we investigate three different network filtering methods to estimate financial market structure: \textit{(i)} Dynamic Asset Graph; \textit{(ii)} Dynamic Threshold Networks and \textit{(iii)} Dynamic Minimal Spanning Tree. We explore these three methods due to their importance for financial analysis, considering that there is a vast literature~\cite{onnela2003dynamic,onnela2003dynamics,meng2014systemic,mantegna1999introduction,onnela2003,Yang2008,onnela2004clustering} that uses these methods to study different characteristics of the structure of financial networks.

%The methods uniquely define selected edges in the filtered network and no topological constraint is imposed in the market structure~\cite{onnela2003}. 
These methods estimate an asset distance matrix through co-movement metrics of daily return prices. Let $P(t)$ be the closing price of an asset at day $t$. We consider assets' daily log-returns $R(t) = \log{P(t)} - \log{P(t-1)}$ that are calculated at time $t$. First, we calculate a distance matrix that measures the co-movement of daily log-returns~\cite{Mantegna1999}, defined as 

\begin{equation}
\label{eq:distance}
    D_{i,j}(t) = \sqrt{2(1 - \rho_t(i,j))}, \,
\end{equation}

%$$D_{i,j}(t) = \sqrt{2(1 - \rho_t(i,j))}, \, \forall (i,j)_t \in E_t,$$ 

\noindent where $\rho_t(i,j)$ is the Pearson's correlation coefficient between the time series of log-returns of assets $i$ and $j$ at time $t$, $\forall i,j \in V$, where $V$ is the set of assets. The distance matrix is constructed by dividing the returns time-series $R(t)$ into rolling windows of size $L$ trading days with $\delta T$ trading days between two consecutive windows (time-step). The choice of window width $L$ and window time-step $\delta T$ is arbitrary, and it is a trade-off between having an analysis that is either too dynamic or too smooth~\cite{tumminello2007correlation}. The smaller the window width and the larger the window steps, the more dynamic the data are. We report results for $L \in \lbrace 126, 252, 504\rbrace$ and $\delta T = 5$ trading days. A dynamic financial network is defined as a temporal network

\begin{equation}
    W = \langle V, E_1, \ldots, E_T  :  E_t \subseteq V \times V, \, \forall t \in \{1, \ldots, T\}  \rangle,
\end{equation}

%$$W = \langle V, E_1, \ldots, E_T  :  E_t \subseteq V \times V, \, \forall t \in \{1, \ldots, T\}  \rangle,$$ 

\noindent where vertices $i \in {V}$ correspond to assets of interest. For every pair $\langle i, j \rangle$ at time-window $t$, $\forall i,j \in {V}$ $\vert$ $ i \neq j$, there is a corresponding edge $(i,j)_t \in {E_t}$ and every edge has a weight $w_{i, j}(t) = D_{i, j}(t)$. Considering the distance matrix $D_{i,j}(t)$ previously defined, we can apply a network filtering method in order to create dynamic networks. The three evaluated methods in this work are described in the next sections.

\subsubsection{Dynamic Asset Graph (DAG) }

A Dynamic Asset Graph~\cite{onnela2003} is a type of filtered financial network modeled by first ranking edges in ascending order of weights $w_1(t), w_2(t), ... , w_{N(N-1)/2}(t)$. The resulting graph is obtained by selecting the edges with the strongest connections. The number of edges are, of course, arbitrary. Here, we select edges with weights in the top quartile, i.e., $w_1(t), w_2(t), ... , w_{\floor{N(N-1)/8}}(t)$, as proposed in Souza et al.~\cite{SOUZA2019122343}. The main idea of this method is to identify the smallest distances in the stock market.

\subsubsection{Dynamic Threshold Networks (DTN) }

Considering the distance matrix $D(t)$ defined in Equation (\ref{eq:distance}), we create a filtered adjacency matrix $A$ to construct the financial network using the following rules~\cite{Yang2008,onnela2004clustering}:

\begin{equation}
    A_{i,j}(t) = \left\{
  \begin{array}{lr}
    1, & \left| D_{i,j}(t) \right| \ge r_c\\
    0, & \left| D_{i,j}(t) \right| < r_c
  \end{array}
\right.
\end{equation}

\noindent where assets $i,j \in V$ and $\forall (i,j)_t \in E_t$. The critical value $r_c$ converts the matrix $D$ into an undirected network, whereby $A_{ij}(t) = 1$ and $A_{ij}(t) = 0$ represents the existence and absence of edges between $i$ and $j$ at time window $t$, respectively. We fixed the $r_c$ value in $0.65$ because for $r_c \leq 0.65$ the network characteristics are submerged in large fluctuations~\cite{Yang2008}. It is important to observe that the DTN method can produce disconnected graphs and the number of edges is dynamic. In general, the main goal of this method is to identify pairs of assets that are highly correlated and above the threshold $r_c$. This is different from DAG, where pairs with a correlation value lower than $r_c$ can be added to the network.

\subsubsection{Dynamic Minimal Spanning Tree (DMST) }

We create a Dynamic Minimal Spanning Tree~\cite{Mantegna1999} based on the smallest asset distance in the previous defined matrix $D(t)$. We use the Kruskal's Algorithm to identify the Minimal Spanning Tree (MST) in the fully connected graph $D$ at time $t$. The number of edges is fixed and calculated as $N - 1$, where $N$ is the number of assets. This method provides the smallest distance to interconnect the market, producing the minimal market structure to connect all assets.

\subsection{Machine Learning Based Approach}
\label{sec:machine-learning-based-approach}

In this section, we describe the proposed machine learning based approach to forecast stock market structure for a given market index. In this study, we address market structure forecasting as a network link prediction problem. Given snapshots of financial networks up to time $t$, we want to accurately predict the edges that will be present in the network at a given future time $t'$. We choose three times $t_0 < t < t'$ and provide an algorithm that accesses $W[t_0, t] = \langle V, E_{t_0}, \ldots, E_t \rangle$ to estimate the likelihood of edges to be present in $W[t']$, where $t' = t + h$ and $h = \lbrace 1, 2, \dots, 20 \rbrace$ trading weeks.

Similarity-based methods and classifier-based methods are two of the most common approaches for link prediction~\cite{martinez2016survey}. In similarity-based methods~\cite{liben2007link}, the algorithm assigns a connection weight $score(x, y)$ to pairs of nodes $\langle x, y \rangle$, based on the input graph $G$, and then produces a ranked list in decreasing order of $score(x, y)$. These algorithms can be viewed as computing a measure of proximity or ``similarity'' between nodes $x$ and  $y$. Common Neighbors, Jaccard Coefficient, Preferential Attachment, Adamic Adar, and Resource Allocation are among the most popular local indices (node-based). Katz, Leicht-Holme-Newman, Average Commute Time, Random Walk, and Local Path represents global indices (path-based). While the local indices are simple in computation, the global indices may provide more accurate predictions. 

\begin{figure*}[ht!]
	\centering
	\includegraphics[trim=4.5cm 29.5cm 11.9cm 0.9cm, clip=true, width=0.7\textwidth]{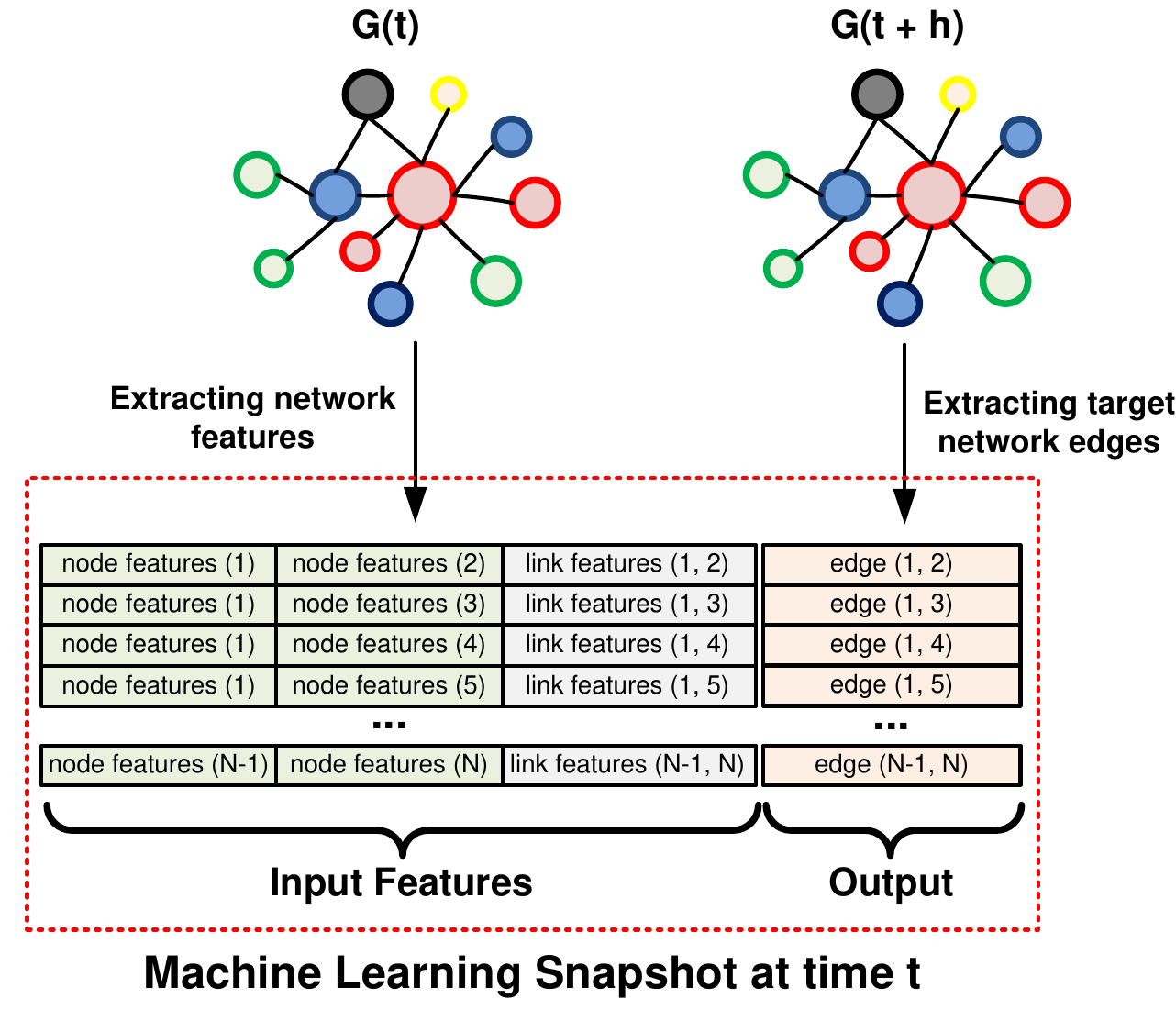}
	\caption{\textbf{Building the machine learning dataset.} We calculate features for each node ranging from $1$ to $N$, where $N$ is the number of assets. We applied a pairwise concatenation of node and link features as input variables for the link prediction, while edges on the network at time $t+h$ are used as the target variable, where $h$ is the number of trading weeks. }
	\label{fig:machine-learning}
\end{figure*}

In classifier-based methods, the link prediction is defined as a binary classification problem. Here, a feature vector is extracted for each pair of nodes and a $1/0$ label should be assigned based on the existence/not-existence of that link in the network. Any similarity-based method could form the required feature vector for a supervised learning method~\cite{al2006link}. Afterwards, any conventional supervised learning algorithm might be applied to train a supervised link predictor. In this article, we applied a classifier-based method to forecast the financial market structure. Our approach uses financial network features as input to a machine learning model in order to create a link prediction method, as presented in Figure~\ref{fig:machine-learning}. 

%\begin{figure*}[ht!]
%	\centering
%	\includegraphics[trim=4cm 10cm 2.3cm 0.9cm, clip=true, width=0.6\textwidth]{fig/methodology/machine-learning.pdf}
%	\caption{\textbf{Machine Learning Approach.} }
%	\label{fig:machine-learning}
%\end{figure*}

\begin{figure*}[ht!]
	\centering
	\includegraphics[trim=0.5cm 21.5cm 9.7cm 13cm, clip=true, width=0.9\textwidth]{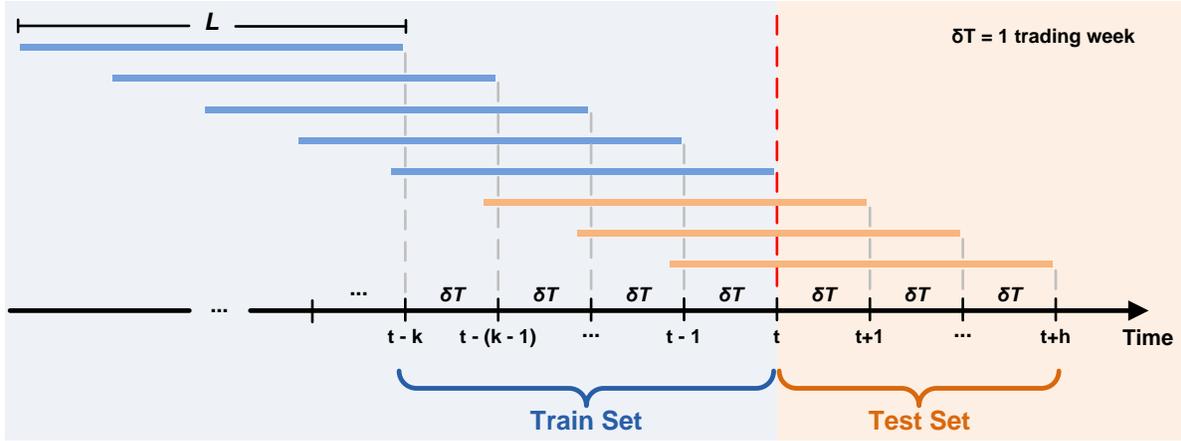}
	\caption{\textbf{Train and test sets used to induce the machine learning model.} Machine learning models were trained and tested using a rolling window approach. Considering $L$ as the size of the log-return time series and $t$ as current time, we create the train set using data from $t-k$ to $t-1$ and the test set using data from $t$. The target of the supervised learning is the network $G(t+h)$, where $h$ is the number of trading weeks. After training and testing the machine learning model, the time-step $\delta T$ is used to move the rolling window forward, in order to restart the process and re-train the machine learning model. The train set includes data from $1$ March $2005$ to $30$ May $2007$ and the test set has data from $30$ May  $2007$ to $18$ December $2019$.}
	\label{fig:machine-learning-train-test}
\end{figure*}

Figure~\ref{fig:machine-learning} presents the process used to create the machine learning database. Assuming $i$ and $j$ as two arbitrary nodes ranging from $1$ to $N$ and $t$ as the current time, an instance of the dataset used in the machine learning algorithm has the following predictive attributes: (a) $i$ node-level features; (b) $j$ node-level features; (c) $(i,j)$ link-level features. As previously described, the target of the supervised machine learning model is to forecast the existence of links in a network $G(t + h)$, where $h = 1, 2, \dots, 20 $ trading weeks. Figure~\ref{fig:machine-learning} presents an illustration of how we build instances to the machine learning model, exemplified as the snapshot at time $t$.

We split the dataset between train and test sets taking into account the temporal sequence of the data. The train set includes data produced in the period from $1$ March $2005$ to $30$ May $2007$ and the test set has data from $30$ May  $2007$ to $18$ December $2019$. Figure~\ref{fig:machine-learning-train-test} presents an illustration explaining how we created the train and test sets. Machine learning models were trained and tested using a rolling window approach. Considering $L$ as the size of the log-return time series, $t$ as current time and $t - k < t < t + h$, we create the train set using network features from $G(t - k)$, where $k = 1, 2, \ldots, 30 $. The test set contains data from the current network $G(t)$, in which $G(t + h)$ is the target, where $h = 1, 2, \dots, 20$ trading weeks. After training the machine learning model and testing it, we move the rolling window forward taking into account the time-step $\delta T = 5$ trading days ($1$ trading week) between two consecutive executions (see Supplementary Material, Section S.$1$ for further details).  

To assess the information rate that a machine learning model can extract from the features set, we applied the XGboost~\cite{chen2016xgboost} algorithm. In this experiment, the algorithm induces a predictive model for stock market structure forecasting. XGboost is a fast, highly effective, interpretable and widely used machine learning model. Further information regarding the experimental setup is described in the Supplementary Material, Section S.$2$.

\subsubsection{Network Features}
\label{sec:network-based-features}

As previously mentioned, we proposed an approach for market structure forecasting based on supervised machine learning. In order to provide information to train this supervised method, we extracted a set of network features at node- and link-level. These features are used as input to the machine learning model. We summarized the network features as follows:

\begin{itemize}
    \item \textbf{Node-Level Features} assess the position of a node within the overall structure of a given graph $G(V,E)$~\cite{oliveira2012overview}. Table~\ref{tab:nodefeatures} presents a set of node-level features related to node/stock $i \in V$ used as input to the machine learning model. 
    \item \textbf{Link-Level Features} examine both the contents and patterns of relationships in a given graph $G(V,E)$ and measure the implications of these relationships~\cite{oliveira2012overview}. Table~\ref{tab:linkfeatures} presents a set link-level features related to link $(i,j) \in E$ used as input to the machine learning model.
\end{itemize}

%\textbf{coisa número 1 - como a gente calcula as features; segunda coisa - para este paper, o que é relevante é a pair-wise correlation based features e non pair-wise correlation based features. por isso colocamos o asterísco, para que a gente compare os dois grupos de features. nem toda link-level feature é non pair-wise correlation feature. a gente pode dizer, for this non pair-wise correlation features, por se tratar específico do paper, a gente denominou essas features como }. nos table 1 e 2 background gray. TAbela 1 somente tem non pair-wise correlation features. 

Researchers in finance, particularly in portfolio management, commonly use asset correlation in important use cases, such as risk management. Given the importance of this information in financial analyses, we also explore them as input feature for market structure forecasting. However, we are interested in analyzing how topological information helps to forecast the market structure itself. For this reason, we separated the feature set into two distinct subsets. We labeled the two subsets according to their source of information: \textit{(i)} pair-wise correlation features, which are attributes based on asset correlation and not derived from any other network information, and \textit{(ii)} non pair-wise correlation features, which are attributes derived from the network topology. While pair-wise correlation features are traditionally used in financial analysis, the importance of non pair-wise correlation features to forecast market structure is a research question investigated in this work. Thus, we can compare their information gain in market structure forecasting. In Table~\ref{tab:nodefeatures}, all features are non pair-wise correlation attributes. In Table~\ref{tab:linkfeatures}, the pair-wise correlation features are marked with ($^\ast$).

%\textbf{Em finanças, em particular portfolio management, quando você vai calcular risco, a principal coisa que é utilizada é a correlação e a covariancia entre dois assets. Não é considerado a importância do asset naquela estrutura, qual a topologia do portfolio, não é considerado nenhuma característica estrutural do porfolio. o que é considerado é a correlação entre cada assets. O que esse paper vai falar é que considerar a topologia da rede não somente ajuda a prever uma estrutura do mercado melhor, mas de fato é mais importante que a correlação em si.} 

%In Table~\ref{tab:linkfeatures}, we also labeled the features according to their source of information, sorting them into two distinct classes: \textit{(i)} topological features, which are attributes derived from network information and \textit{(ii)} pair-wise correlation based, where features are based on correlation information and they are not derived from any information about the network.  While pair-wise correlation features are traditionally used in financial analysis, the importance of topological features to forecast stock market structure is a research question investigated in this work. 

%``Correlation Value'' and ``Link Existence in G(t)'' link-level features shown in Table~\ref{tab:linkfeatures} are labeled as pair-wise correlation based. The remaining attributes are labeled as topological features.

\begin{table}[ht!]
  \centering
  %\scriptsize
  \caption{\textbf{Node-Level Features:} Features were calculated to node $i$, $ \forall \text{ } i \in V$ for a given graph $G(V,E)$. Consider $N_i$ as the set of adjacent vertices (neighborhood) of node $i$. This set contains only non pair-wise correlation features.}
    \begin{tabular}{C{4.0cm} C{12.0cm}}
    \hline
    \multicolumn{1}{c}{\textbf{Name}} & \textbf{Definition} \\
    \hline
    \multicolumn{1}{C{4.0cm}}{\textit{Node Degree}} & $$deg(i) = \vert i \vert$$ \\
     \hline
    \multicolumn{1}{C{4.0cm}}{\textit{Weighted Node Degree}} & $$deg_w(i) = \sum_{j \in N_i }{w_{ < i, j > }},$$ where $w_{ < i, j > }$ is the weight of the edge $e(i,j)$ \\
    \hline
    \multicolumn{1}{C{4.0cm}}{\textit{Average Neighbor Degree}} & $$avg (i) = \frac{ \sum_{j \in N_i }{ \vert j \vert } }{ \vert i \vert}$$ \\
    \hline
    \multicolumn{1}{C{4.0cm}}{\textit{Propensity of $i$ to Increase its Degree}} & $$\gamma (i) = \frac{\vert i \vert}{deg_w(i)} $$ \\
    \hline
    \multicolumn{1}{C{4.0cm}}{\textit{Node Betweenness}} &  $$b(v) = \sum_{i,j \in V \setminus v}{ \frac{ \sigma_{ij}(v)}{\sigma_{ij}} },$$ where $\sigma_{ij}(v)$ is the number of shortest paths between $i$ and $j$ passing through node $v$ and $\sigma_{ij}$ the total number of shortest paths from $i$ to $j$\\ 
    \hline
    \multicolumn{1}{C{4.0cm}}{\textit{Node Closeness}} & $$nc(i) = \frac{n - 1}{ \sum_{j \in V \setminus i}{d(i,j)}}, $$ where $d(i,j)$ represents the distance between $i$ and $j$ and $n$ is the number of nodes in the graph\\
    \hline
    \multicolumn{1}{C{4.0cm}}{\textit{Node Eigenvector}} & $$ne(i) =  x_i \frac{1}{ \lambda } \sum_{j=1}^{n}{d_{ij}x_j}, $$ where $d_{ij}$ represents an entry of the adjacency matrix $D$ ($0$ or $1$), $\lambda$ denotes the largest eigenvalue, $x_i$ and $x_j$ denotes the centrality of node $i$ and $j$, respectively \\
    \hline
    \multicolumn{1}{C{4.0cm}}{\textit{Node Clustering Coefficient}} & $$cc(i) = \frac{2 \left| e_{jk} \right| }{\vert i \vert * ( \vert i \vert - 1 )} : j, k \in N_i, e_{jk} \in E$$ \\
    \hline
    \end{tabular}%
  \label{tab:nodefeatures}%
\end{table}%

\begin{table}[ht!]
  %\scriptsize
  \centering
  \caption{\textbf{Link-Level Features:} Features were calculated between nodes $i$ and $j$, $ \forall \text{ } (i, j) \in E$ for a given graph $G(V,E)$. Pair-wise correlation features are marked with $(^\ast)$, while the remaining are features based on non pair-wise correlation.
  %topological ones.
  Consider $N_i$ and $N_j$ as the set of adjacent vertices of node $i$ and $j$, respectively.}
    \begin{tabular}{C{4.0cm} C{12.0cm}}
    \hline
    \multicolumn{1}{c}{\textbf{Name}} & \textbf{Defition} \\
    \hline
    \multicolumn{1}{C{4.0cm}}{\textit{Link Existence in $G(t)$} ($^\ast$)} &         \begin{equation*}
        E(i,j) = \begin{cases}
        1, & \quad \text{exists link}, \\[0ex]
        0, & \quad \text{not exists link}.
        \end{cases}
        \end{equation*} \\
     \hline
    \multicolumn{1}{C{4.0cm}}{\textit{Correlation Value} ($^\ast$)} & $$C (i,j) = \rho_{ij},$$ where $\rho_{i,j}$ is the Pearson’s correlation coefficient between time series of log-returns of assets $i$ and $j$ \\
    \hline
    \multicolumn{1}{C{4.0cm}}{\textit{Common neighbors}} & $$CN (i,j) = \vert  N_i  \cap  N_j  \vert$$ \\
    \hline
    \multicolumn{1}{C{4.0cm}}{\textit{Jaccard Coefficient}} & $$JC (i,j) = \frac{\vert  N_i  \cap  N_j \vert}{\vert  N_i \cup N_j  \vert}$$ \\
    \hline
    \multicolumn{1}{C{4.0cm}}{\textit{Adamic-Adar Coefficient}} & $$AA (i,j) = \sum_{k \in N_i \cap N_j }{ \frac{1}{ \log{\vert N_k \vert} } }, $$ where $N_k$ is the set of adjacent vertices of node $k$\\
    \hline
    \multicolumn{1}{C{4.0cm}}{\textit{Sorenson-Dice Coefficient}} & $$SDC (i,j) = \frac{2 * \vert  N_i \cap  N_j  \vert}{\vert  i \vert  + \vert j \vert}$$ \\[0ex]
    \hline
    \multicolumn{1}{C{4.0cm}}{\textit{Edge Betweenness}} &  $$B (i,j) = \sum_{i,j \in V}{ \frac{ \sigma_{ij}(e)}{\sigma_{ij}} },$$ where $\sigma_{ij}(e)$ is the number of shortest paths between $i$ and $j$ crossing the edge $e$ and $\sigma_{i,j}$ is the total number of shortest paths from $i$ to $j$
    \\
    \hline
    \multicolumn{1}{C{4.0cm}}{\textit{Same Community}~\cite{blondel2008fast}} & 
        \begin{equation*}
        SC(i,j) = \begin{cases}
        1, & \quad \text{if $i$ and $j$ $\in$ same community}, \\
        0, & \quad \text{if $i$ and $j$ $\notin$ same community}.
        \end{cases}
        \end{equation*} \\
    \hline
    \multicolumn{1}{C{4.0cm}}{\textit{Preferential Attachment}} & $$PA(i,j) = \vert i \vert * \vert j \vert,$$ where $\vert i \vert$ and $\vert j \vert$ represent the node degree of vertex $i$ and $j$\\
    \hline
    \end{tabular}%
  \label{tab:linkfeatures}%
\end{table}%

\subsubsection{Model Evaluation}

We calculate the \textit{Area Under the ROC curve} (AUC) to evaluate the predictive performance of the link prediction methods. This metric is largely applied in binary classification and unbalanced problems and ranges from $0.5$ to $1$, where $0.5$ represents a random naive algorithm and $1$ represents the highest result. The AUC measure gives a summary metric for the algorithm’s overall performance with different prediction set sizes, while a detailed look into the shape of the ROC curve reveals the predictive performance of the algorithm at each prediction set size~\cite{huang2009time}. %Link prediction problem can be described as a binary classification problem and evaluating both presence or absence of links ($1/0$) is important in the market structure prediction problem~\cite{davis2006relationship}. 

%Further, it is known that the information given in graph $G(t)$ may contain the most critical temporal dependency information for predicting $G(t+h)$~\cite{huang2009time}. 

To verify the performance of the proposed method, we compared it against the following similarity-based methods commonly used in literature for link prediction, separated into three categories as follows~\cite{mutlu2019review}: 

\begin{enumerate}
    \item \textbf{Local Similarity Methods}
    \begin{itemize}
        \item Common Neighbors~\cite{liben2007link} (CN): This is a simple and effective link prediction method based on common neighbors shared by two nodes. Pairs of nodes with high number of common neighbors tend to establish a link;

        \item Preferential Attachment~\cite{barabasi1999emergence} (PA): This method defines that new links are formed between nodes with higher degrees rather than nodes with lower degrees;

        \item Jaccard Coefficient~\cite{mutlu2019review} (JC): This method is based on similarity Jaccard's coefficient, taking into account the number of common neighbors shared by two nodes, but normalized by the total number of neighbors of both nodes;

        \item Adamic-Adar~\cite{adamic2003friends} (AA): This method is also based on common neighbors shared by two nodes. Instead of using the raw number of common neighbors as CN, it is defined using the sum of the inverse of the logarithmic degree of each shared neighbor.  
    \end{itemize}
    
    \item \textbf{Quasi-Local Similarity Method}
    \begin{itemize}
        \item Local Path Index~\cite{zhou2009predicting} (LP): Similar to CN, this method uses information from the next $2$ and $3$ nearest neighbors instead of using only information of the neighbors shared by two nodes.
    \end{itemize}
    
    \item \textbf{Global Similarity Method}
    \begin{itemize}
        \item Random Walk with Restart~\cite{brin1998anatomy} (RW): Based on Random Walk, 
        it is a special case of following the Markov chain, starting from a given node and randomly reaching a selected neighbor. The restart looks for the probability of a random walker starting from node $x$ visits node $y$ and comes back to the initial state node $x$~\cite{mutlu2019review}.
    \end{itemize}

\end{enumerate}

In addition to these methods, we included a naive Time Invariant (TI) baseline benchmark in our experiments. This algorithm uses the link occurrence in graph $G(t)$ as the prediction of link occurrence in graph $G(t+h)$, assuming that market structure is time invariant. This assumption is traditionally used in risk management algorithms, which commonly rely on a backward-looking covariance matrix to estimate portfolio risk~\cite{markowitz1952,SOUZA2019122343}.  

%In effect, the link prediction problem asks: to what extent can the evolution of a social network be modeled using features intrinsic to the network itself?
\subsection{Market Data}

In this study, we used data from six different stock market indices spread across the American, European and Asian markets. The stock indices were chosen to measure the performance of the proposed approach in different scenarios, given the diversity of the stock markets. Moreover, it is important to mention that they represent the stock market of the region or country where they are listed. We considered the following indices and associated countries/regions:

\begin{itemize}
    \item \textbf{DAX30} (Germany): This is a stock market index that consists of the $30$ largest and most liquid German companies trading on the Frankfurt Stock Exchange. 
    \item \textbf{EUROSTOXX50} (Eurozone): This is a list of the $50$ companies that are leaders in their respective sectors from eleven Eurozone countries, including Austria, Belgium, Finland, France, Germany, Ireland, Italy, Luxembourg, the Netherlands, Portugal and Spain. 
    \item \textbf{FTSE100} (United Kingdom): This is an index listed in the London Stock Exchange. The Financial Times Stock Exchange Index (FTSE) is Britain's main asset indicator, managed by the independent organization and calculated based on the $100$ largest companies in the United Kingdom.
    \item \textbf{HANGSENG50} (Hong Kong): This is an index listed in the Stock Exchange of Hong Kong. This stock market index has the $50$ constituent companies with the highest market capitalization. It is the main indicator of the market performance in Hong Kong.
    \item \textbf{NASDAQ100} (United States): This is an index composed of the $100$ non-financial largest companies listed in NASDAQ.
    \item \textbf{NIFTY50} (India): This is a stock market index listed in the National Stock Exchange of India based on the $50$ largest Indian companies.
\end{itemize}

Each financial index has a daily price time series for each one of its constituent stocks. Price time series are constructed using daily closing prices collected from \textit{Thomson Reuters}. The list of company constituents of each stock market index is not static and may change over time. In this article, we only consider companies that were part of the underlying indices across the entire period analyzed, as commonly used in other studies, when node prediction is out-of-scope~\cite{SOUZA2019122343, 10.1007/978-3-030-22744-9_27}. %Otherwise, we also should to predict whether a node will appear/disappear in future networks. 
We consider prices ranging from $1$ March $2005$ to $18$ December $2019$.

% ----------------------------------------------------------
% Experiments and Results
% ----------------------------------------------------------
\section{Results and Discussion}
\label{sec:results-and-discussion}

In this section, we present the experimental results for financial market structure forecasting. Initially, we present a set of descriptive analyses on evolution of financial networks and a brief discussion about the impact of the different network filtering methods in the financial market structure. Afterwards, we present a set of predictive analyses related to the machine learning approach and the benchmark methods. Finally, we present a discussion about the interpretability of the machine learning models. 

\subsection{Descriptive Analysis}

We present a set of descriptive analyses of temporal financial networks created across different market indices. Our first descriptive analysis describes financial network persistence, considering $L = 252$ trading days to create each graph (results regarding $L \in \lbrace 126, 504 \rbrace$ trading days can be found in Supplementary Material, Section S.$3$). This analysis allows us to measure how the financial networks change their structure over time. We estimate the network persistence by calculating pair-wise network similarity between $G(t)$ and $G(t')$ using the Jaccard Distance, defined as follows:
\begin{equation}
    sim (G(t), G(t')) = \frac{ \left| G(t) \cap G(t')\right|}{\left| G(t) \cup G(t)\right|},
\end{equation}

\noindent where $t$ and $t'$ range from $12$ May $2006$ to $18$ December $2019$.

\begin{figure*}[t!]
	\centering
	\subfigure[DAX30]{\includegraphics[trim=0cm 3.8cm 0.1cm 0cm, clip=true, width=0.30\textwidth]{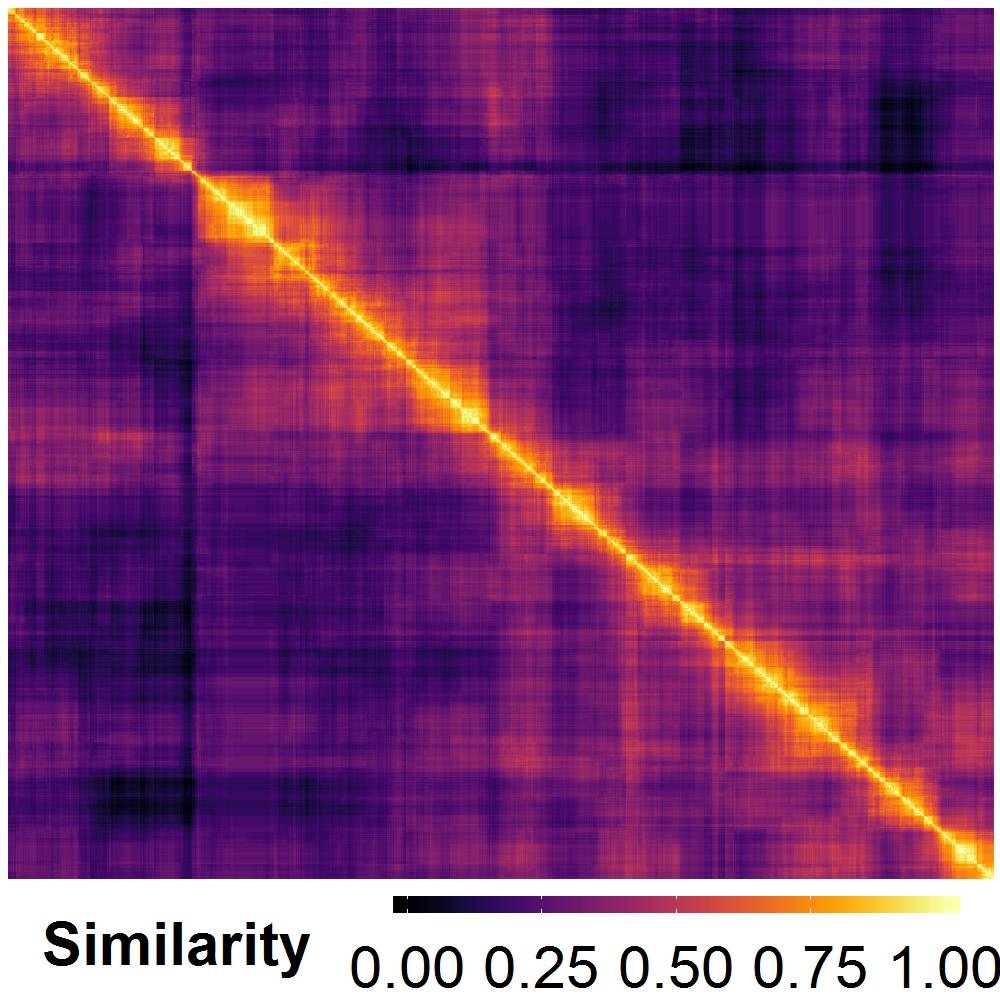}}
	\subfigure[EUROSTOXX50]{\includegraphics[trim=0cm 3.8cm 0.1cm 0cm, clip=true, width=0.30\textwidth]{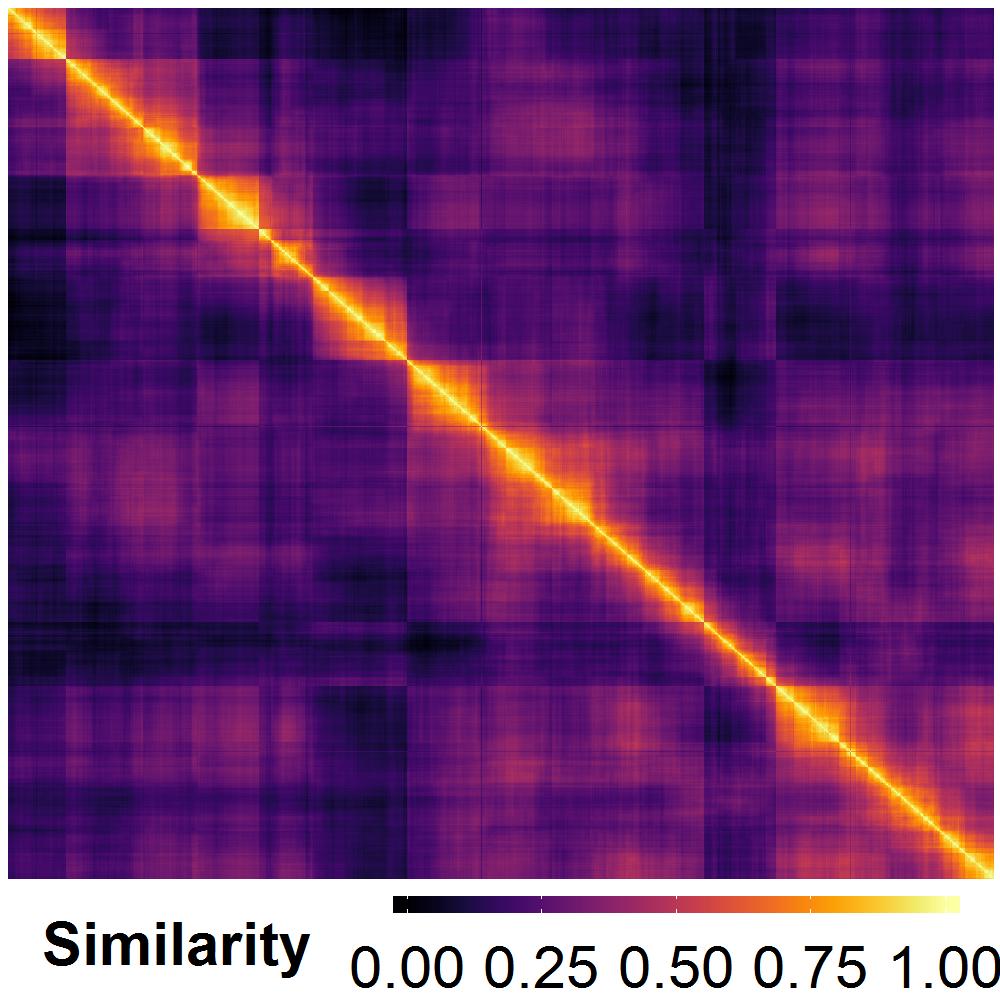}}
	\subfigure[FTSE100]{\includegraphics[trim=0cm 3.8cm 0.1cm 0cm, clip=true, width=0.30\textwidth]{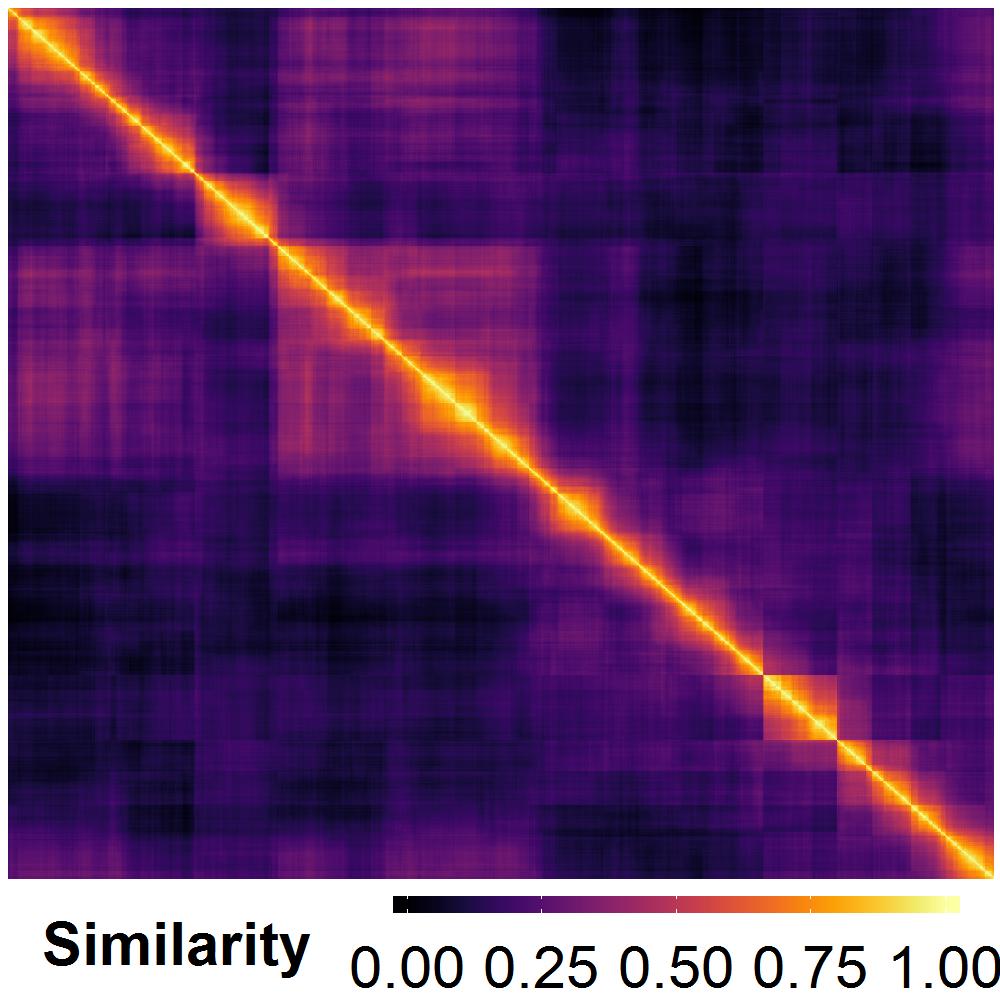}}
	\subfigure[HANGSENG50]{\includegraphics[trim=0cm 3.8cm 0.1cm 0cm, clip=true, width=0.30\textwidth]{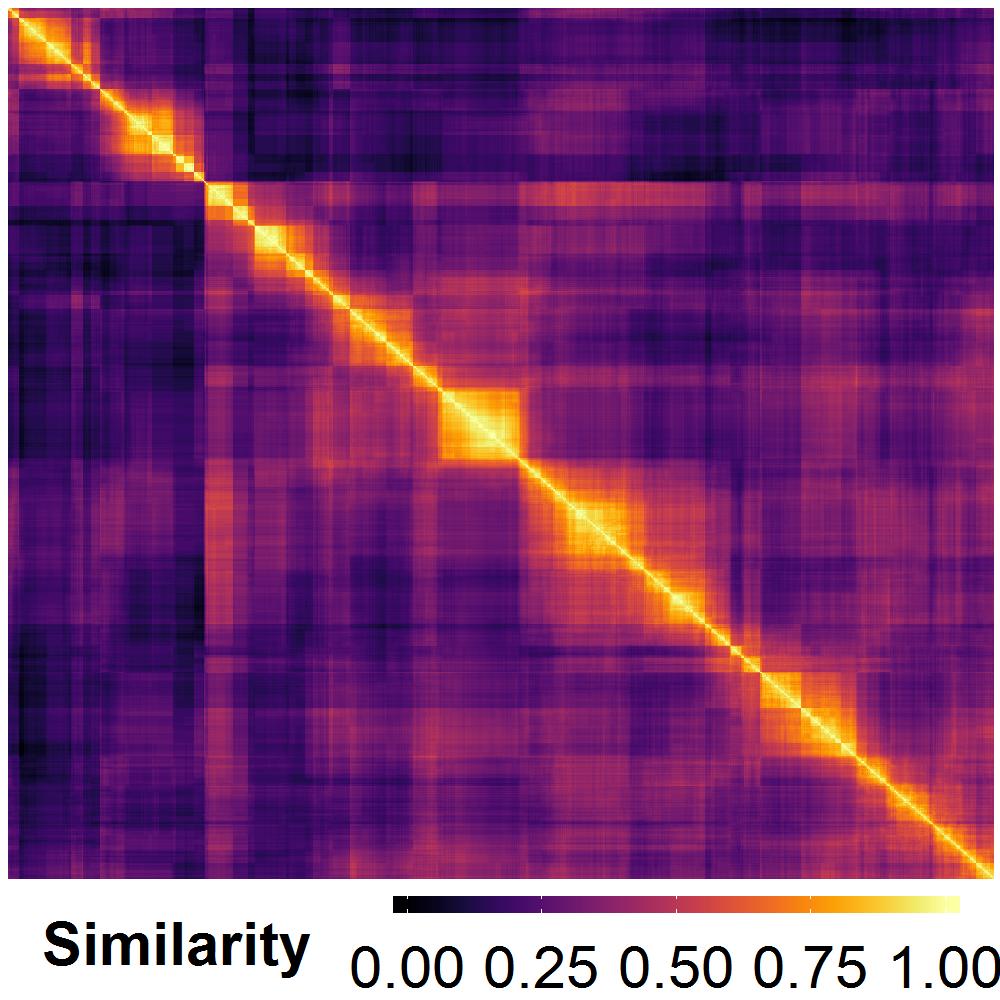}}
	\subfigure[NASDAQ100]{\includegraphics[trim=0cm 0.2cm 0.1cm 0cm, clip=true, width=0.30\textwidth]{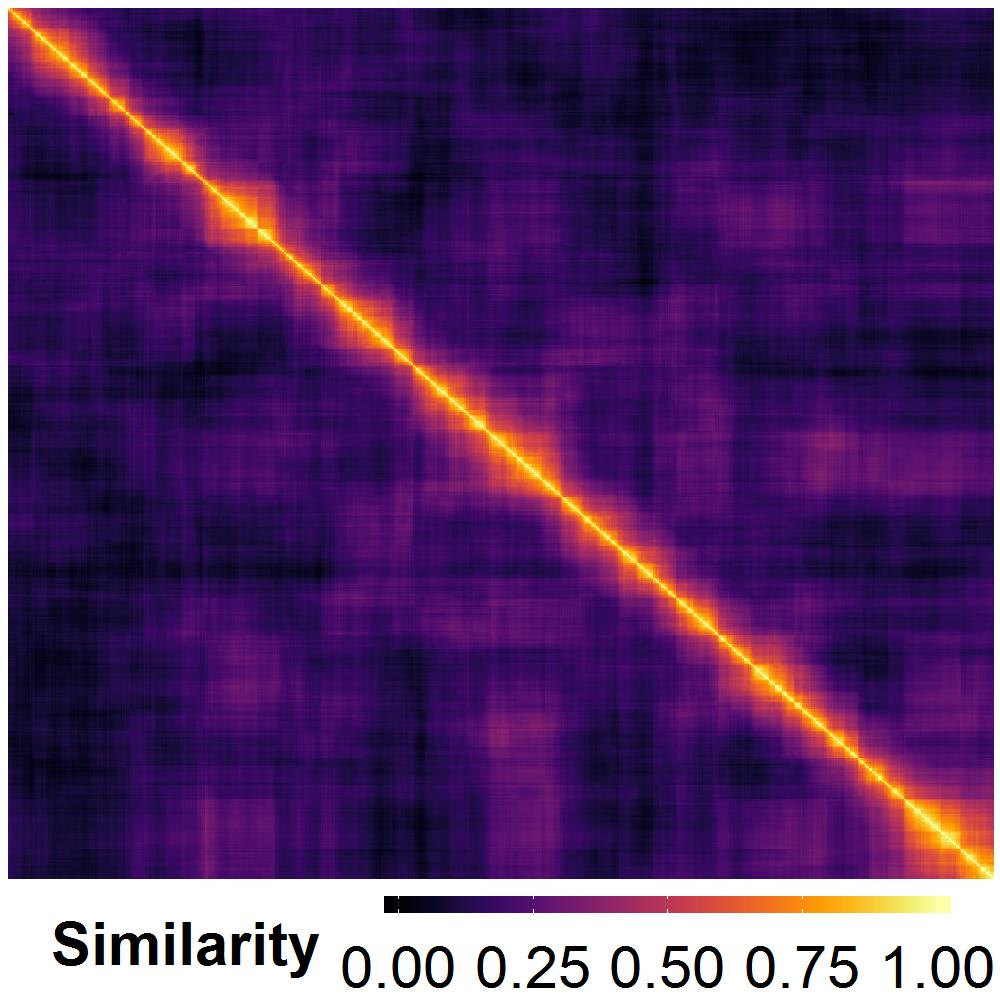}}
	\subfigure[NIFTY50]{\includegraphics[trim=0cm 3.8cm 0.1cm 0cm, clip=true, width=0.30\textwidth]{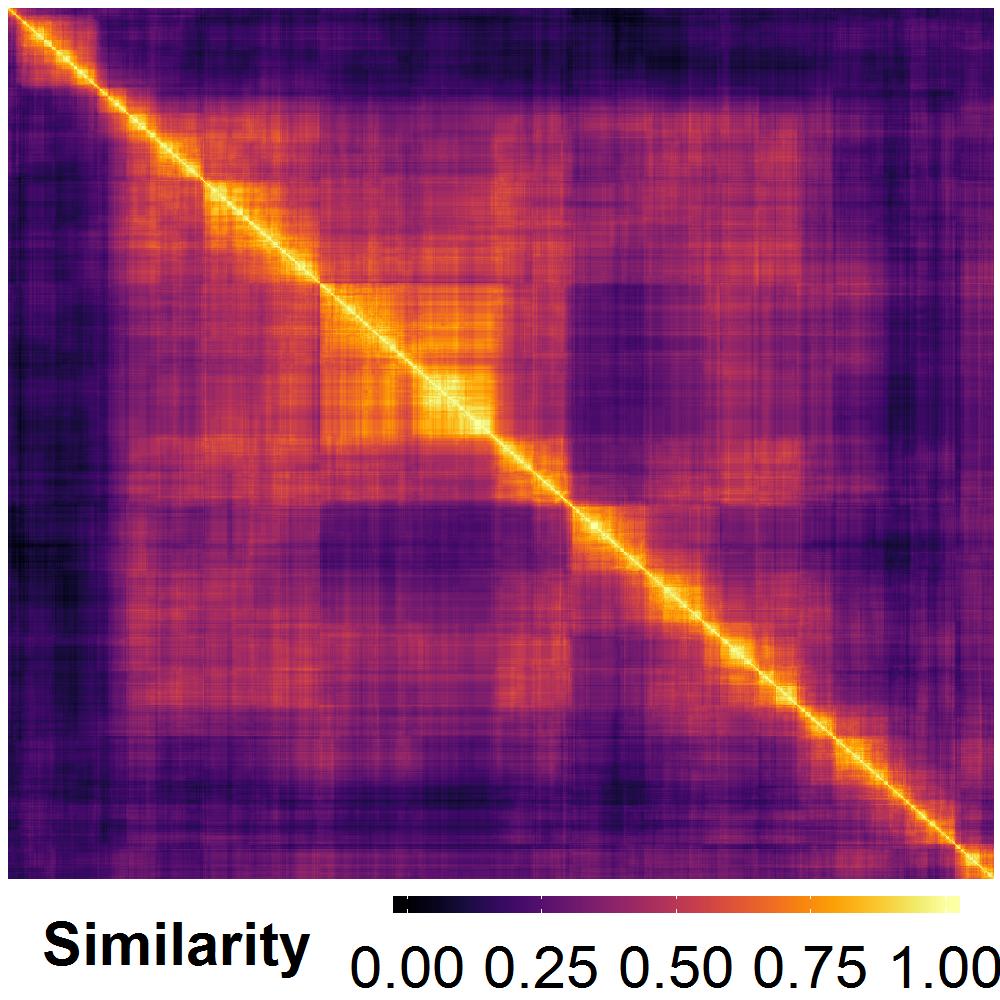}}
	\caption{\textbf{DAG - Cross-similarity matrix for each market index.} We calculate the pair-wise Jaccard Distance across all financial networks $G(t)$ and $G(t')$ ranging from $12$ May $2006$ to $18$ December $2019$, related to a given market index. For each market index figure, the first network on $12$ May $2006$ is represented in the top-left and the last network on $18$ December $2019$ in the bottom-right corner of each individual figure.}
	\label{fig:cross-similarity-dag}
\end{figure*}

\begin{figure*}[h!]
	\centering
	\subfigure[DAX30]{\includegraphics[trim=0cm 4.2cm 0.1cm 0cm, clip=true, width=0.30\textwidth]{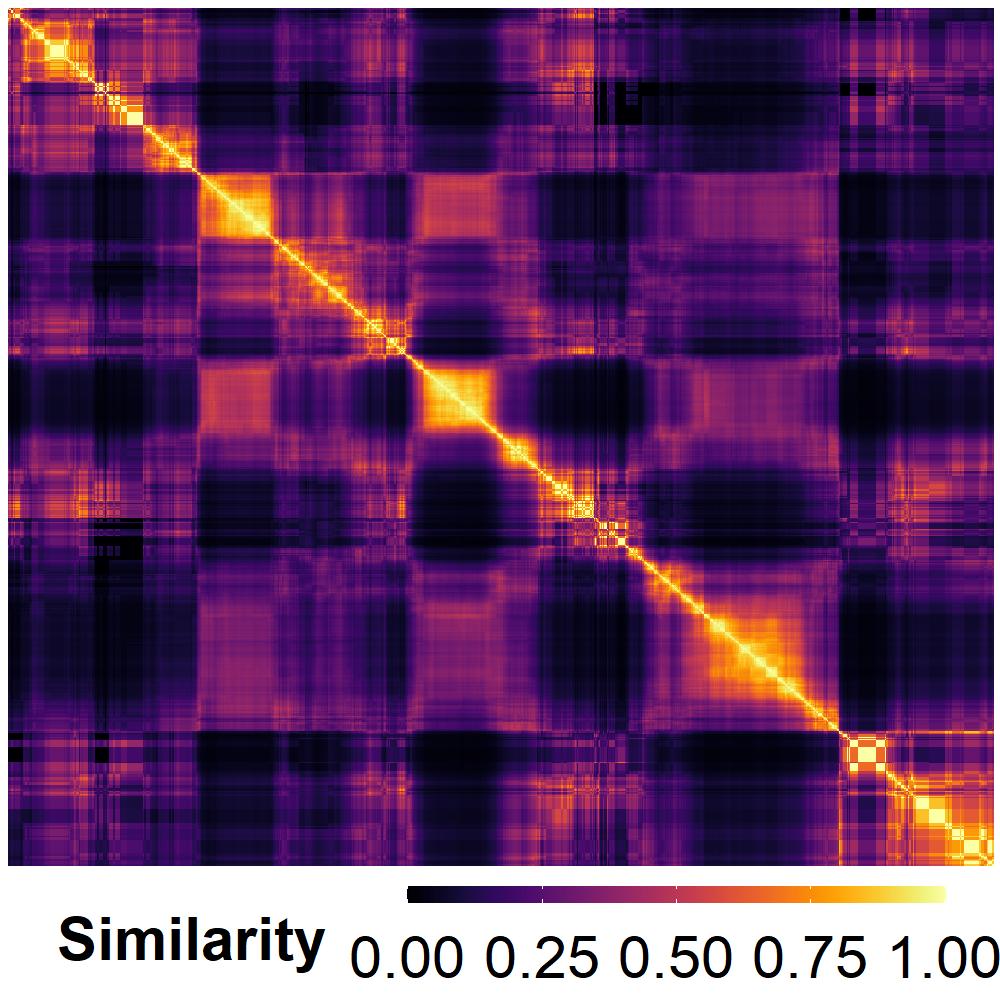}}
	\subfigure[EUROSTOXX50]{\includegraphics[trim=0cm 4.2cm 0.1cm 0cm, clip=true, width=0.30\textwidth]{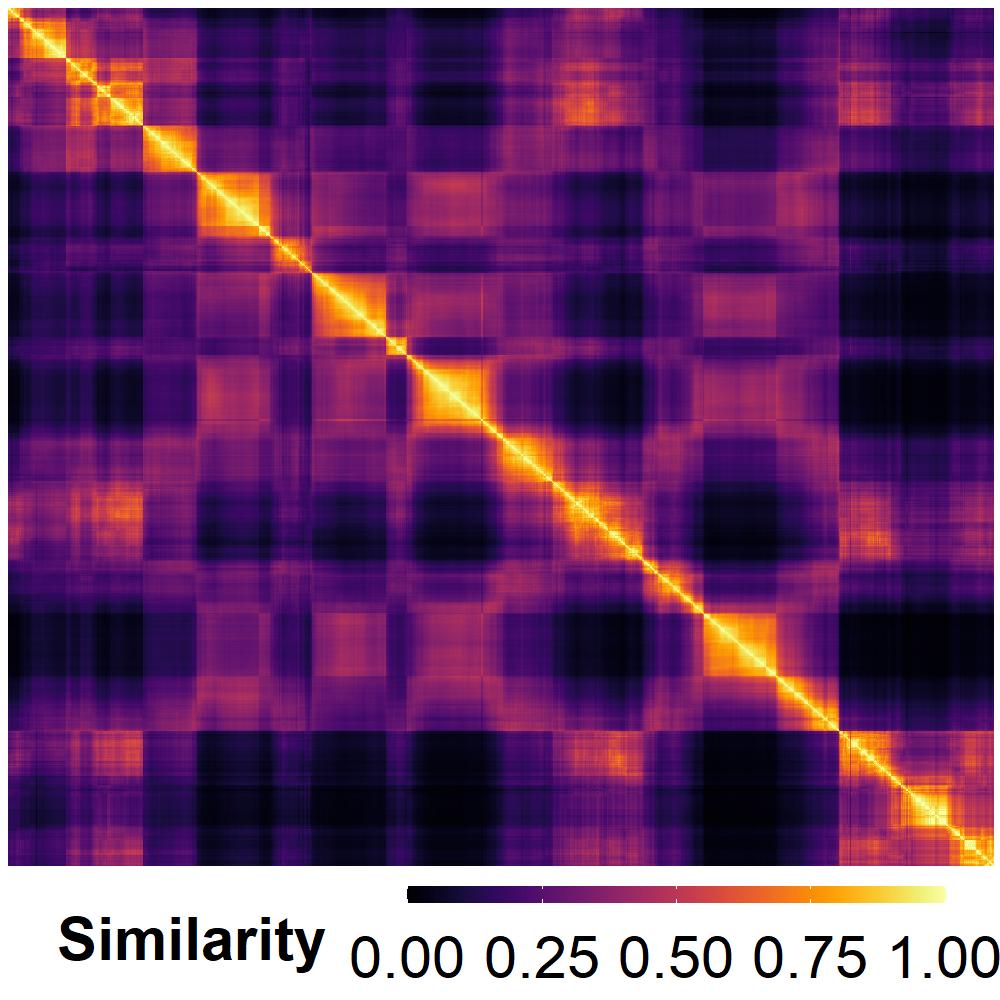}}
	\subfigure[FTSE100]{\includegraphics[trim=0cm 4.2cm 0.1cm 0cm, clip=true, width=0.30\textwidth]{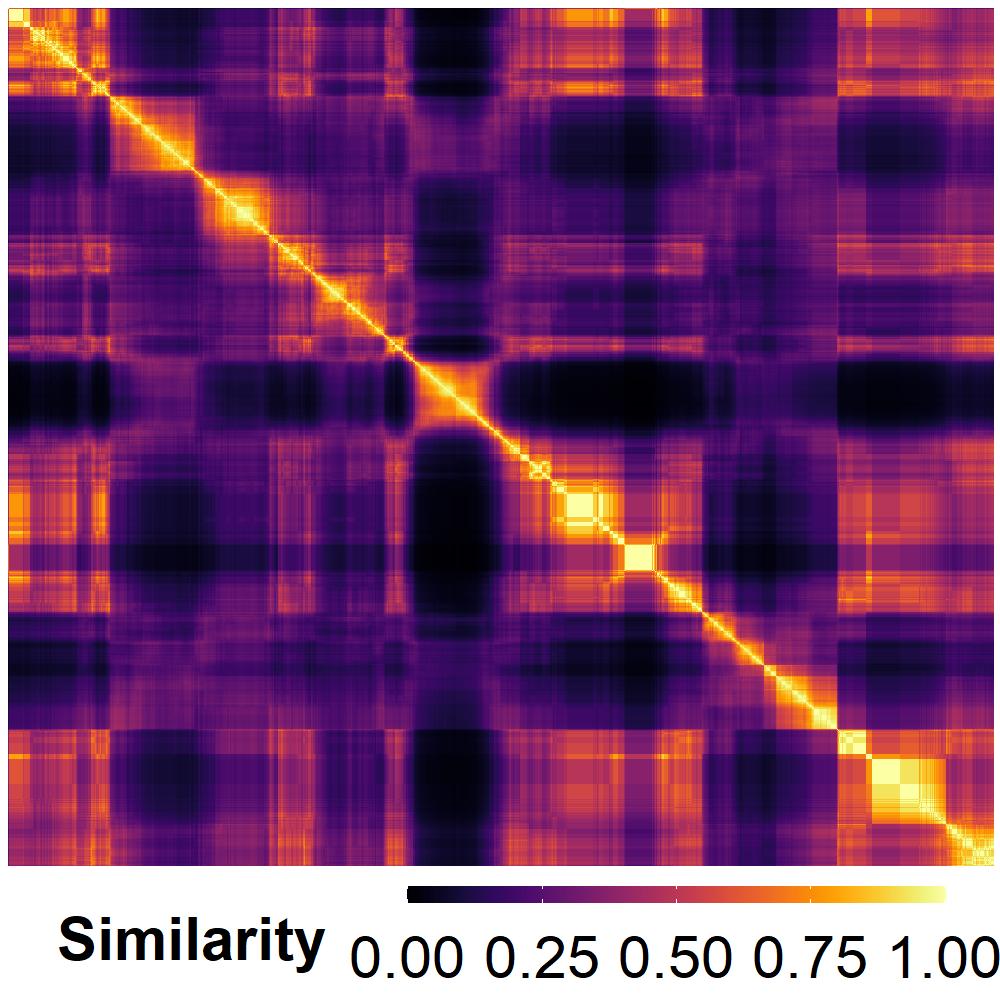}}
	\subfigure[HANGSENG50]{\includegraphics[trim=0cm 4.2cm 0.1cm 0cm, clip=true, width=0.30\textwidth]{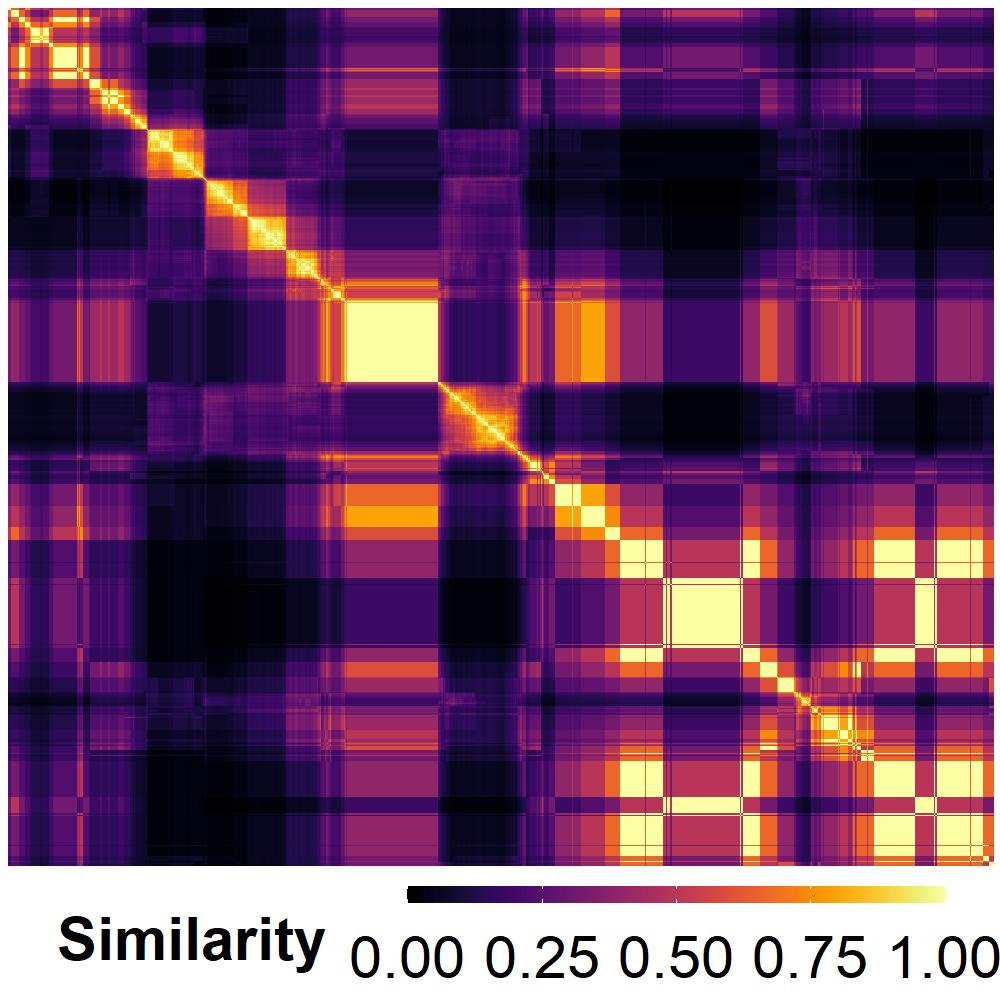}}
	\subfigure[NASDAQ100]{\includegraphics[trim=0cm 0.2cm 0cm 0cm, clip=true, width=0.30\textwidth]{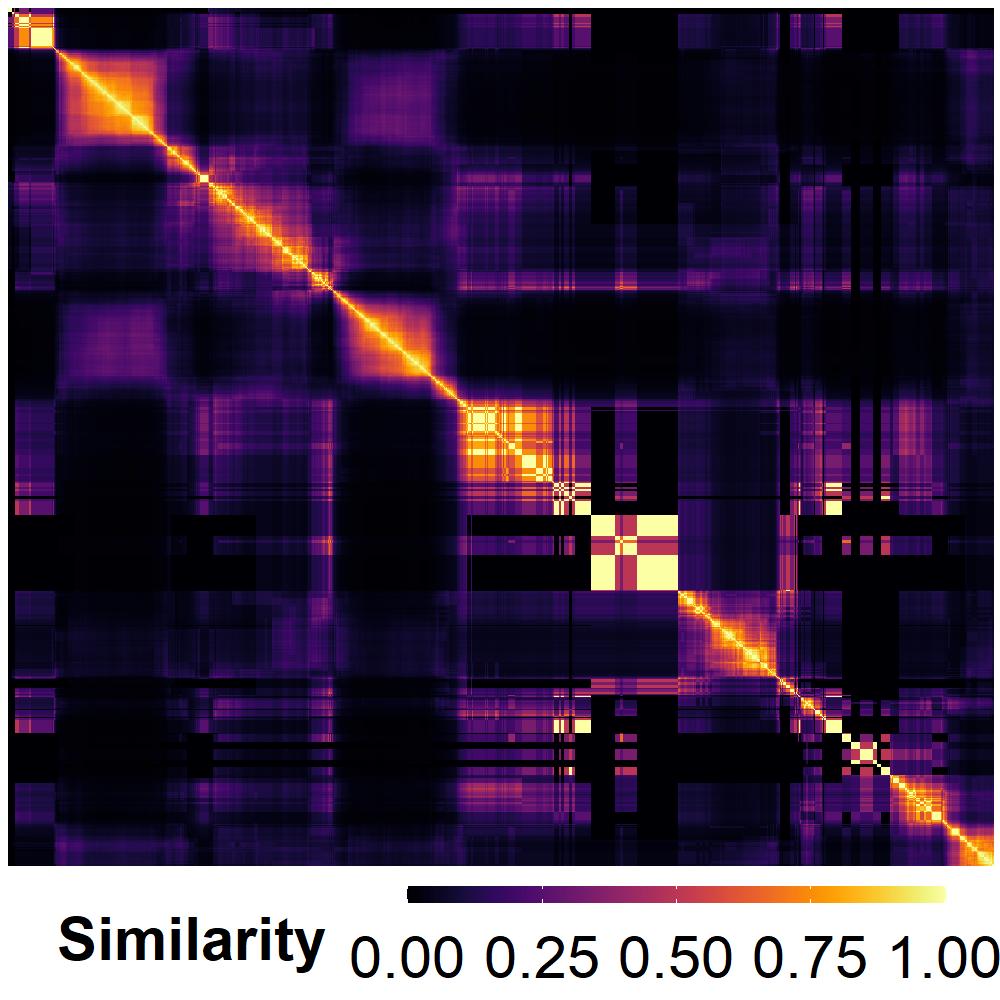}}
	\subfigure[NIFTY50]{\includegraphics[trim=0cm 4.2cm 0.1cm 0cm, clip=true, width=0.30\textwidth]{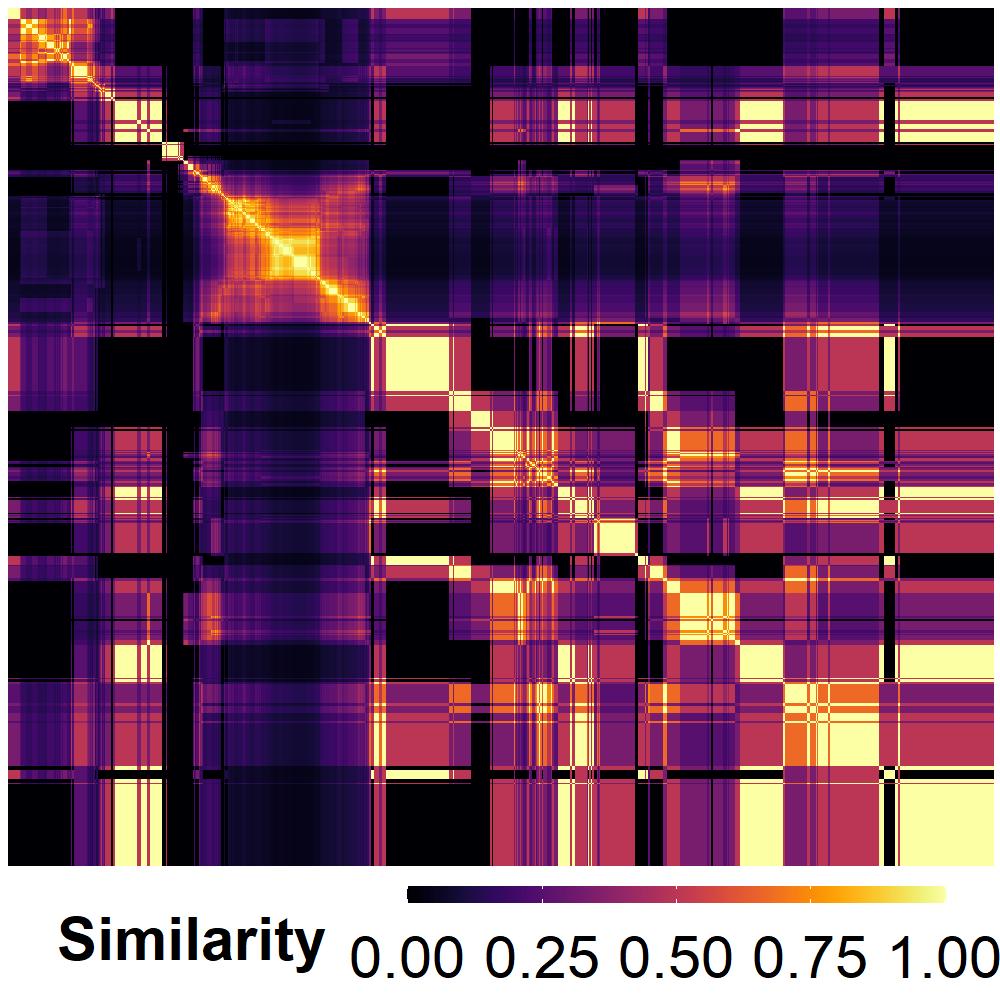}}
	\caption{\textbf{DTN - Cross-similarity matrix for each market index.} We calculate the pair-wise Jaccard Distance across all financial networks $G(t)$ and $G(t')$ ranging from $12$ May $2006$ to $18$ December $2019$, related to a given market index. For each market index figure, the first network on $12$ May $2006$ is represented in the top-left and the last network on $18$ December $2019$ in the bottom-right of each individual figure.}
	\label{fig:cross-similarity-dtn}
\end{figure*}

\begin{figure*}[h!]
	\centering
	\subfigure[DAX30]{\includegraphics[trim=0cm 4.2cm 0.1cm 0cm, clip=true, width=0.30\textwidth]{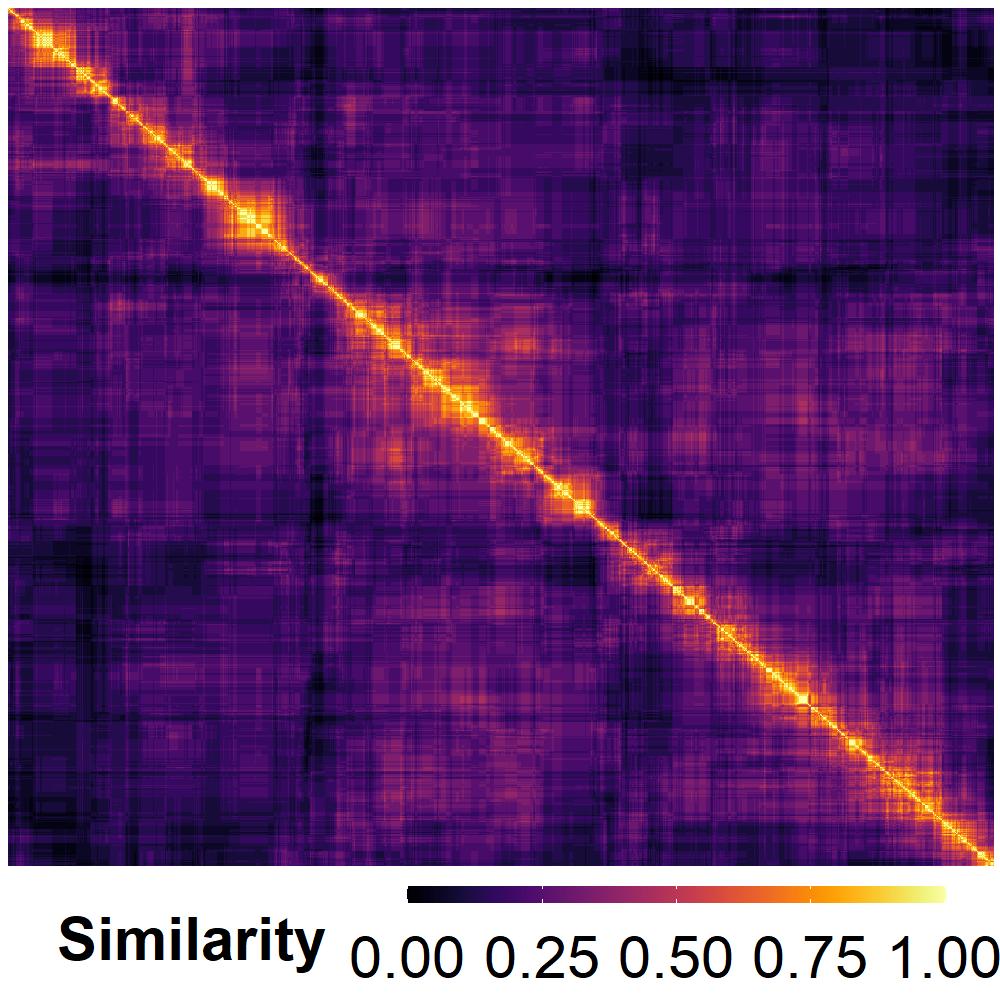}}
	\subfigure[EUROSTOXX50]{\includegraphics[trim=0cm 4.2cm 0.1cm 0cm, clip=true, width=0.30\textwidth]{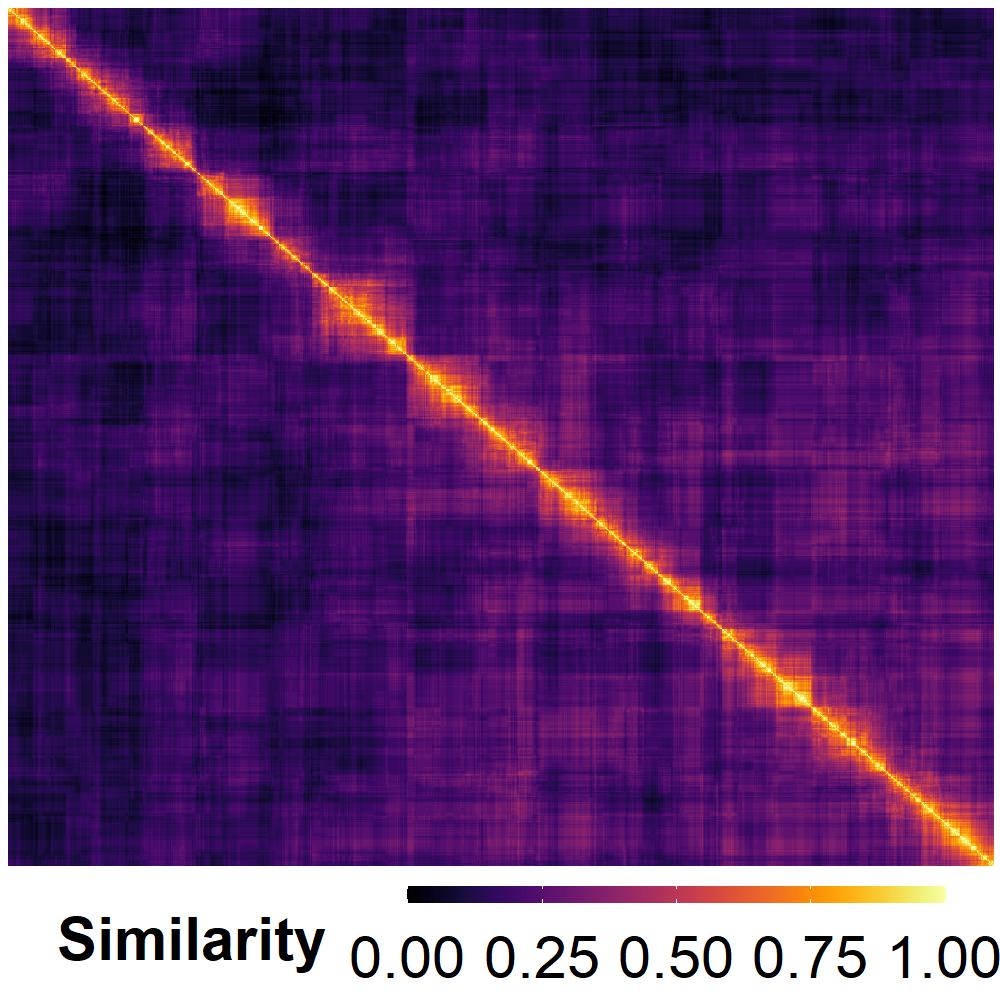}}
	\subfigure[FTSE100]{\includegraphics[trim=0cm 4.2cm 0.1cm 0cm, clip=true, width=0.30\textwidth]{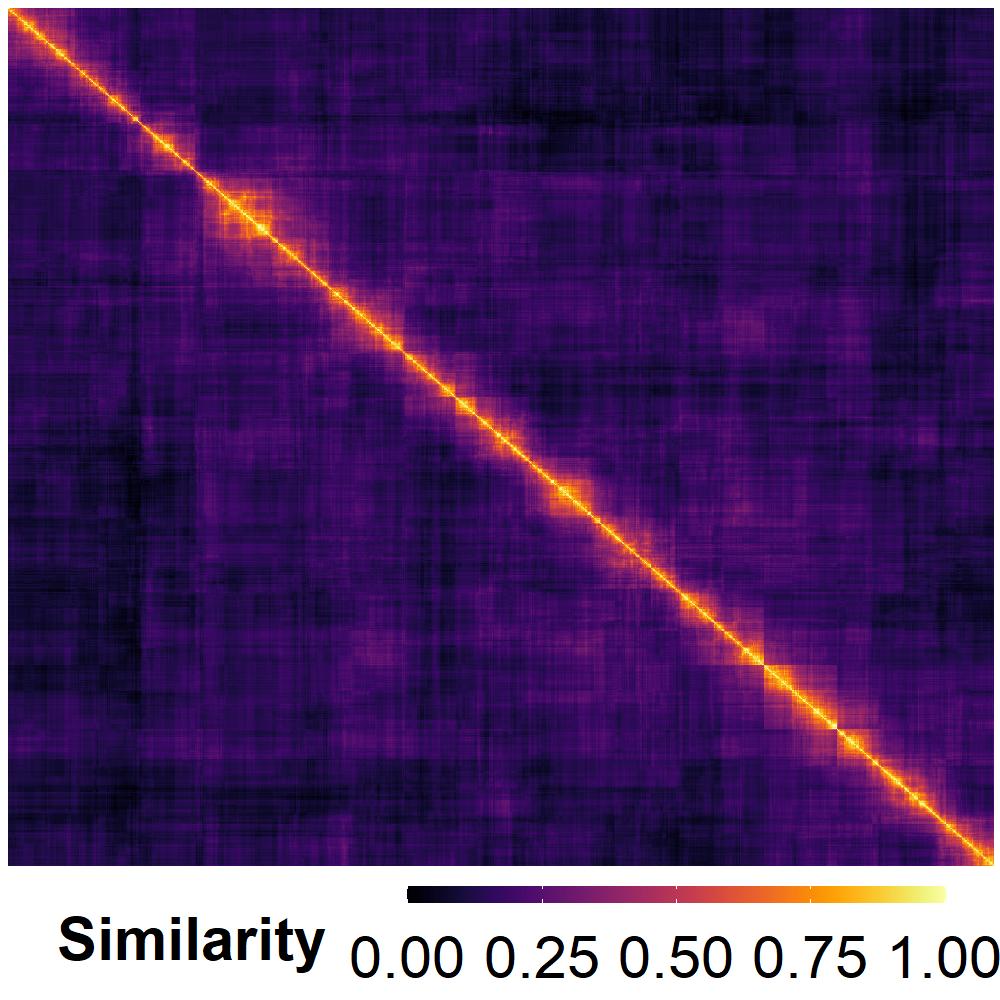}}
	\subfigure[HANGSENG50]{\includegraphics[trim=0cm 4.2cm 0.1cm 0cm, clip=true, width=0.30\textwidth]{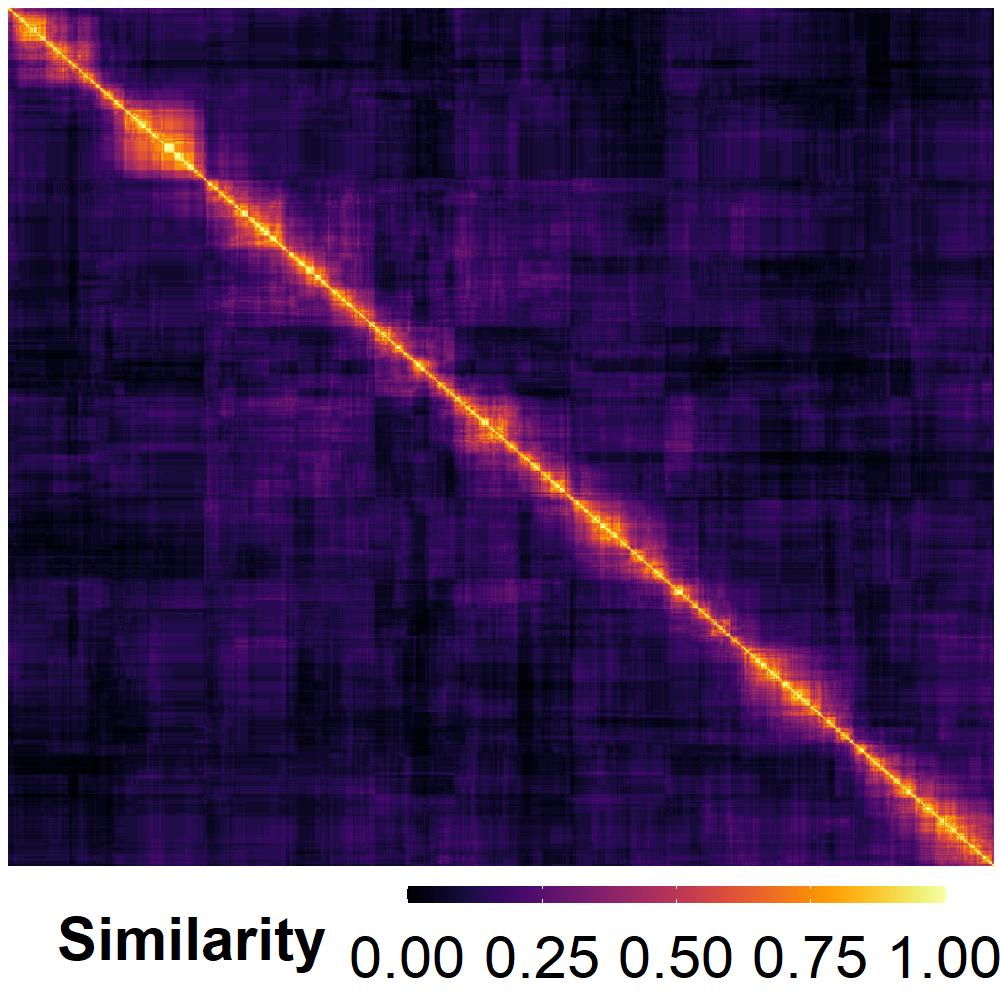}}
	\subfigure[NASDAQ100]{\includegraphics[trim=0cm 0.2cm 0cm 0cm, clip=true, width=0.30\textwidth]{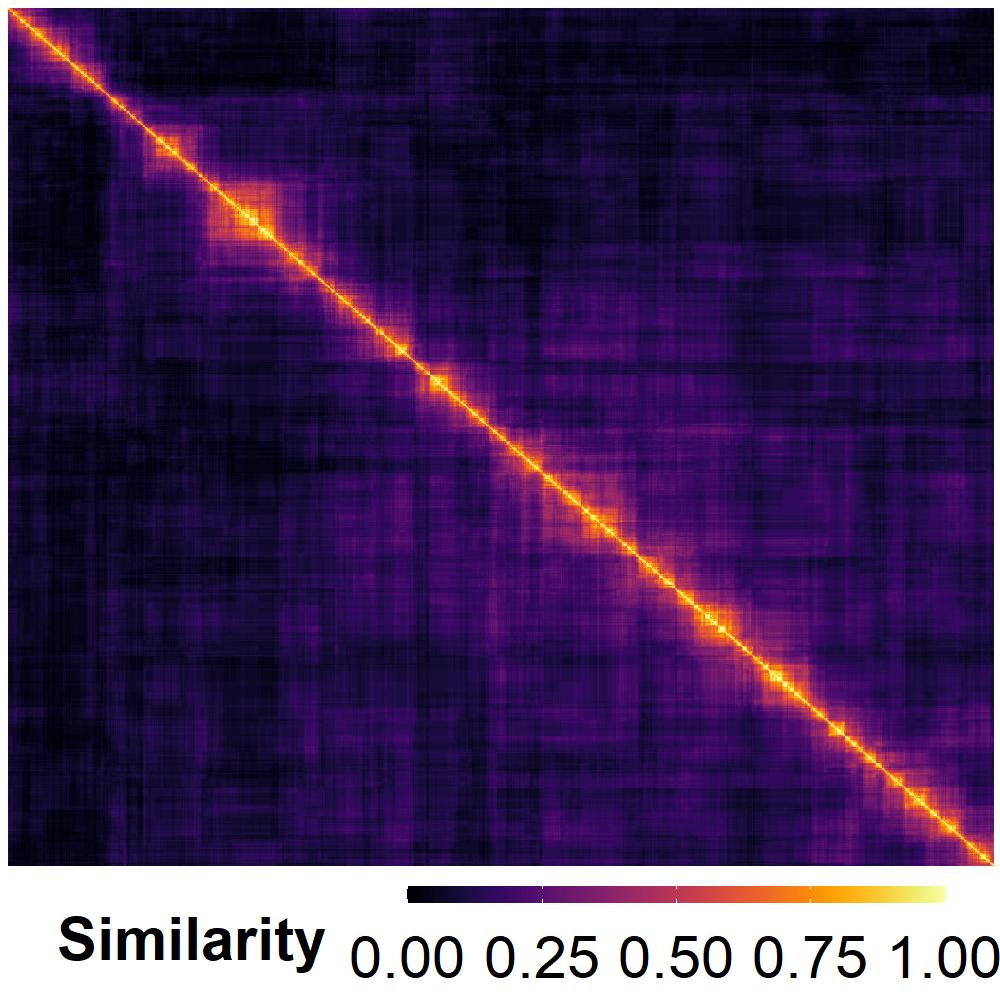}}
	\subfigure[NIFTY50]{\includegraphics[trim=0cm 4.2cm 0.1cm 0cm, clip=true, width=0.30\textwidth]{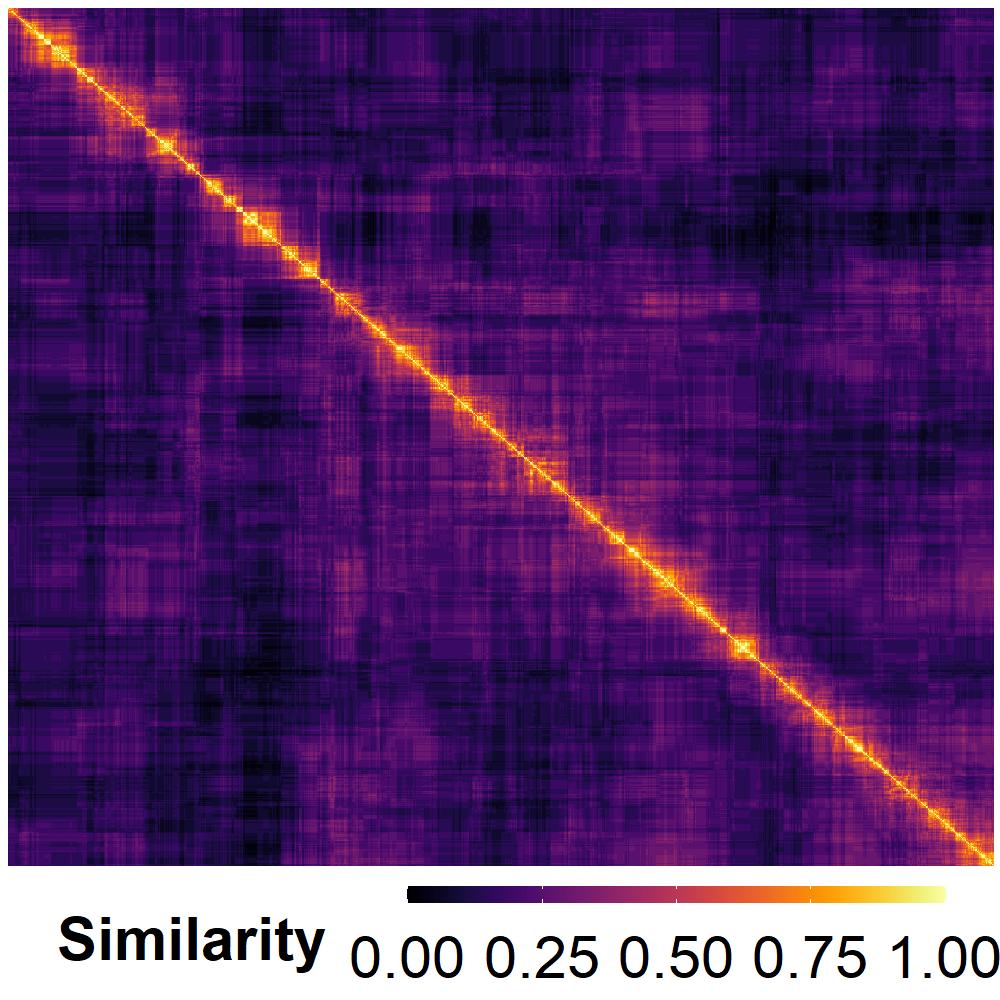}}
	\caption{\textbf{DMST - Cross-similarity matrix for each market index.} We calculate the pair-wise Jaccard Distance across all financial networks $G(t)$ and $G(t')$ ranging from $12$ May $2006$ to $18$ December $2019$, related to a given market index. For each market index figure, the first network on $12$ May $2006$ is represented in the top-left corner and the last network on $18$ December $2019$ in the bottom-right of each individual figure.}
	\label{fig:cross-similarity-dmst}
\end{figure*}

Figures~\ref{fig:cross-similarity-dag},~\ref{fig:cross-similarity-dtn} and~\ref{fig:cross-similarity-dmst} present the cross-similarity analysis for DAG, DTN and DMST of each stock market index, respectively. In the individual figure of each stock market index, the first network is represented in the top-left and the last network is represented in the bottom-right, where the first network is $12$ May $2006$ and the last network is $18$ December $2019$. In general, we can observe that the structure consistently changes over time, which emphasizes the importance of tools to forecast market structure.

%we did this analysis to justify why predicting the future is a problem... if the networks didn't change, we wouldn't need to do the article. with the cross similarity matrix, we see that the studied markets, the structure changes a lot, ´mainly for more developed markets, ftse, nasdaq, they are darker and have a greater dissimiality, that is, they change a lot. The assumption of using today's matrix to calculate future risk is accurate because the structure of these markets changes a lot. thus, tools like the one we are proposing are relevant.

%\textbf{nós fizemos essa análise para justificar o porque prever o futuro é um problema.. se as networks não mudassem, a gente não precisaria fazer o artigo. com a cross similarity matrix, a gente ve que os mercados estudados a estrutura muda bastante, ´rincipalmente para mercados mais desenvolvidos, ftse, nasdaq, são mais escuros e possuem uma dissimialridade maior, ou sejam, mudam muito. A assumption de usar a matriz de hojae para calcular o risco do futuro, isso está acurado porque a estrutura desses mercado muda muito. assim, ferramentas como essa que estamos propondo são relevantes.}

DAG results in Figure~\ref{fig:cross-similarity-dag} show network structure changes considerably throughout the time in all stock market indices. Figure~\ref{fig:cross-similarity-dtn} presents results from the DTN network filtering method. We can observe the similarity among networks tends to be noisier than the previous DAG method. In some periods, the similarity among the networks is maximum, while at other times it reaches zero, as can be seen in NASDAQ100 and NIFTY50. The DTN network filtering method can produce disconnected or even empty graphs, which may cause these similarity oscillations. DMST results are shown in Figure~\ref{fig:cross-similarity-dmst}. This figure shows that there is low similarity for long-range comparisons among trees created by the DMST filtering method for all market indices, suggesting low stability as reported by other authors~\cite{carlsson2010characterization,marti2015proposal}. 

Given the cross-similarity matrices of each market, we calculate the distance among all matrices to measure the market similarity in terms of network evolution. This analysis allows us to identify which markets have similar behavior considering the persistence of networks. To do this, we use the cosine similarity, calculated using the following formula:

\begin{equation}
cosine\_sim (a,b) = \frac{\sqrt{\sum_{}^{}{(a-b)^{2}  }  }  }{\sqrt{\sum_{}^{}{a^2}}  * \sqrt{\sum_{}^{}{b^2} } } ,
\end{equation}

\noindent where $a$ and $b$ are two non-zero numeric vectors and represents the upper triangle of two distinct cross-similarity matrices. This metric ranges from $0$ to $1$ and it is defined as the angular distance from two vectors. Table~\ref{tab:cosine-distance} presents the pairwise cosine similarity for DAG, DTN and DMST. DAX30 and EUROSTOXX50 have the highest cosine similarity for DAG and DTN. For DMST, the highest value is between FTSE100 and EUROSTOXX50. This analysis demonstrates that the network persistence among markets from Europe are higher than markets from other regions of the world, given the three network filtering methods. 

% Table generated by Excel2LaTeX from sheet 'Plan3'
\begin{table}[hb!]
  \centering
  \small
  \caption{\textbf{Cosine distance from cross-similarity results.} We calculate the cosine similarity from cross-similarity matrices. We use the upper triangle of each matrix as the input vector. European markets have the highest similarity.}
    \begin{tabular}{|c|l|ccccc|}
\cmidrule{3-7}    \multicolumn{1}{r}{} &       & \multicolumn{1}{l}{\textit{\textbf{EUROSTOXX50}}} & \multicolumn{1}{l}{\textit{\textbf{FTSE100}}} & \multicolumn{1}{l}{\textit{\textbf{HANGSENG50}}} & \multicolumn{1}{l}{\textit{\textbf{NASDAQ100}}} & \multicolumn{1}{l|}{\textit{\textbf{NIFTY50}}} \\
    \midrule
    \multirow{5}[2]{*}{\textbf{DAG}} & \textit{\textbf{DAX30}} & \textbf{0.9532} & 0.9435 & 0.9472 & 0.9341 & 0.9257 \\
          & \textit{\textbf{EUROSTOXX50}} &       & 0.9228 & 0.9403 & 0.9420 & 0.9070 \\
          & \textit{\textbf{FTSE100}} &       &       & 0.9150 & 0.9358 & 0.8978 \\
          & \textit{\textbf{HANGSENG50}} &       &       &       & 0.9297 & 0.9302 \\
          & \textit{\textbf{NASDAQ100}} &       &       &       &       & 0.9137 \\
    \midrule
    \multirow{5}[2]{*}{\textbf{DTN}} & \textit{\textbf{DAX30}} & \textbf{0.9338} & 0.8367 & 0.7573 & 0.6209 & 0.5795 \\
          & \textit{\textbf{EUROSTOXX50}} &       & 0.8755 & 0.7873 & 0.6143 & 0.6000 \\
          & \textit{\textbf{FTSE100}} &       &       & 0.8331 & 0.5479 & 0.5503 \\
          & \textit{\textbf{HANGSENG50}} &       &       &       & 0.5892 & 0.5531 \\
          & \textit{\textbf{NASDAQ100}} &       &       &       &       & 0.4269 \\
    \midrule
    \multirow{5}[2]{*}{\textbf{DMST}} & \textit{\textbf{DAX30}} & 0.9486 & 0.9354 & 0.8967 & 0.9011 & 0.9200 \\
          & \textit{\textbf{EUROSTOXX50}} &       & \textbf{0.9500} & 0.9058 & 0.9294 & 0.9312 \\
          & \textit{\textbf{FTSE100}} &       &       & 0.9253 & 0.9400 & 0.9338 \\
          & \textit{\textbf{HANGSENG50}} &       &       &       & 0.9169 & 0.9080 \\
          & \textit{\textbf{NASDAQ100}} &       &       &       &       & 0.9160 \\
    \bottomrule
    \end{tabular}%
  \label{tab:cosine-distance}%
\end{table}%

\begin{figure*}[ht!]
	\centering
	\subfigure[Dynamic Asset Graph]{\includegraphics[trim=0cm 1cm 0cm 0cm, clip=true, width=0.65\textwidth]{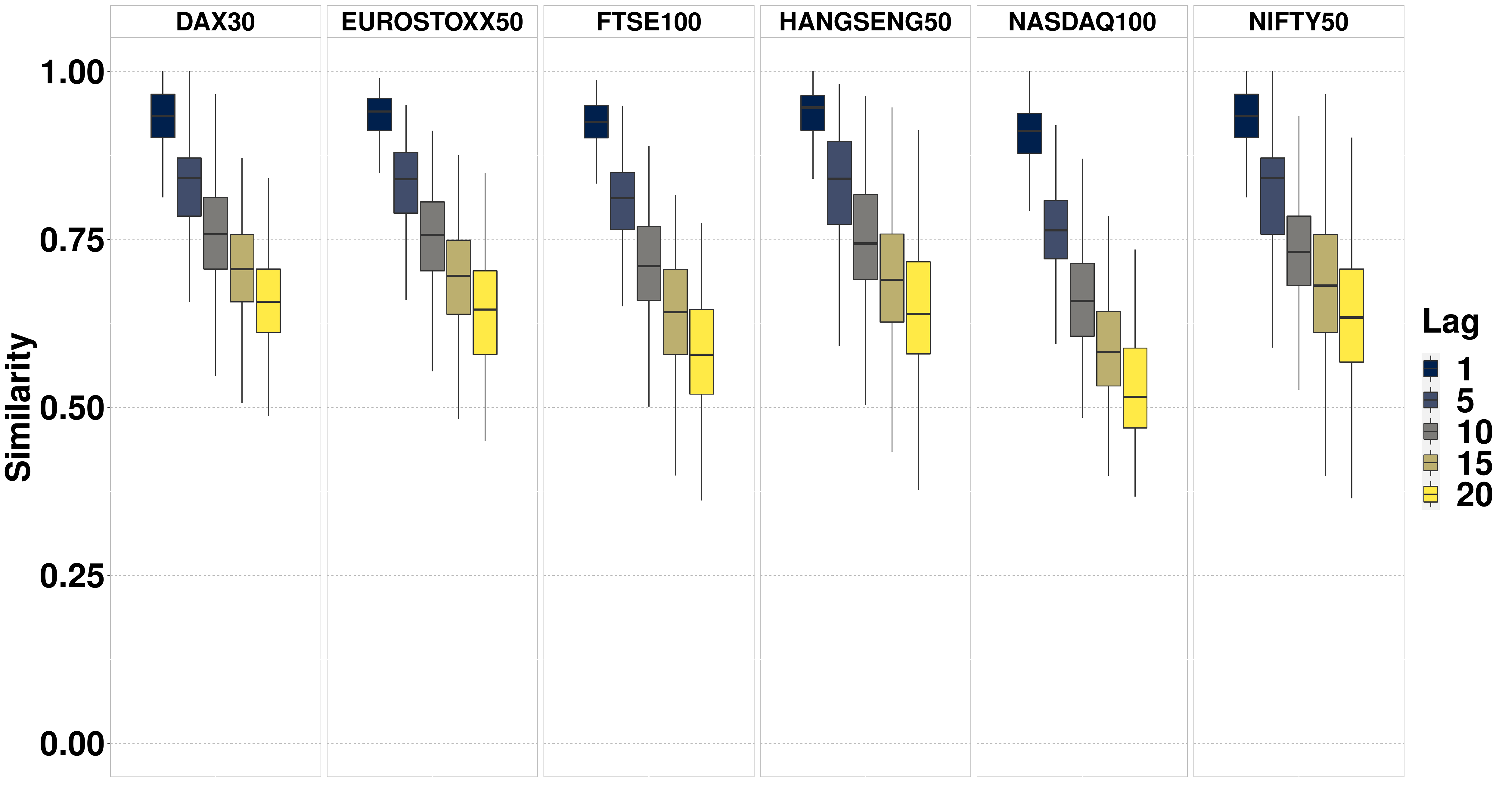} \label{subfig:similarity-lag-dag}}
	\subfigure[Dynamic Threshold Networks]{\includegraphics[trim=0cm 1cm 0cm 0cm, clip=true, width=0.65\textwidth]{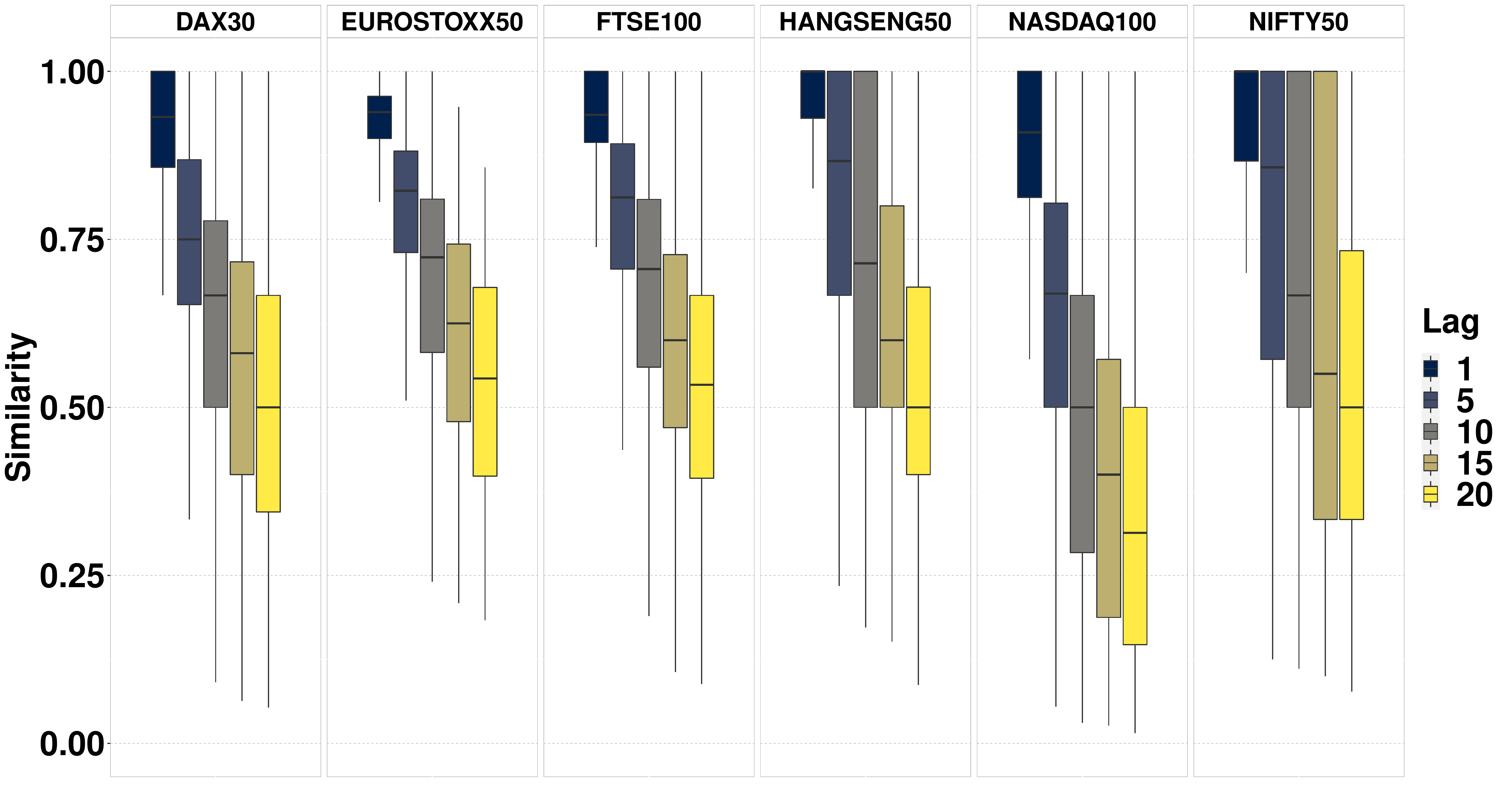} \label{subfig:similarity-lag-dtn}}
	\subfigure[Dynamic Minimal Spanning Tree]{\includegraphics[trim=0cm 1cm 0cm 0cm, clip=true, width=0.65\textwidth]{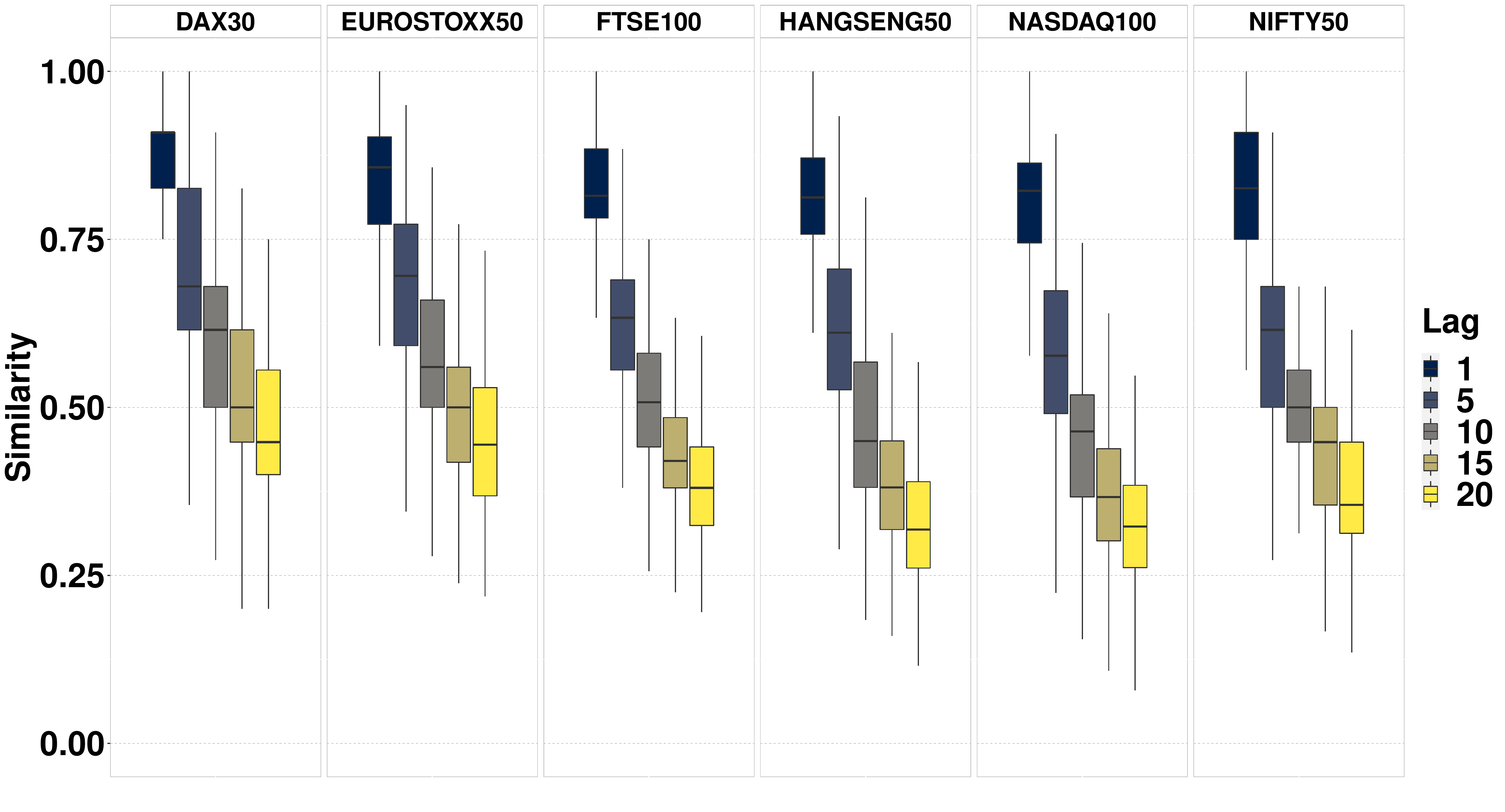} \label{subfig:similarity-lag-dmst}}
	\caption{\textbf{Networks Similarity vs. Time Lag.} Figure shows the  distribution of networks persistence considering $h = \lbrace 1, 5, 10, 15, 20 \rbrace$ trading weeks ahead related to the three network filtering methods: DAG, DTN and DMST. Network similarity is quantified using the Jaccard Distance between graphs $G(t)$ and $G(t+h)$.}
	\label{fig:similarity-lag}
\end{figure*}

The second descriptive analysis is the similarity between the current financial network  $G(t)$ and the future network $G(t + h)$, where $h$ is the time lag, $\forall$  $h \in \lbrace 1, 5, 10, 15, 20 \rbrace$ trading weeks. This analysis provides an accurate point of view concerning how the current network changes in the near future - if they do not change, we do not need to forecast them. We quantify the changes in the network structure using the Jaccard Distance between $G(t)$ and $G(t+h)$, considering $L = 252$ trading days to create each graph. Figure~\ref{fig:similarity-lag} presents the distribution of networks similarity related to the three network filtering methods DAG, DTN and DMST of each stock market index. Experimental results suggest a high similarity distribution among networks considering $h = 1$ step ahead to all network filtering methods. However, the similarity distribution decreases with $h$, mainly in the DMST method. Considering $h = 20$, DMST presents a mean similarity lower than $25\%$ in all markets. In general, financial networks tend to have a certain margin of similarity for low $h$, but as $h$ increases, they become more and more dissimilar, hence justifying the importance of forecasting future market structures, particularly in high-horizon forecasting scenarios. Analyzing the DTN method, NIFTY50 and HANGSENG50 present a different behavior for larger $h$, where the distribution of the similarity behaves differently from other markets, oscillating between the maximum value and almost zero for larger $h$, as shown in $h=5$, $h=10$ and $h=15$. This amplitude can be explained by the analysis presented in Figure~\ref{fig:cross-similarity-dtn}, which shows that for some periods the similarity among networks is high, but it is also very low for other periods. The smallest similarity values are presented for the DMST method considering $L = 20$.

\begin{figure*}[ht!]
	\centering
	\subfigure[DAG ($L = 126$)]{\includegraphics[trim=0.4cm 0.0cm 2.0cm 1.6cm, clip=true, width=0.28\textwidth]{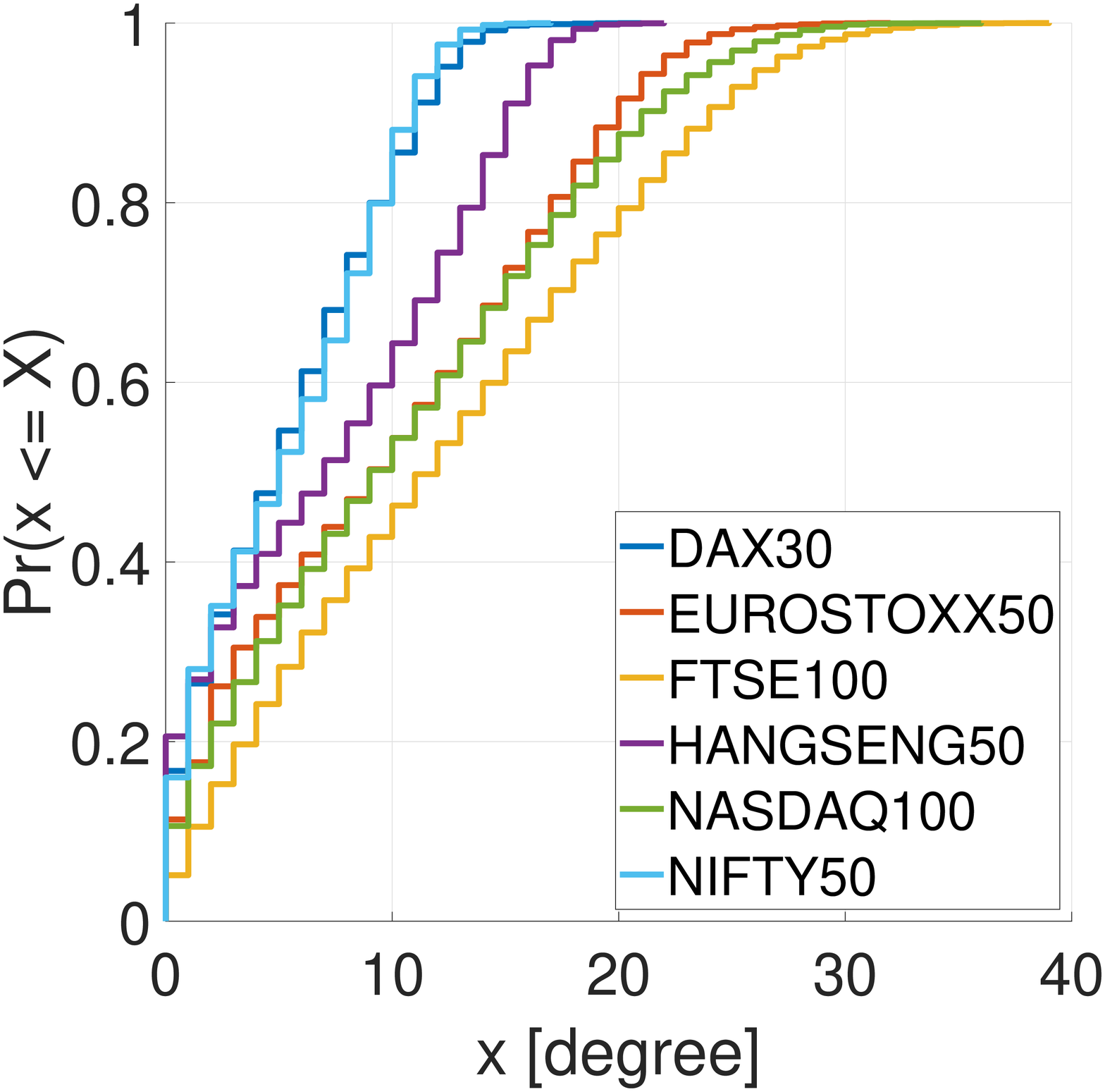} \label{subfig:degree-cdf-126-dag}}
	\subfigure[DAG ($L = 252$)]{\includegraphics[trim=0.4cm 0.0cm 2.0cm 1.6cm, clip=true, width=0.28\textwidth]{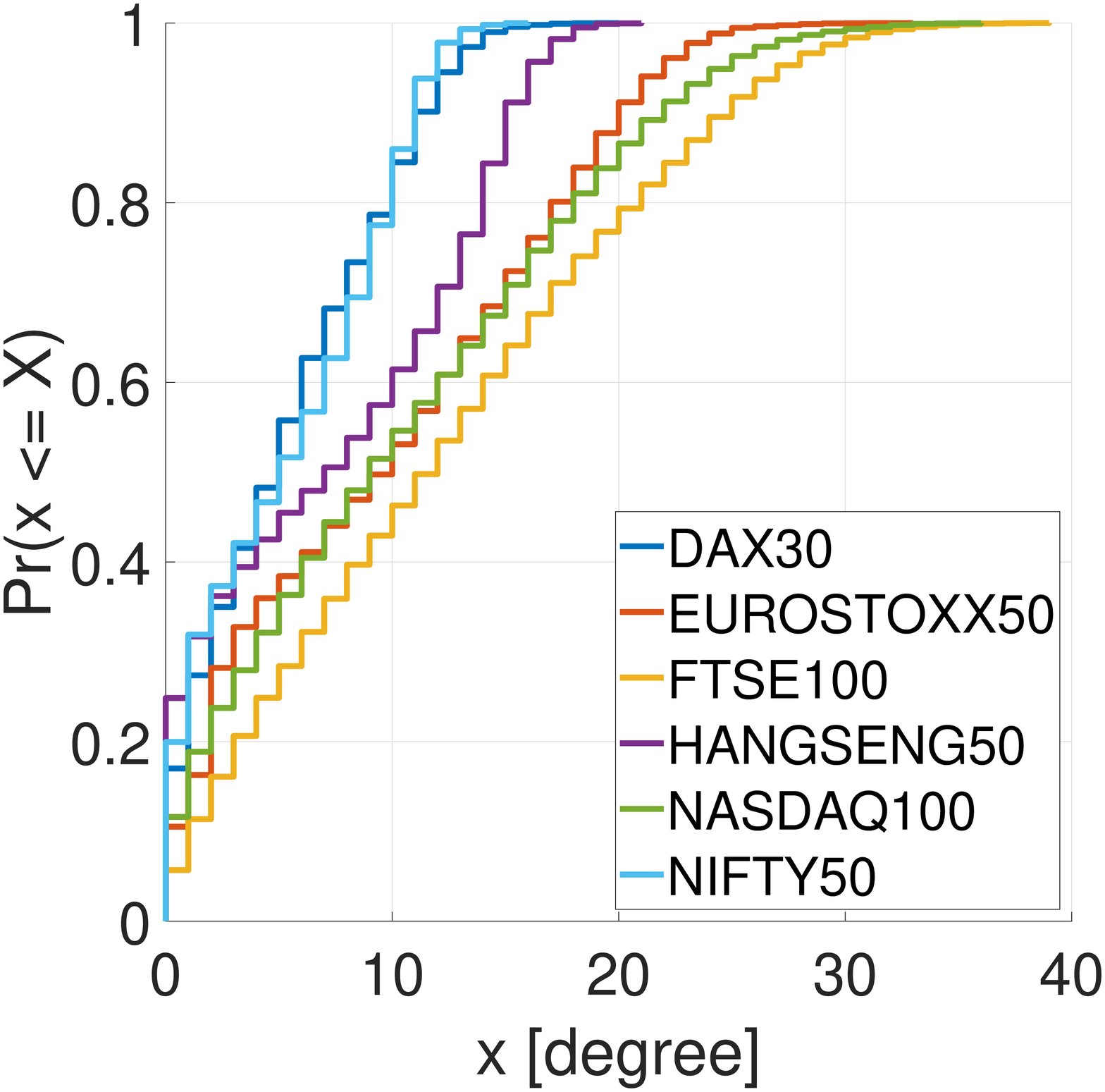} \label{subfig:degree-cdf-252-dag}}
	\subfigure[DAG ($L = 504$)]{\includegraphics[trim=0.4cm 0.0cm 2.0cm 1.6cm, clip=true, width=0.28\textwidth]{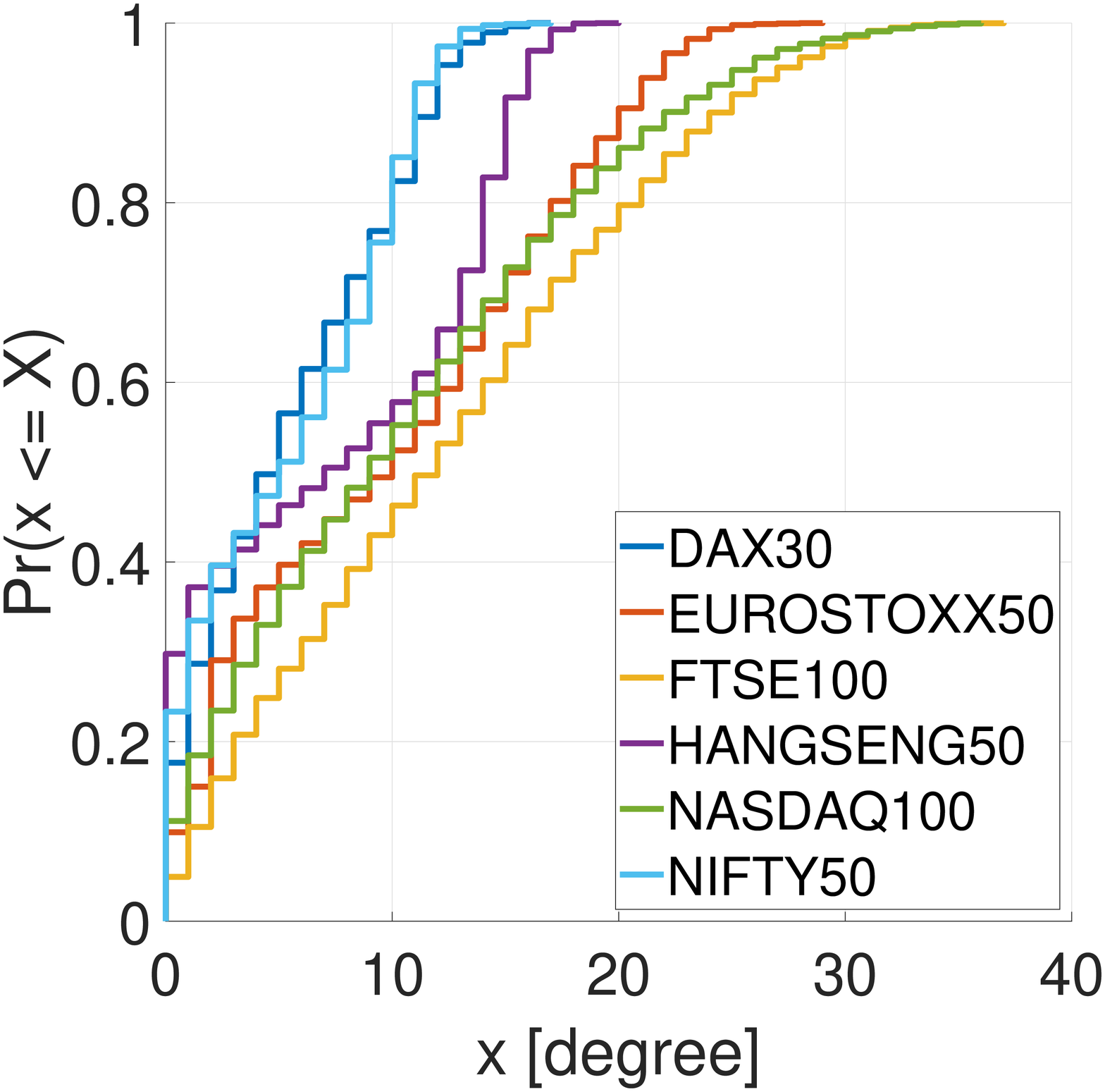} \label{subfig:degree-cdf-504-dag}}
	\subfigure[DTN ($L = 126$)]{\includegraphics[trim=0.4cm 0.0cm 2.0cm 1.6cm, clip=true, width=0.28\textwidth]{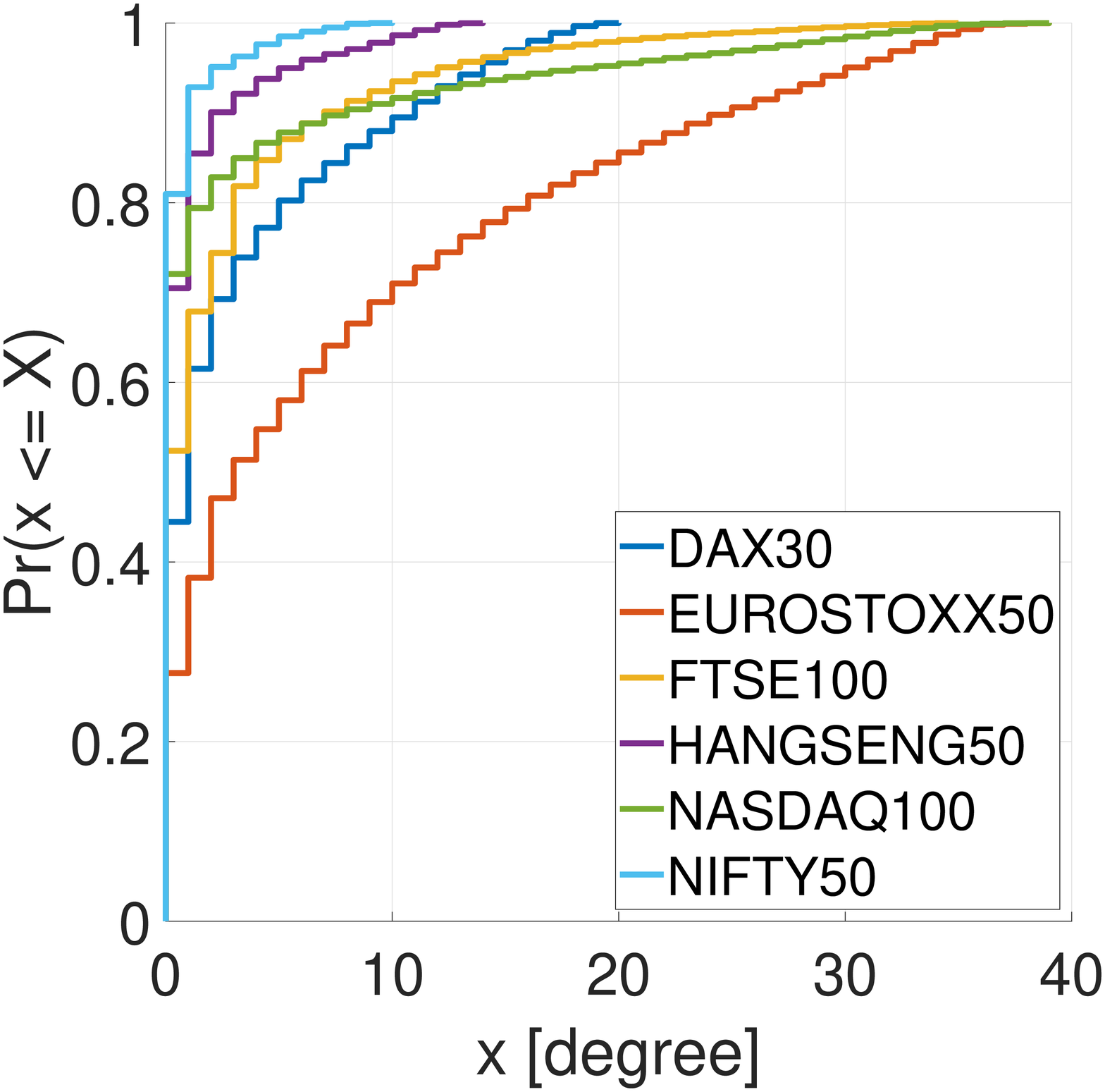} \label{subfig:degree-cdf-126-dtn}}
	\subfigure[DTN ($L = 252$)]{\includegraphics[trim=0.4cm 0.0cm 2.0cm 1.6cm, clip=true, width=0.28\textwidth]{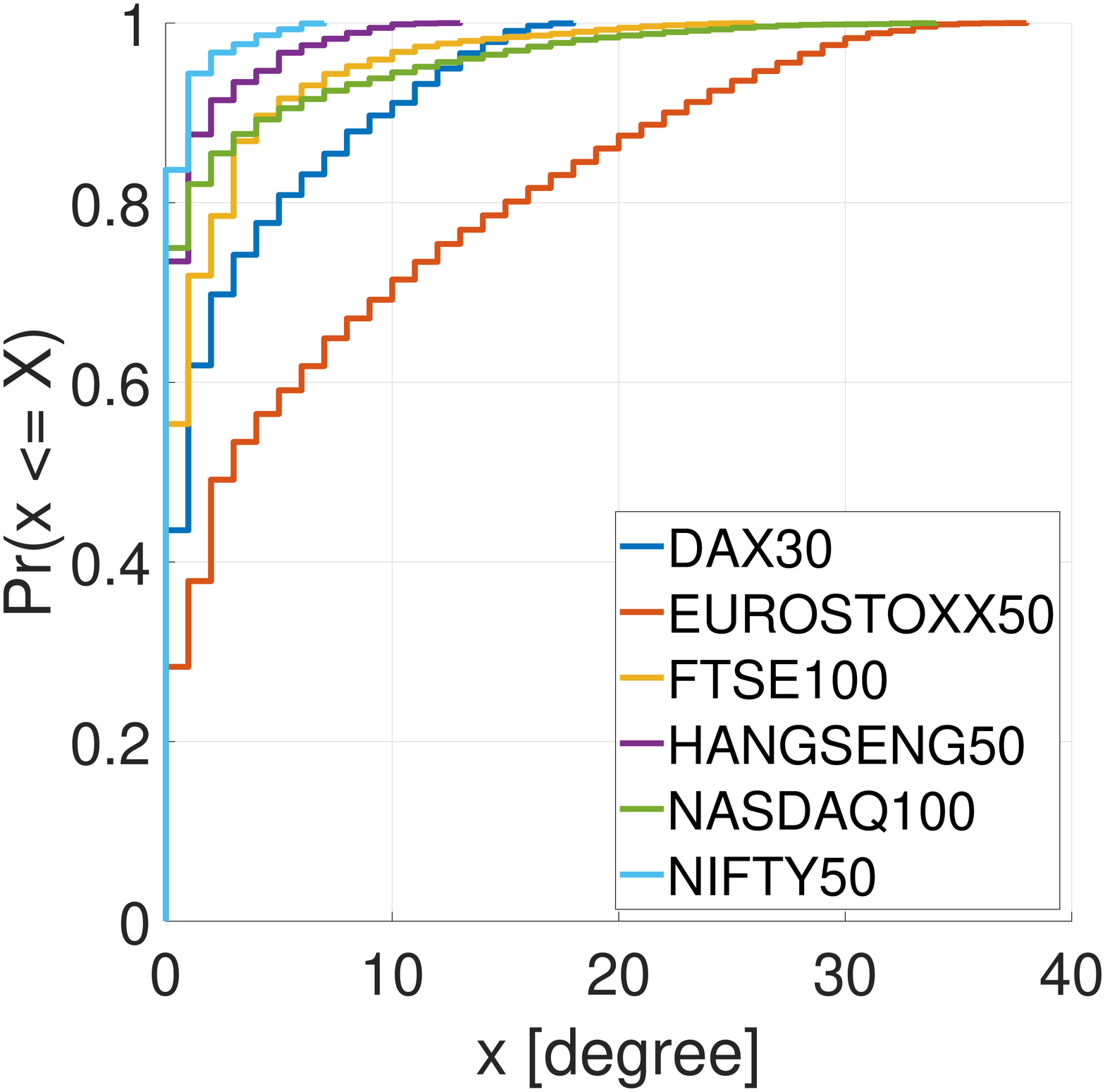} \label{subfig:degree-cdf-252-dtn}}
	\subfigure[DTN ($L = 504$)]{\includegraphics[trim=0.4cm 0.0cm 2.0cm 1.6cm, clip=true, width=0.28\textwidth]{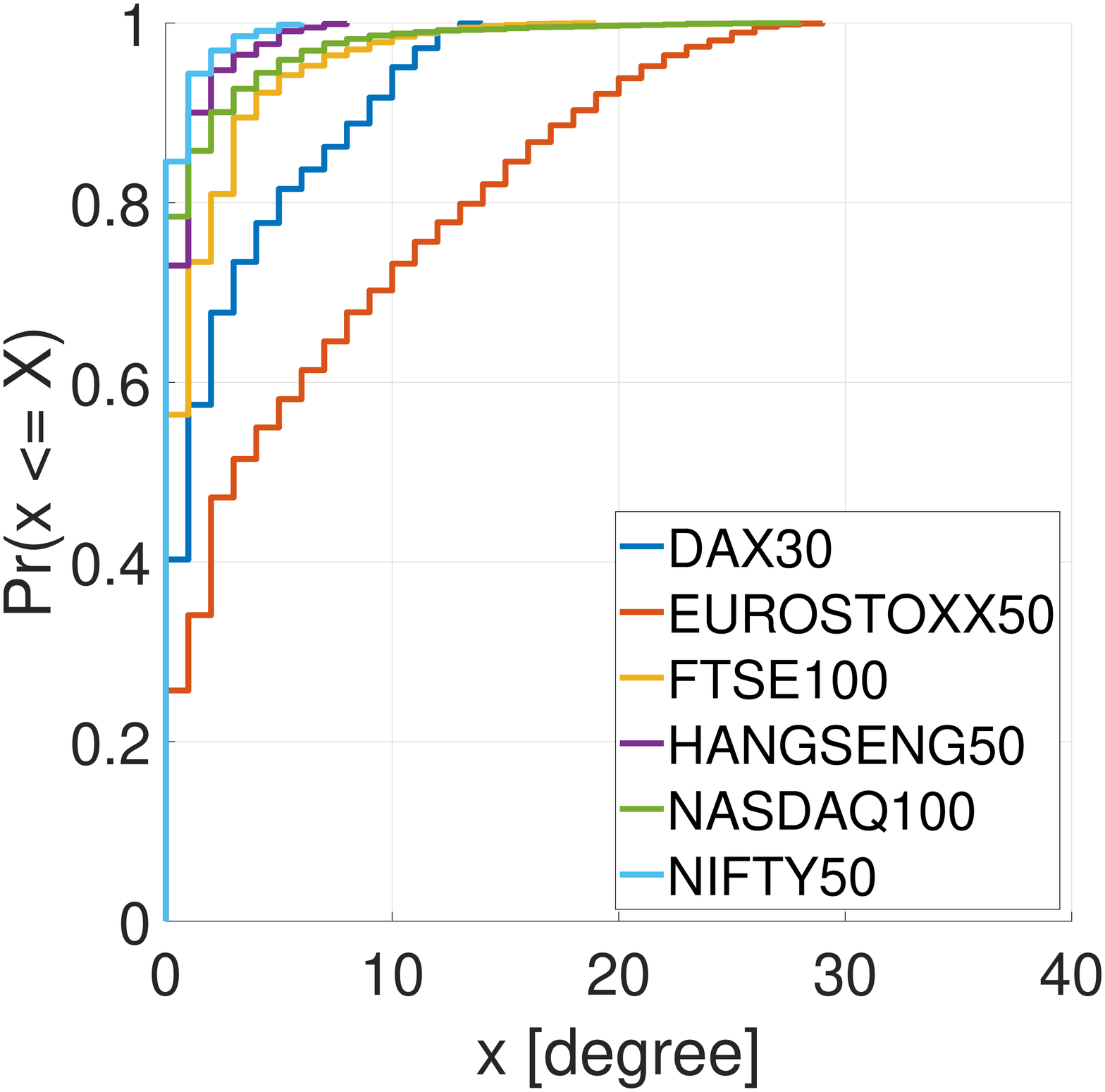} \label{subfig:degree-cdf-504-dtn}}
	\subfigure[DMST ($L = 126$)]{\includegraphics[trim=0.4cm 0.0cm 2.0cm 1.6cm, clip=true, width=0.28\textwidth]{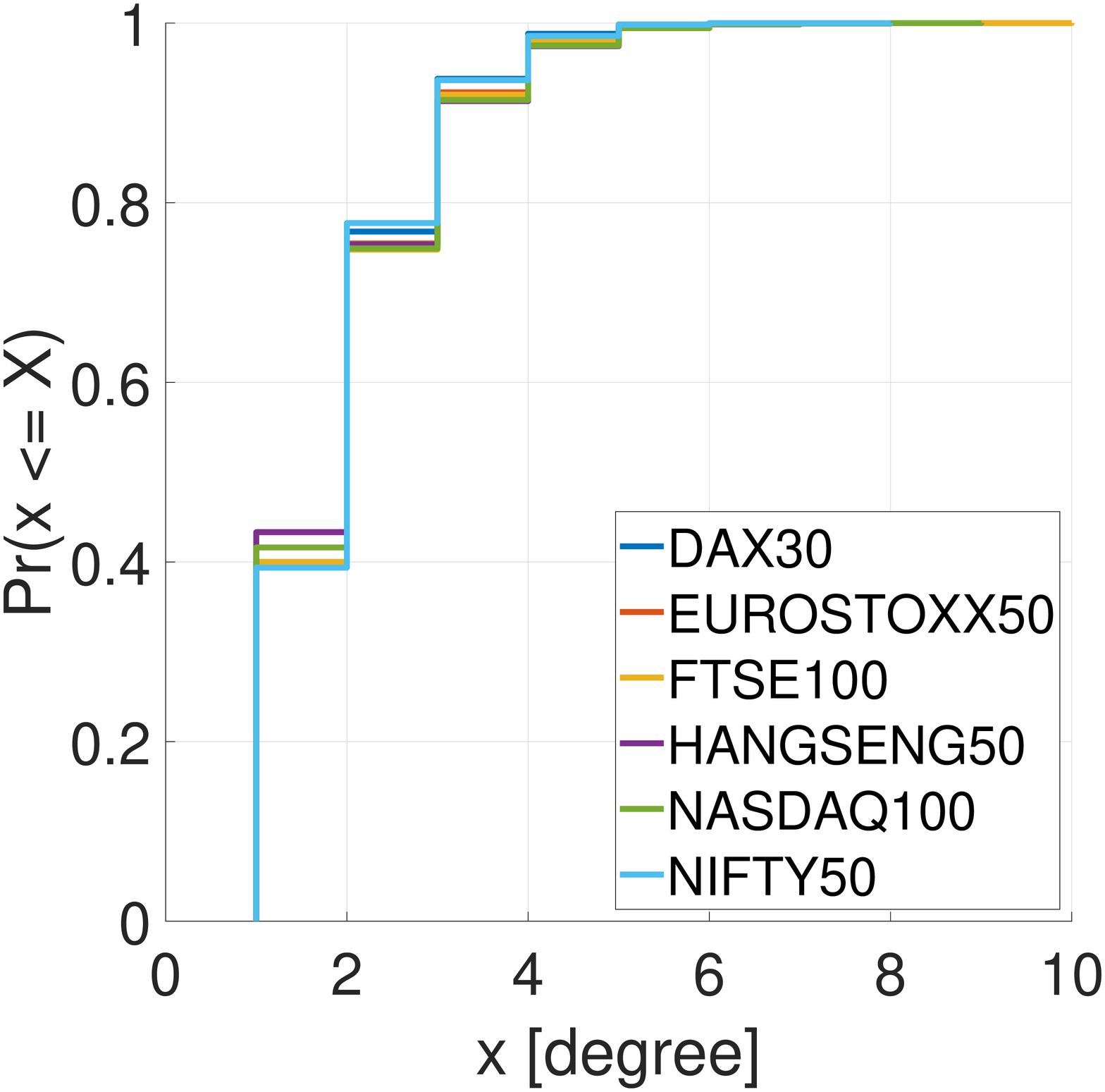} \label{subfig:degree-cdf-126-dmst}}
	\subfigure[DMST ($L = 252$)]{\includegraphics[trim=0.4cm 0.0cm 2.0cm 1.6cm, clip=true, width=0.28\textwidth]{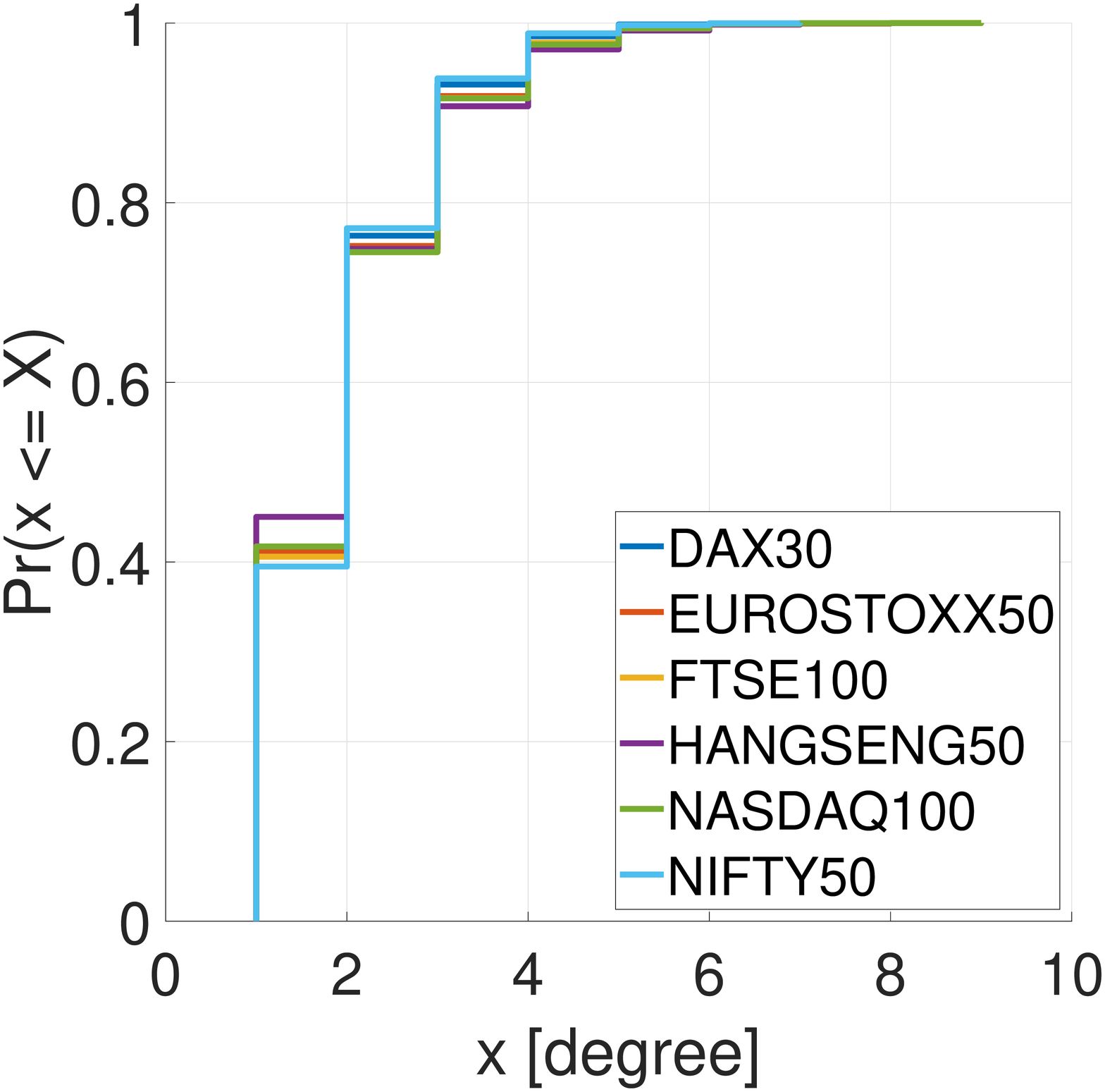} \label{subfig:degree-cdf-252-dmst}}
	\subfigure[DMST ($L = 504$)]{\includegraphics[trim=0.4cm 0.0cm 2.0cm 1.6cm, clip=true, width=0.28\textwidth]{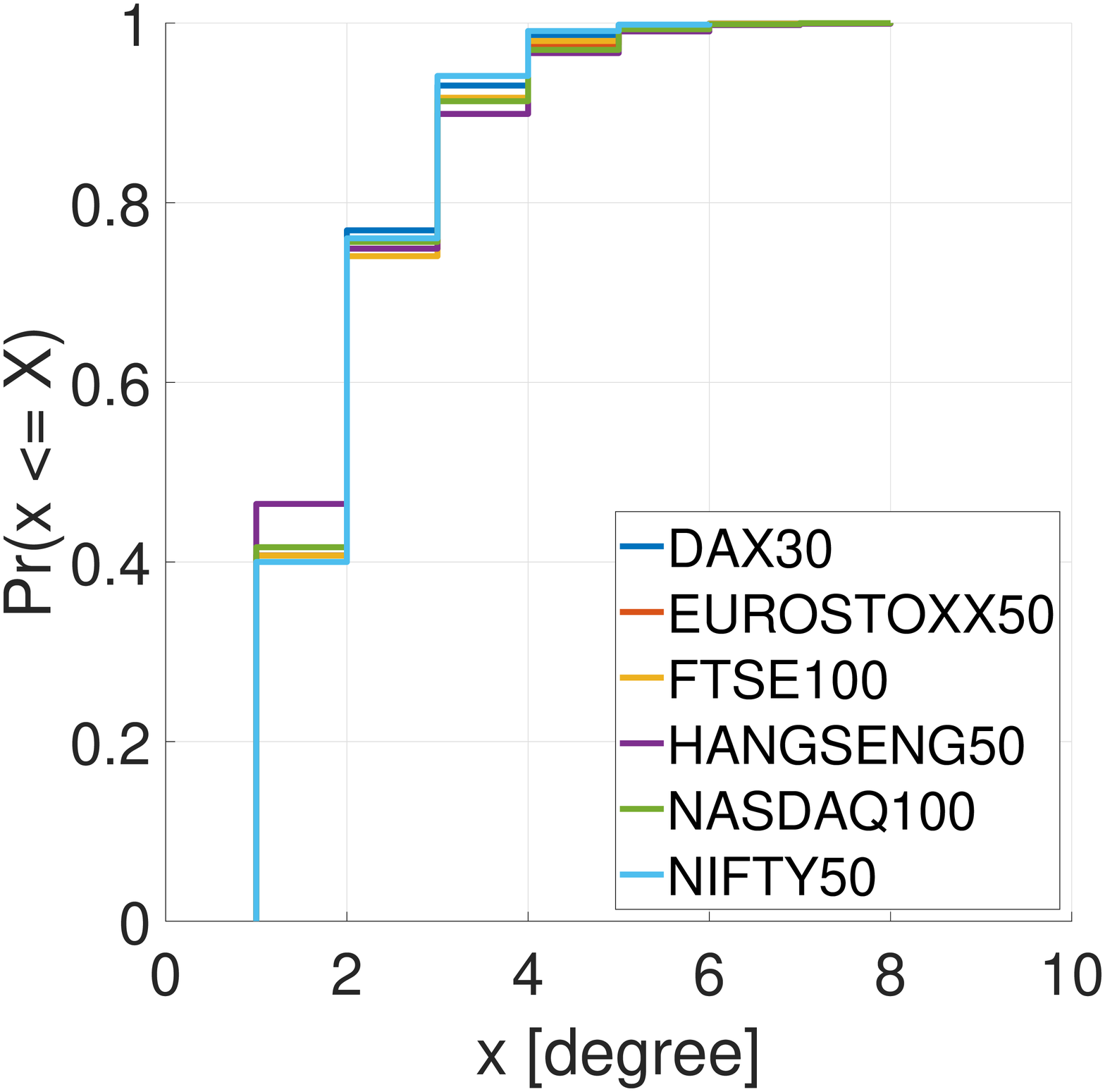} \label{subfig:degree-cdf-504-dmst}}
	\caption{\textbf{CDF of node degree across networks using DAG, DTN and DMST network filtering methods.} We calculate the cumulative distribution function of node degree across all stock networks using the size of rolling window $L = 126, 252$ and $504$ trading days. The period of the experiments ranges from $3$ March $2007$ to $18$ December $2019$. Market indices with the smallest number of constituents present a similar behaviour in the DAG network filtering method. The DTN method presents the highest probability of nodes without edges, mainly on NIFTY50, NASDAQ100 and HANGSENG50. EUROSTOXX50 presents a distinct shape compared with the other market indices in DTN with the smallest number of nodes without connection. Results also suggest the degree distribution of the market indices are similar for $L = 126, 252 \text{ and } 504$ trading days in all network filtering methods.}
	\label{fig:degree-cdf}
\end{figure*}

The third descriptive analysis is presented in Figure~\ref{fig:degree-cdf}. We present the Cumulative Distribution Function (CDF) of the node degree across networks of each index using the DAG, DTN and DMST network filtering methods. This analysis provides information concerning the node degree according to three main aspects: \textit{(i)} the impact of time series size $L$; \textit{(ii)} network filtering method and \textit{(iii)} size of the market index, considering the number of constituents. We calculated the node degree distribution across all financial networks ranging from $3$ March $2007$ to $18$ December $2019$. Results using $L \in \lbrace 126, 252, 504 \rbrace$ trading days as rolling window size are presented. We observe in Figure~\ref{fig:degree-cdf} that market indices with the smallest number of constituents present a similar behaviour in terms of node degree when we use the DAG network filtering method. Besides, DAG nodes are prone to have a higher occurrence of node with no connections. The DTN method also presents high probability of nodes without edges, mainly on NIFTY50, NASDAQ100 and HANGSENG50. EUROSTOXX50 presents a distinct shape compared with the other market indices in DTN with the smallest number of nodes without a connection - more than $75\%$ of nodes has a degree greater than $1$ edge. On the other hand, for all market indices, at least $50\%$ of the nodes have $4$ or more connections in DAG. Considering the number of stocks in each market index, we can also conclude that there are no nodes connecting to all other vertices in any network filtering method because the largest degree distribution of each market index. Results also suggest the degree distribution of the market indices are similar for $L = 126, 252 \text{ and } 504$ trading days in all network filtering methods, indicating that the size of $L$ does not affect the degree distribution of stock networks of each market index. 

\subsection{Predictive Analysis}

In this section, we present a set of experimental results related to market structure forecasting using machine learning. First, we investigate the predictive performance of the proposed method in different scenarios, comparing it against the benchmark methods. Then, we present a qualitative analysis concerning the model interpretability and its implications. 

\subsubsection{Performance Results}

%We present several experiments designed to address the three main research questions proposed in this work. The experiments were made using data from seven stock market indices.  

%\subsection{\textit{Q1} - Machine Learning vs. Static Methods}

\begin{figure*}[h!]
	\centering
	\subfigure[DAX30]{\includegraphics[trim=0.2cm 6.2cm 0.1cm 0.2cm, clip=true, width=0.28\textwidth]{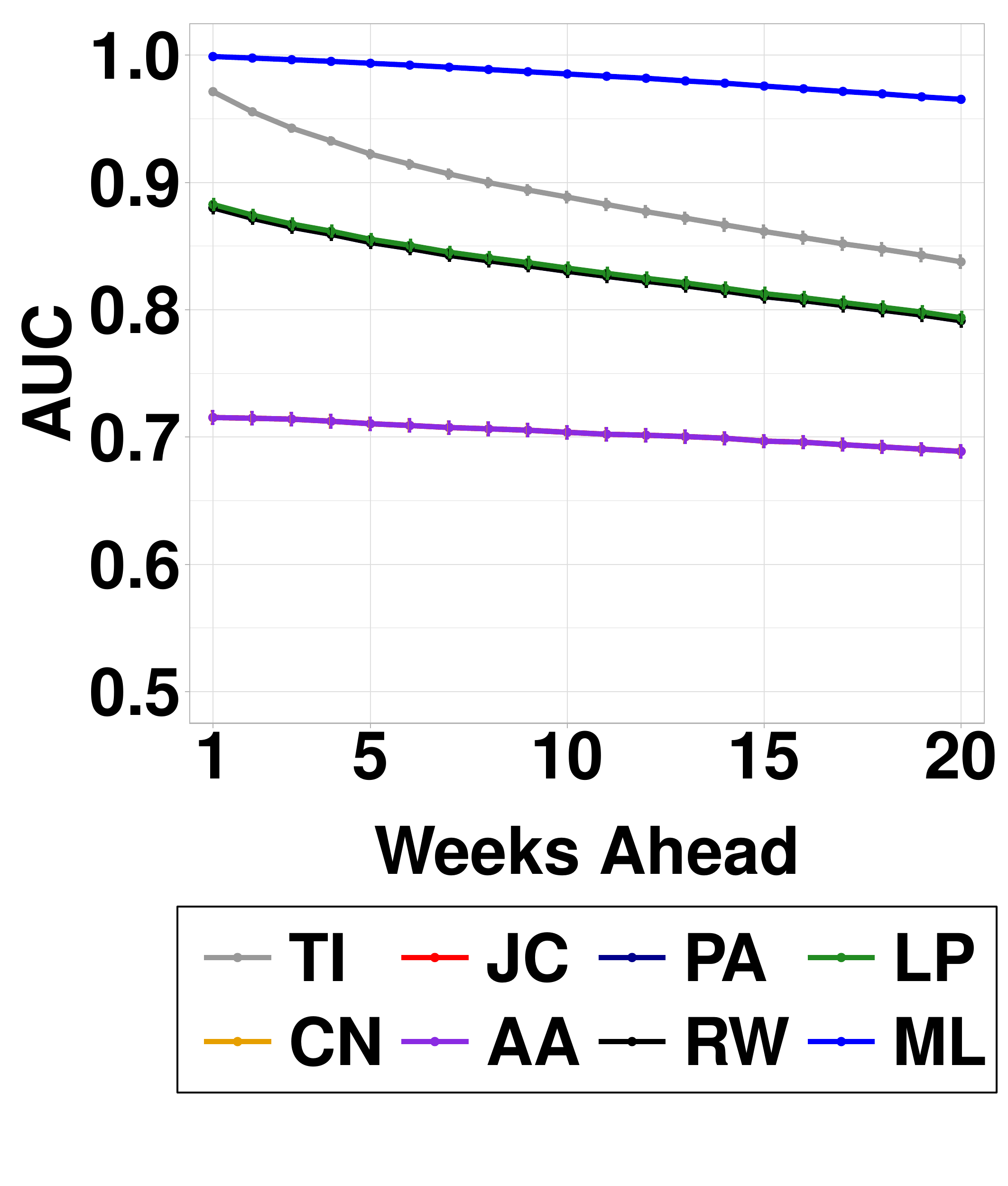}}
	\subfigure[EUROSTOXX50]{\includegraphics[trim=0.2cm 6.2cm 0.1cm 0.2cm, clip=true, width=0.28\textwidth]{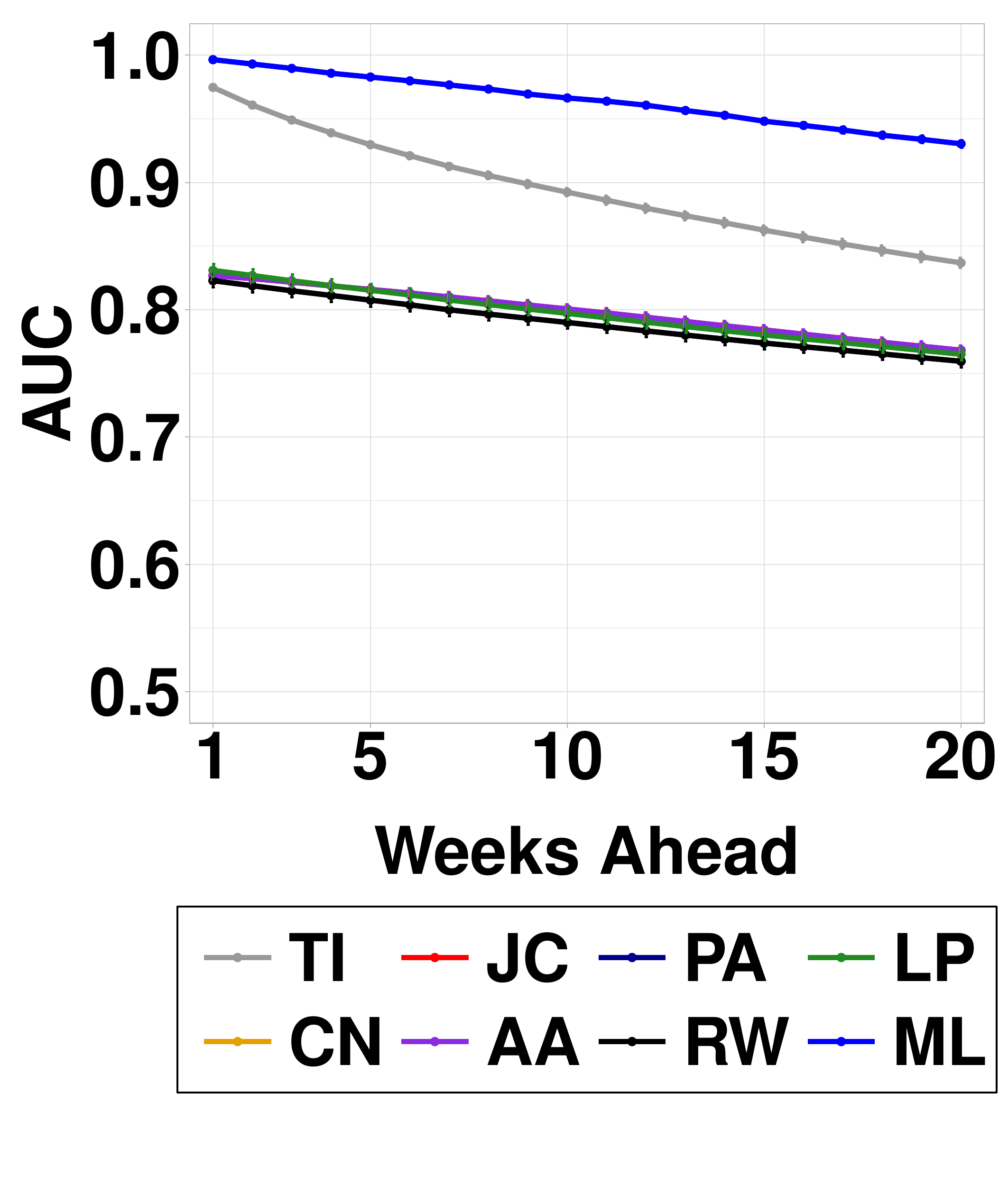}}
	\subfigure[FTSE100]{\includegraphics[trim=0.2cm 6.2cm 0.1cm 0.2cm, clip=true, width=0.28\textwidth]{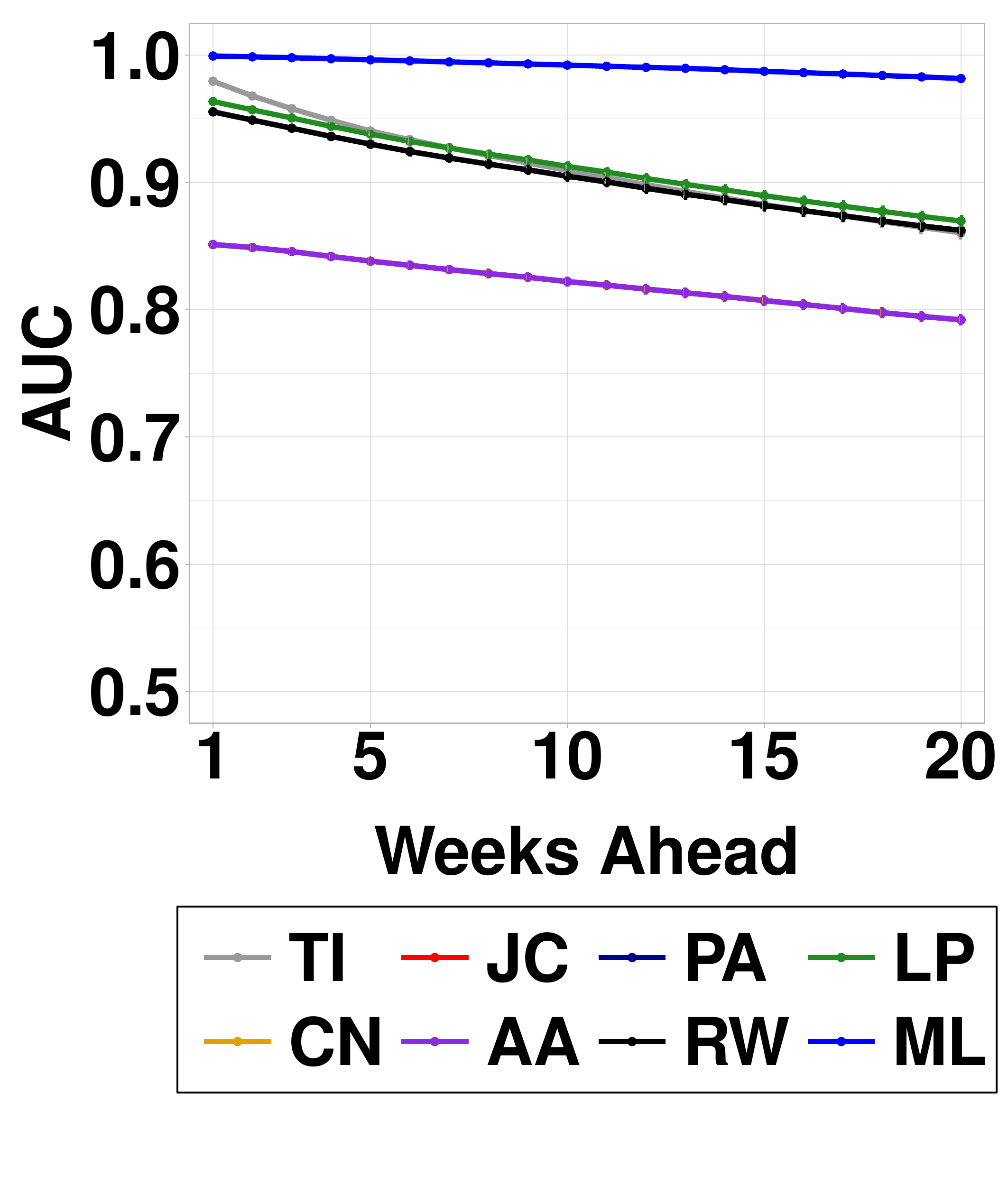}}
	\subfigure[HANGSENG50]{\includegraphics[trim=0.2cm 6.2cm 0.1cm 0.2cm, clip=true, width=0.28\textwidth]{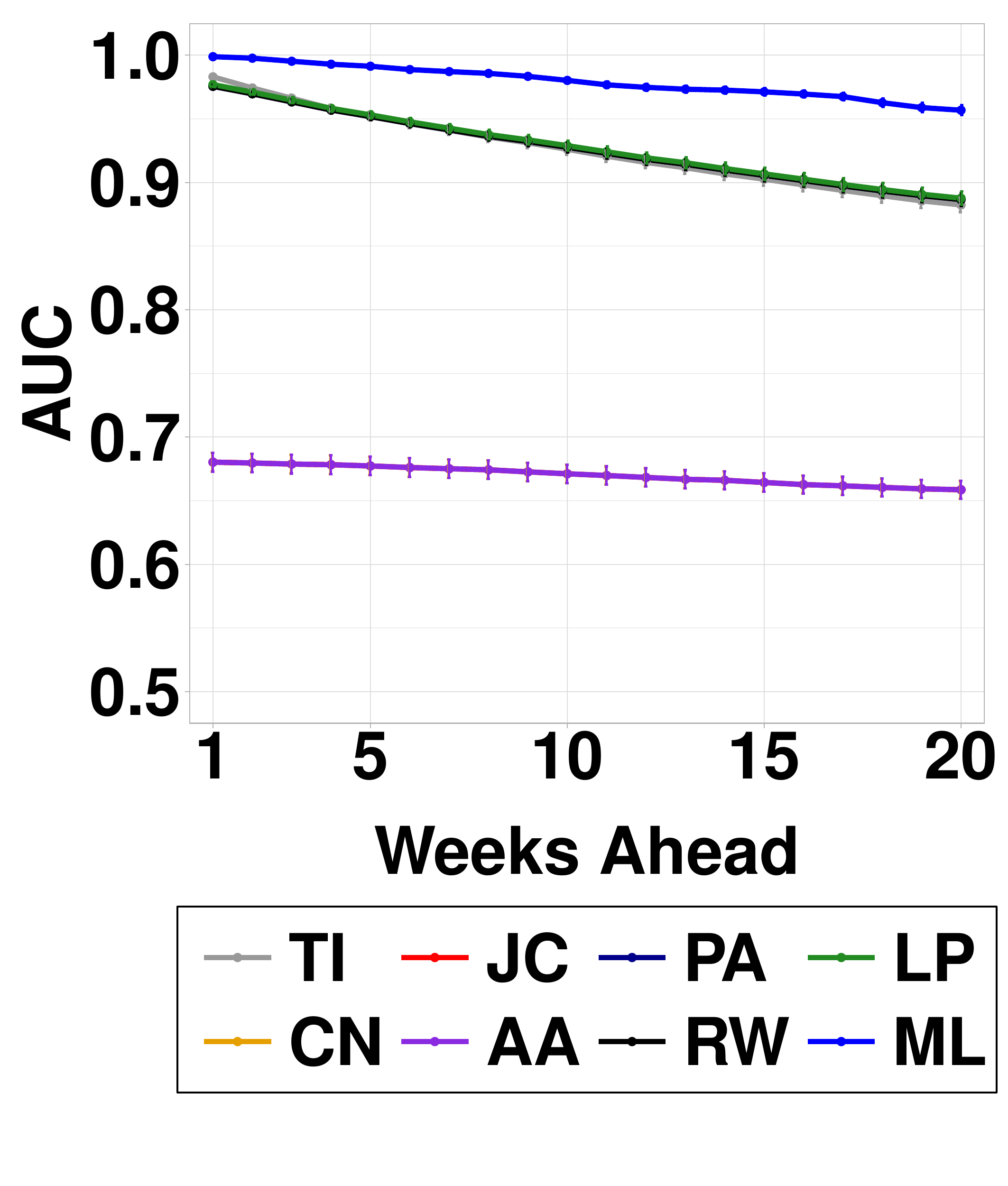}}
	\subfigure[NASDAQ100]{\includegraphics[trim=0.2cm 1.8cm 0.1cm 0.2cm, clip=true, width=0.2804\textwidth]{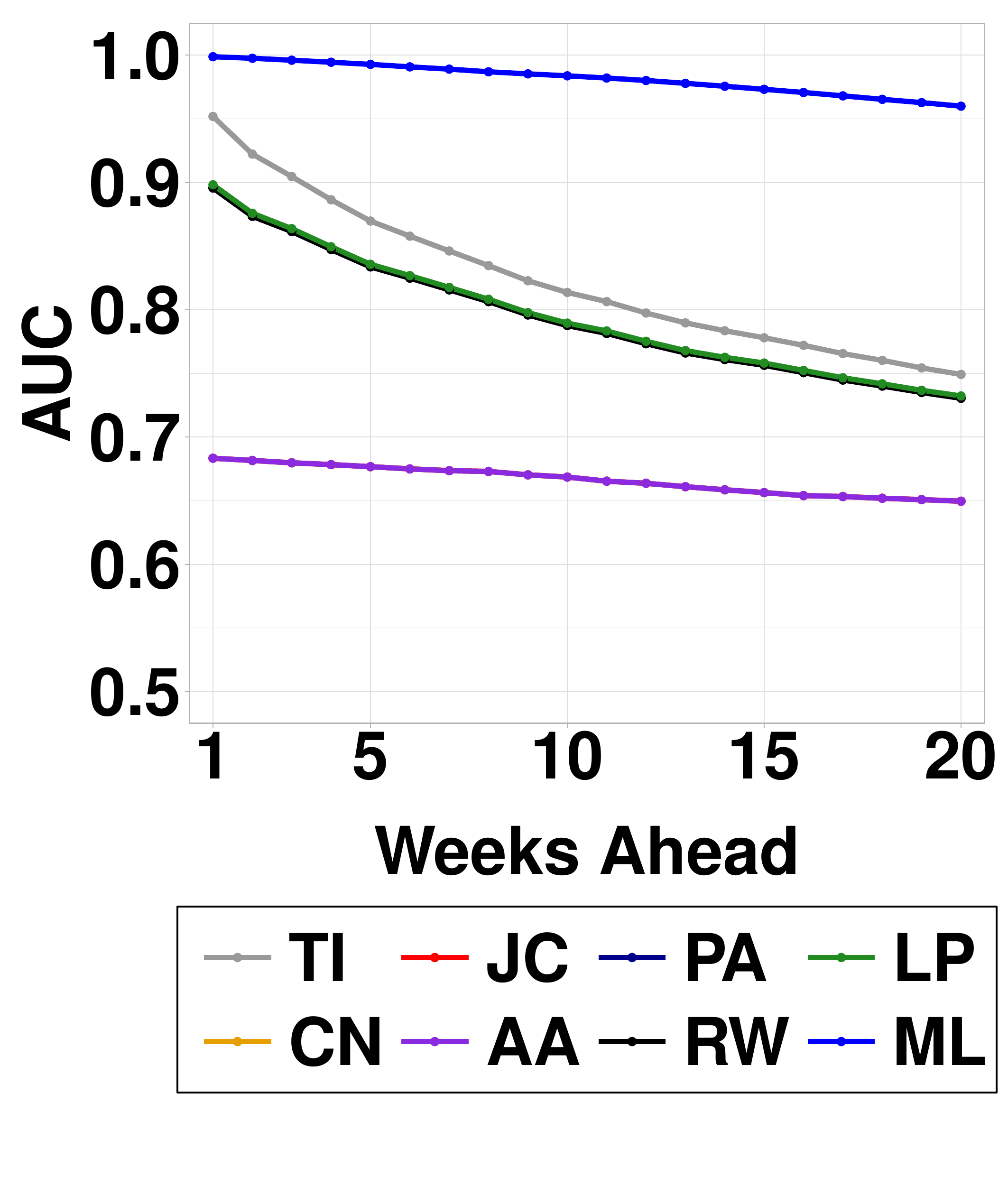}}
	\subfigure[NIFTY50]{\includegraphics[trim=0.2cm 6.2cm 0.1cm 0.2cm, clip=true, width=0.28\textwidth]{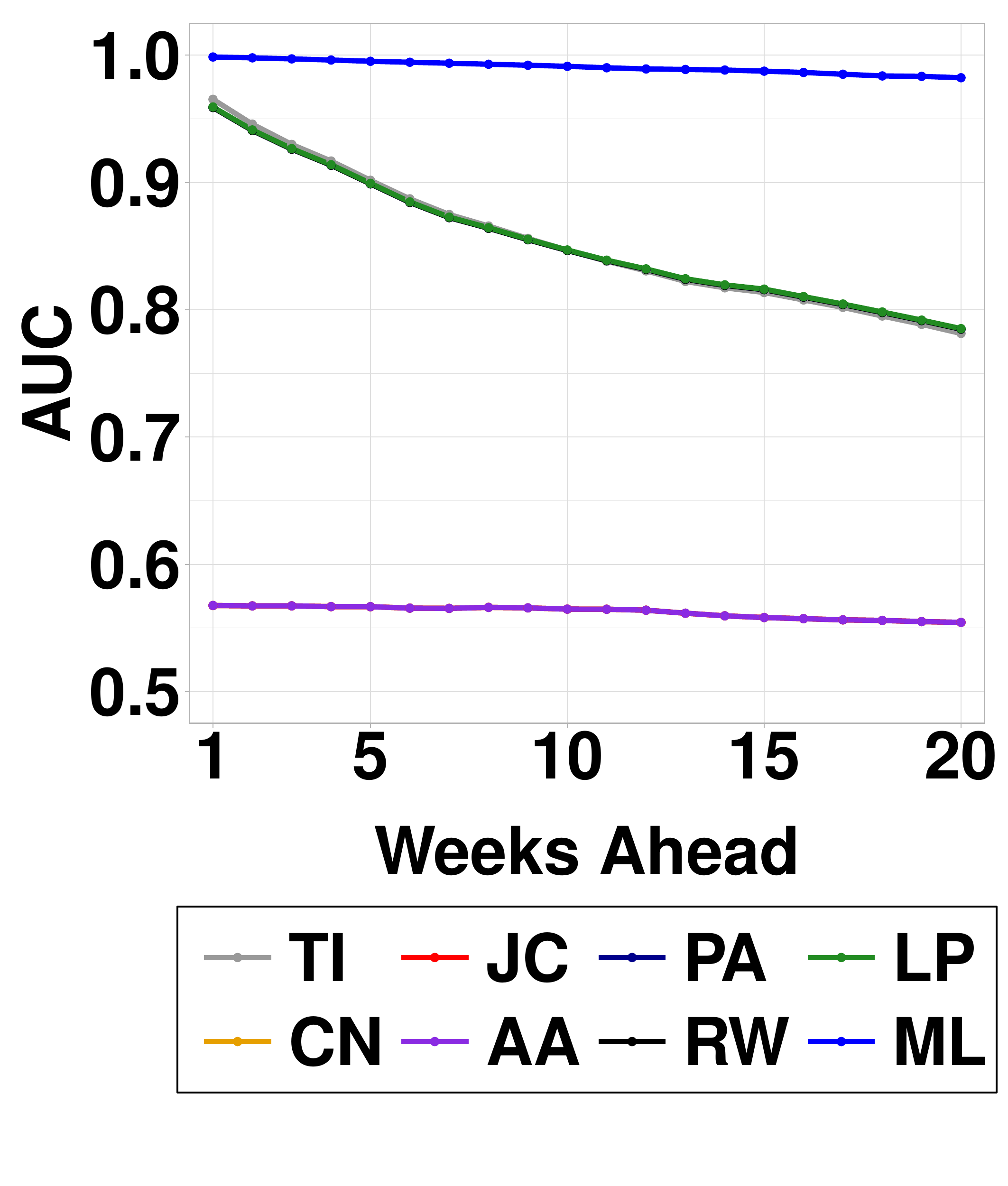}}
	\caption{\textbf{DTN - Predictive performance comparison of all methods.} This figure shows the AUC measure of the machine learning method compared against the baseline methods. For each time step, we calculate the AUC average of each method and its respective standard error over the entire test period. The machine learning method outperforms the baseline methods in all market indices.}
	\label{fig:auc-all-methods-dtn}
\end{figure*}

\begin{figure*}[h!]
	\centering
	\subfigure[DAX30]{\includegraphics[trim=0.2cm 6.2cm 0.1cm 0.2cm, clip=true, width=0.28\textwidth]{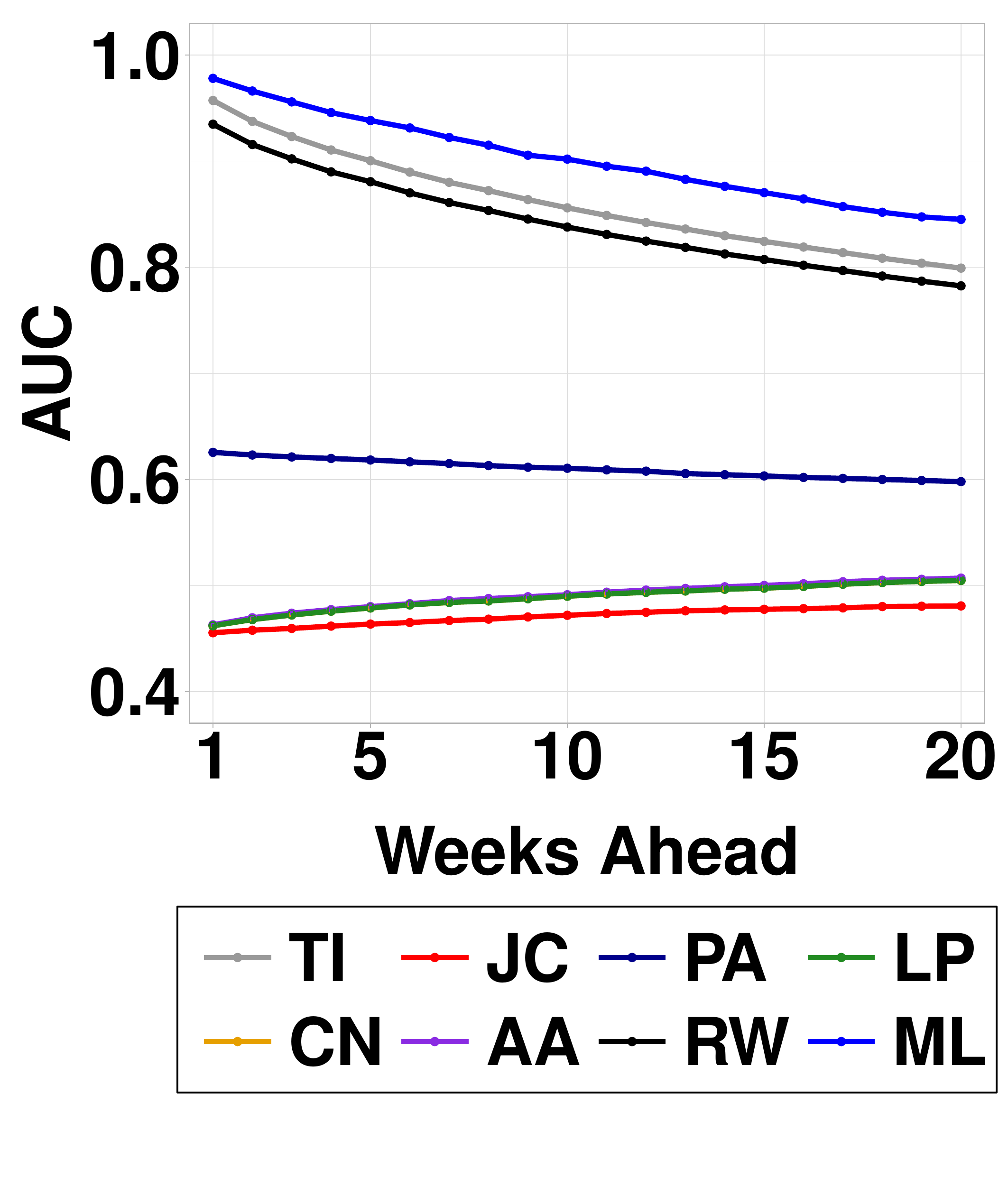}}
	\subfigure[EUROSTOXX50]{\includegraphics[trim=0.2cm 6.2cm 0.1cm 0.2cm, clip=true, width=0.28\textwidth]{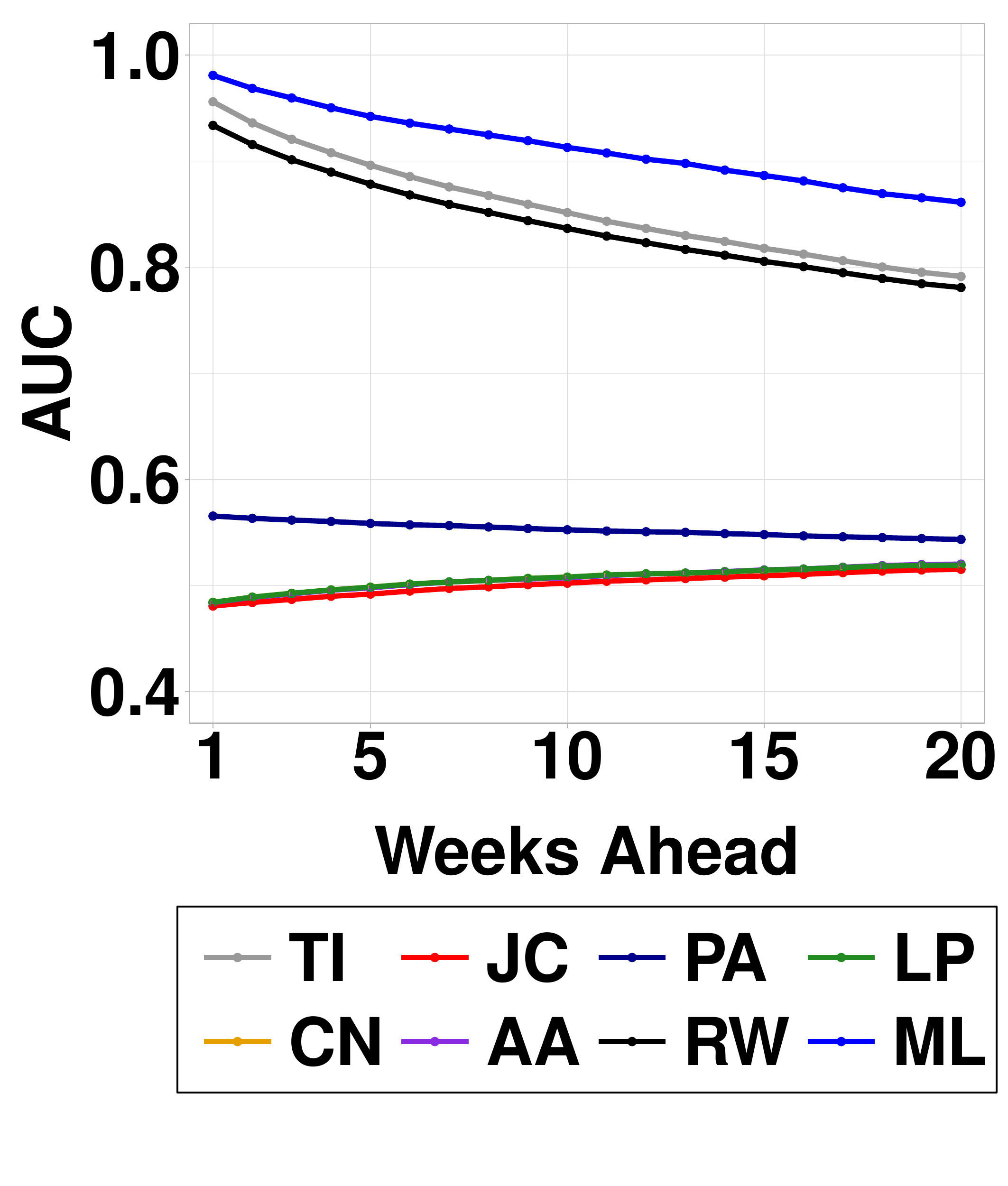}}
	\subfigure[FTSE100]{\includegraphics[trim=0.2cm 6.2cm 0.1cm 0.2cm, clip=true, width=0.28\textwidth]{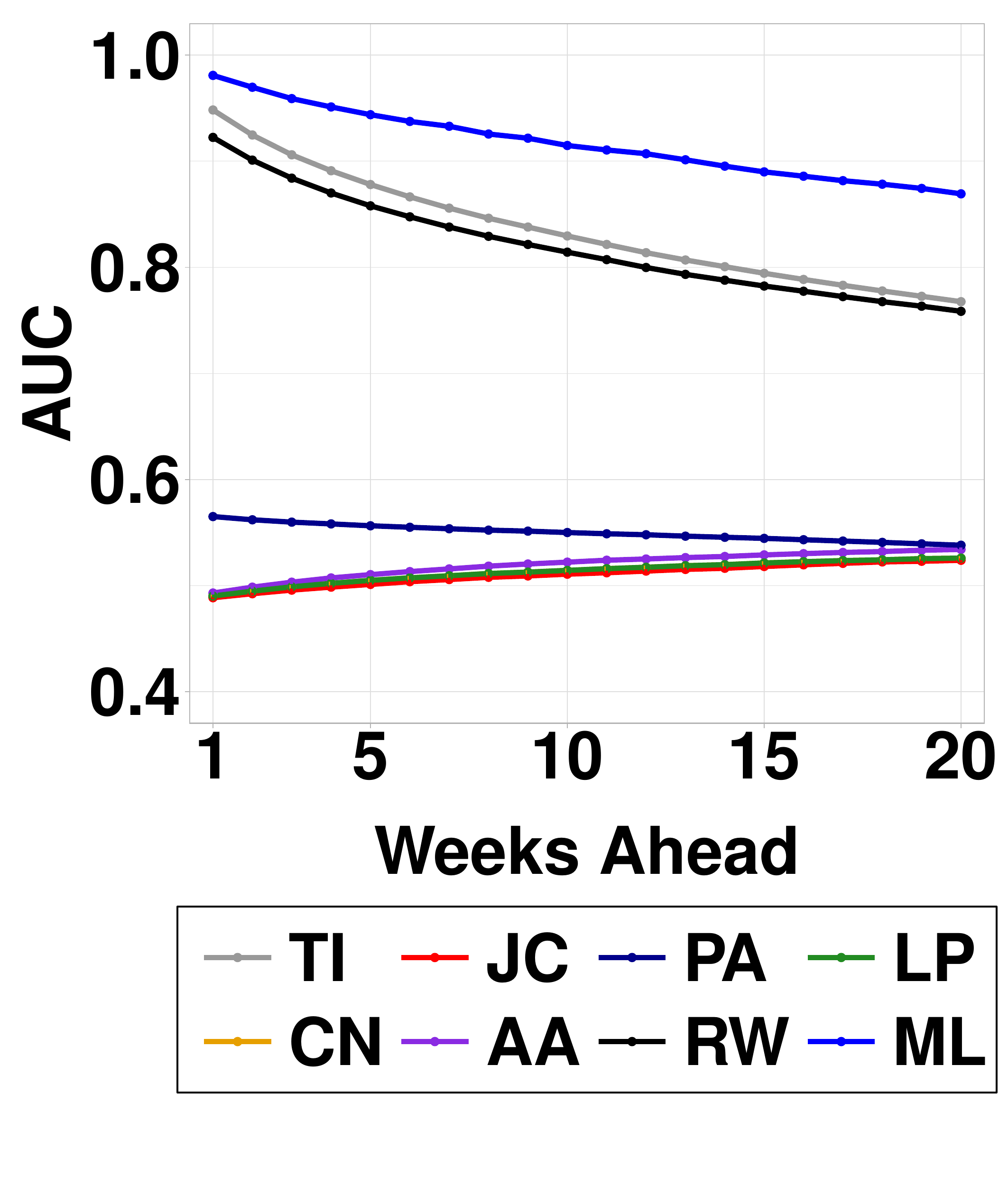}}
	\subfigure[HANGSENG50]{\includegraphics[trim=0.2cm 6.2cm 0.1cm 0.2cm, clip=true, width=0.28\textwidth]{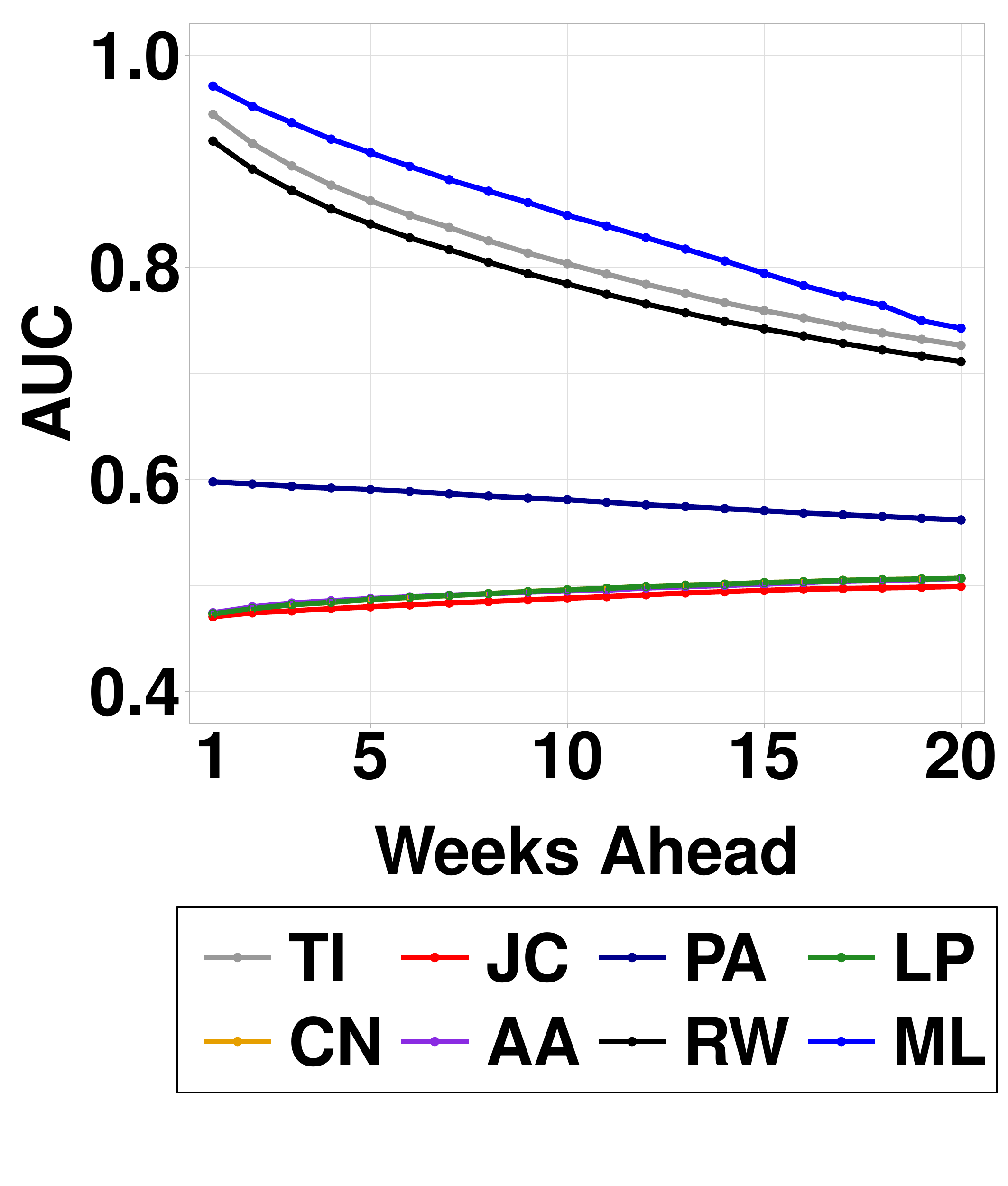}}
	\subfigure[NASDAQ100]{\includegraphics[trim=0.2cm 1.8cm 0.1cm 0.2cm, clip=true, width=0.2804\textwidth]{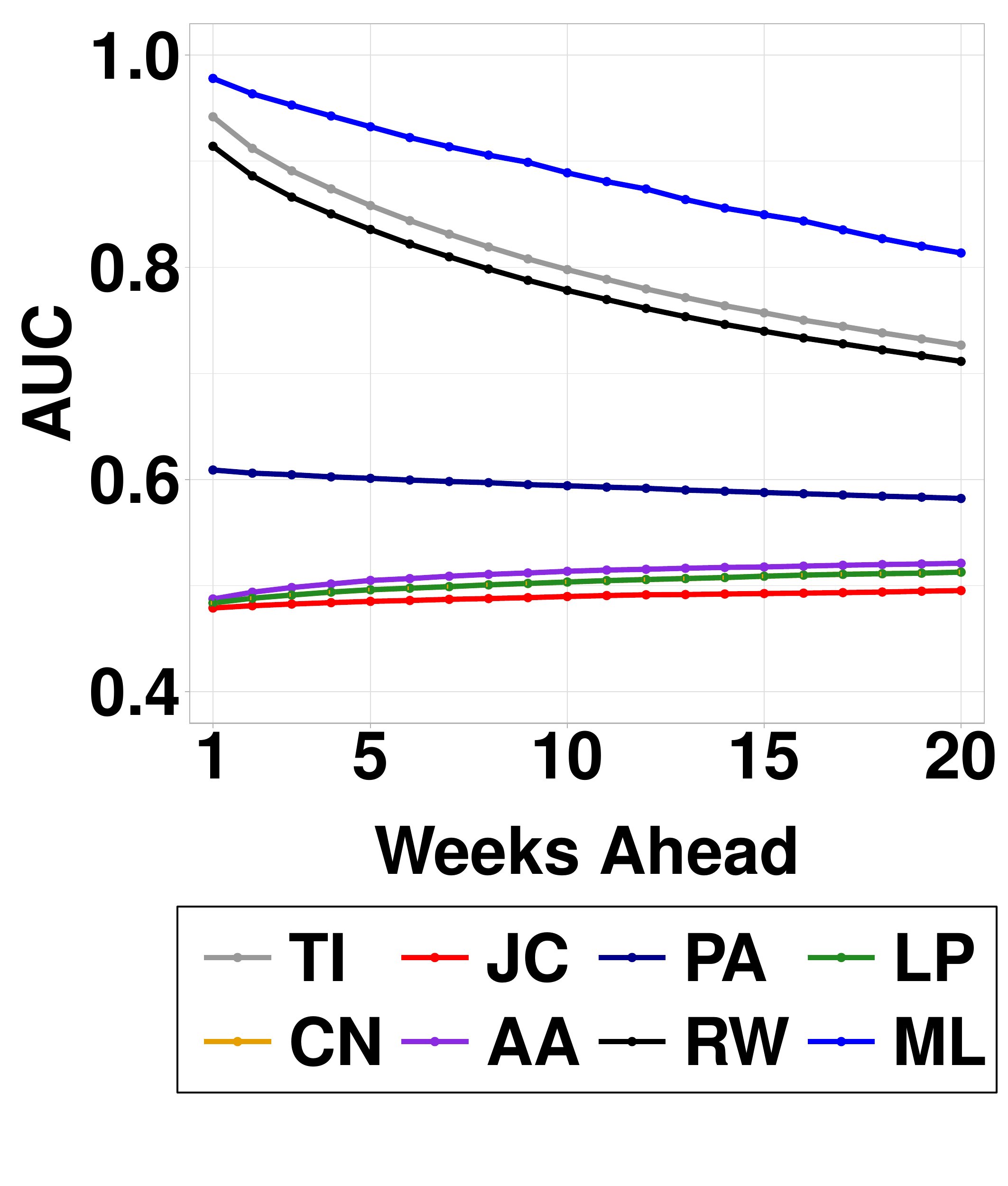}}
	\subfigure[NIFTY50]{\includegraphics[trim=0.2cm 6.2cm 0.1cm 0.2cm, clip=true, width=0.28\textwidth]{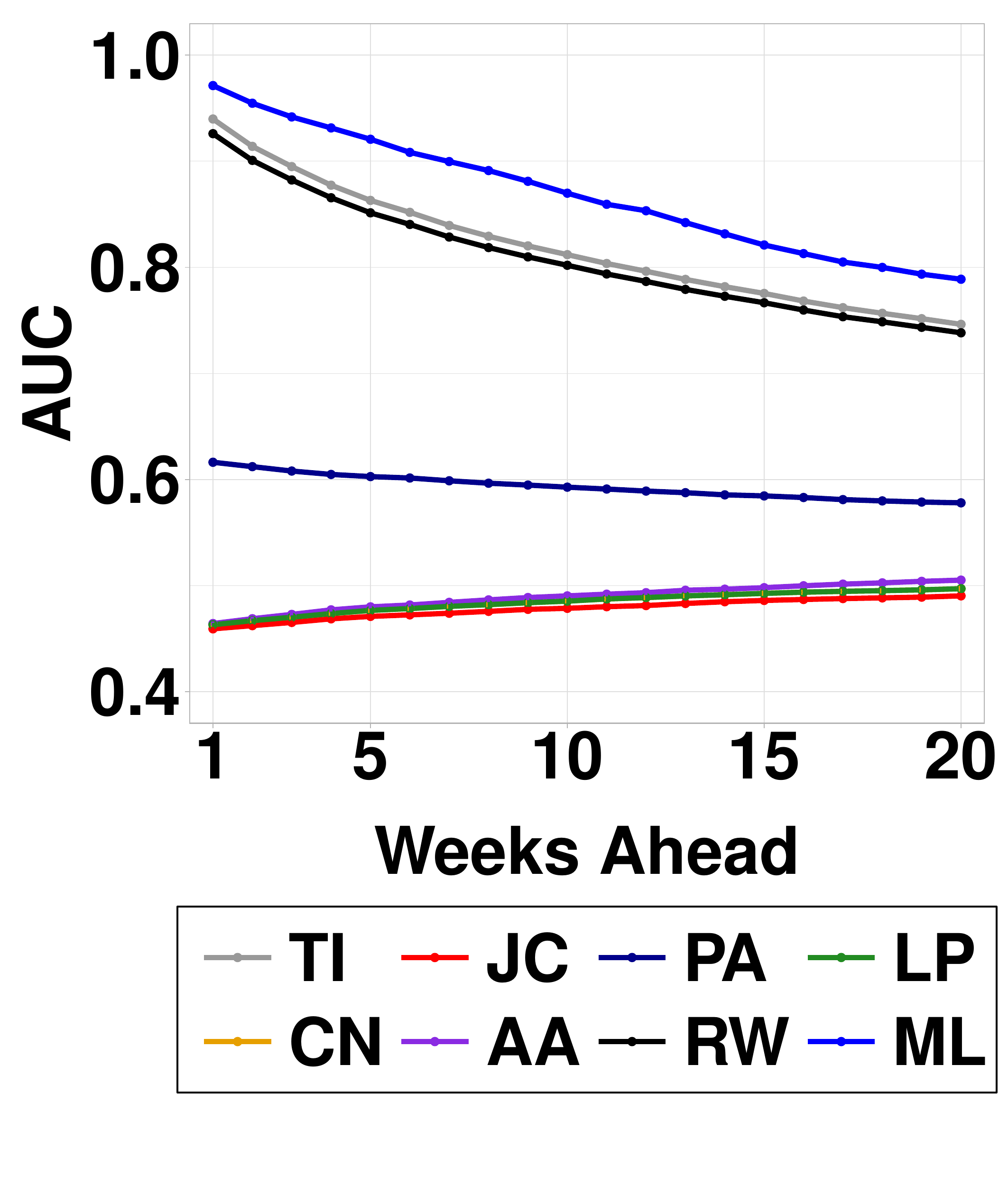}}
	\caption{\textbf{DMST - Predictive performance comparison of all methods.} This figure shows the AUC measure of the machine learning method compared against the baseline methods. For each time step, we calculate the AUC average of each method and its respective standard error over the entire test period. The machine learning method outperforms the baseline methods in all market indices.}
	\label{fig:auc-all-methods-dmst}
\end{figure*}

%In this section, we present results that address the research question \textbf{Q1}. 
We used a machine learning approach to forecast the financial network $G(t + h)$, where $h$ is the number of weeks ahead, $h = 1, 2, \dots, 20$ trading weeks. We discuss and report results using the size of rolling windows $L = 252$ trading days to construct the financial networks. Results regarding $L \in \lbrace 126, 504 \rbrace$ trading days can be found in the Supplementary Material, Section S.$4$. Figures~\ref{fig:auc-all-methods-dag},~\ref{fig:auc-all-methods-dtn} and~\ref{fig:auc-all-methods-dmst} show the AUC measure of the proposed machine learning method compared to baseline algorithms for DAG, DTN and DMST network filtering methods. For each time step ahead $h$, we calculated the average AUC of each method and its respective standard error over the test period, ranging from $5$ May $2007$ to $18$ December $2019$. 

\begin{figure*}[h!]
	\centering
	\subfigure[DAX30]{\includegraphics[trim=0.2cm 6.2cm 0.1cm 0.2cm, clip=true, width=0.28\textwidth]{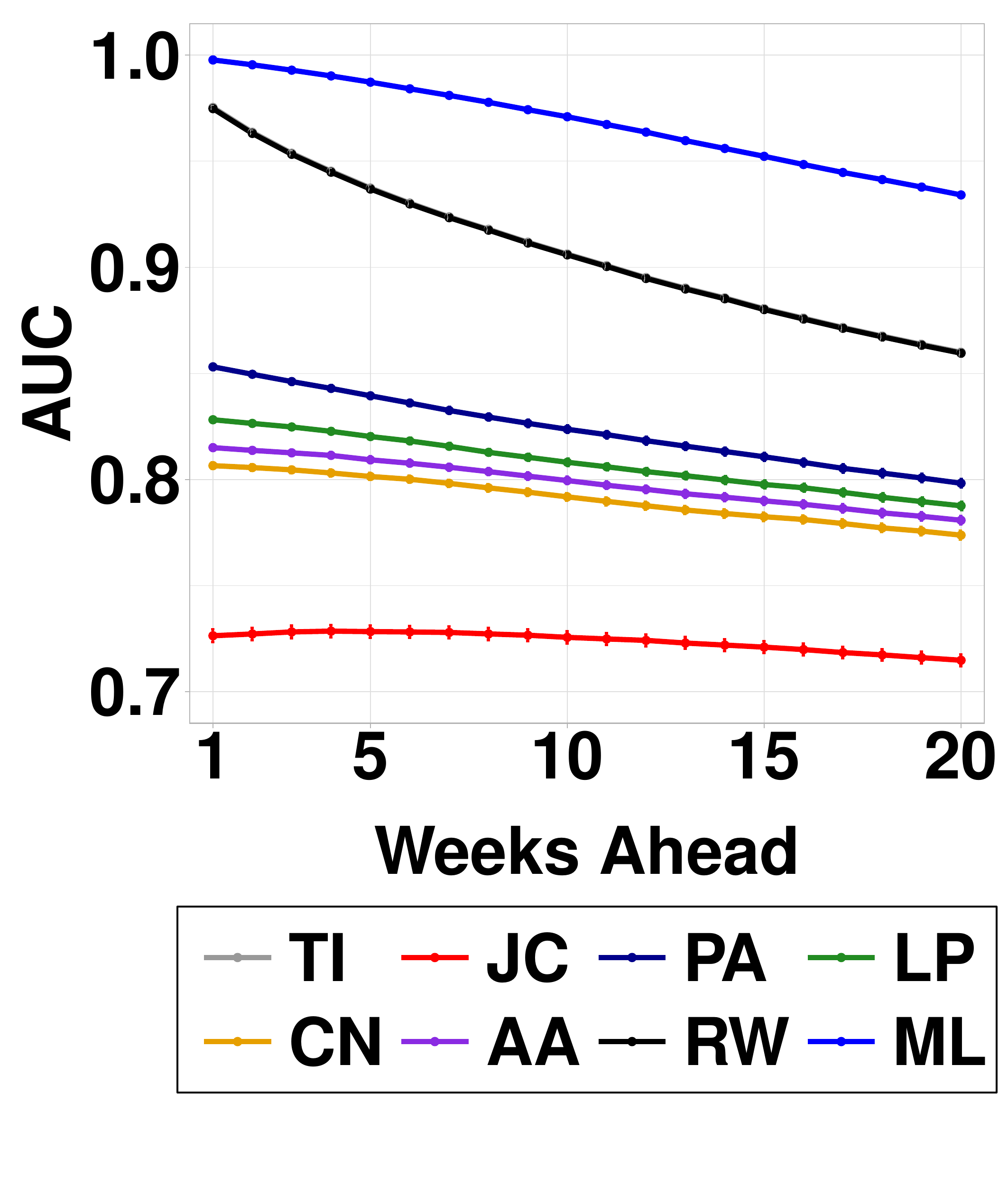}}
	\subfigure[EUROSTOXX50]{\includegraphics[trim=0.2cm 6.2cm 0.1cm 0.2cm, clip=true, width=0.28\textwidth]{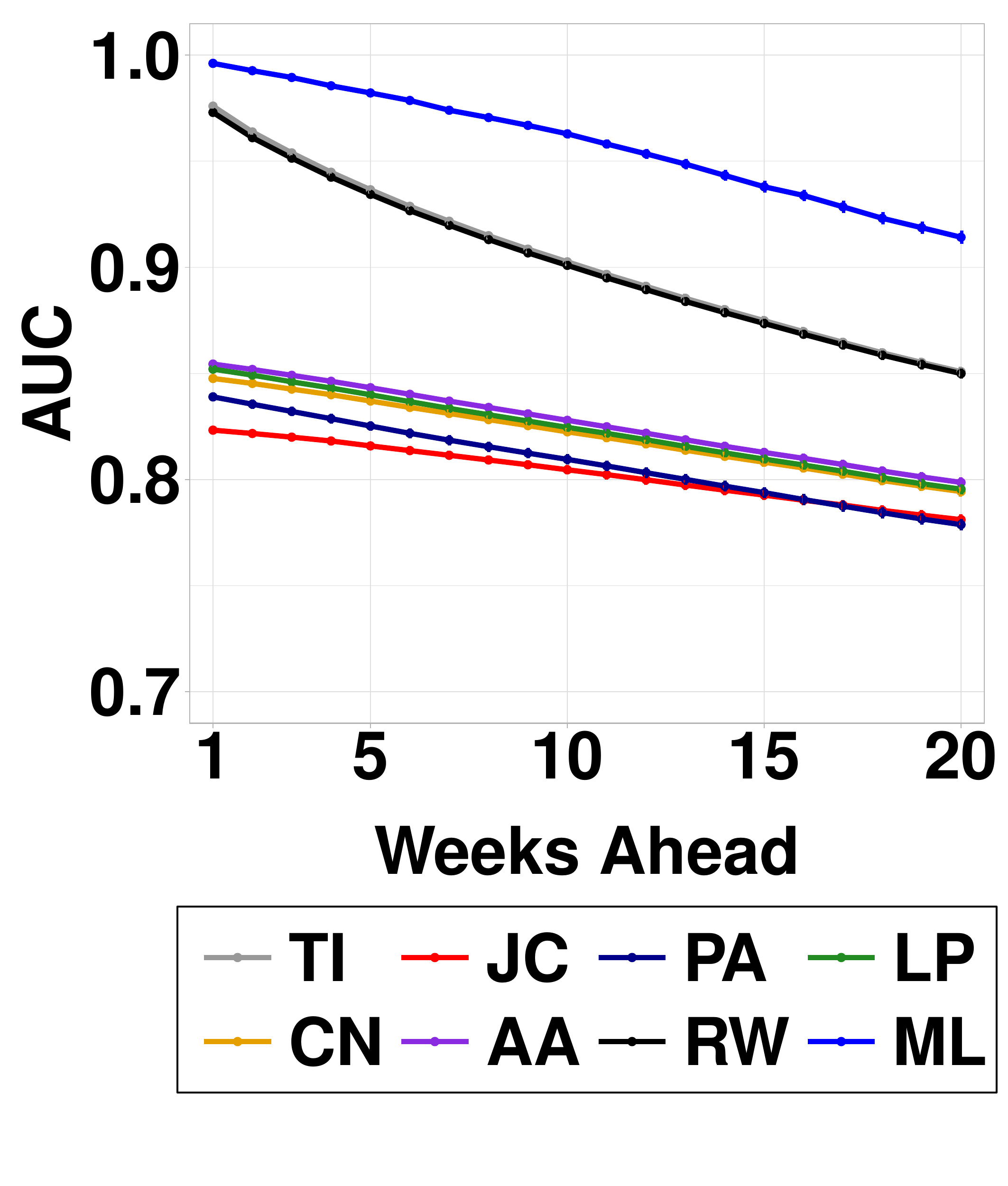}}
	\subfigure[FTSE100]{\includegraphics[trim=0.2cm 6.2cm 0.1cm 0.2cm, clip=true, width=0.28\textwidth]{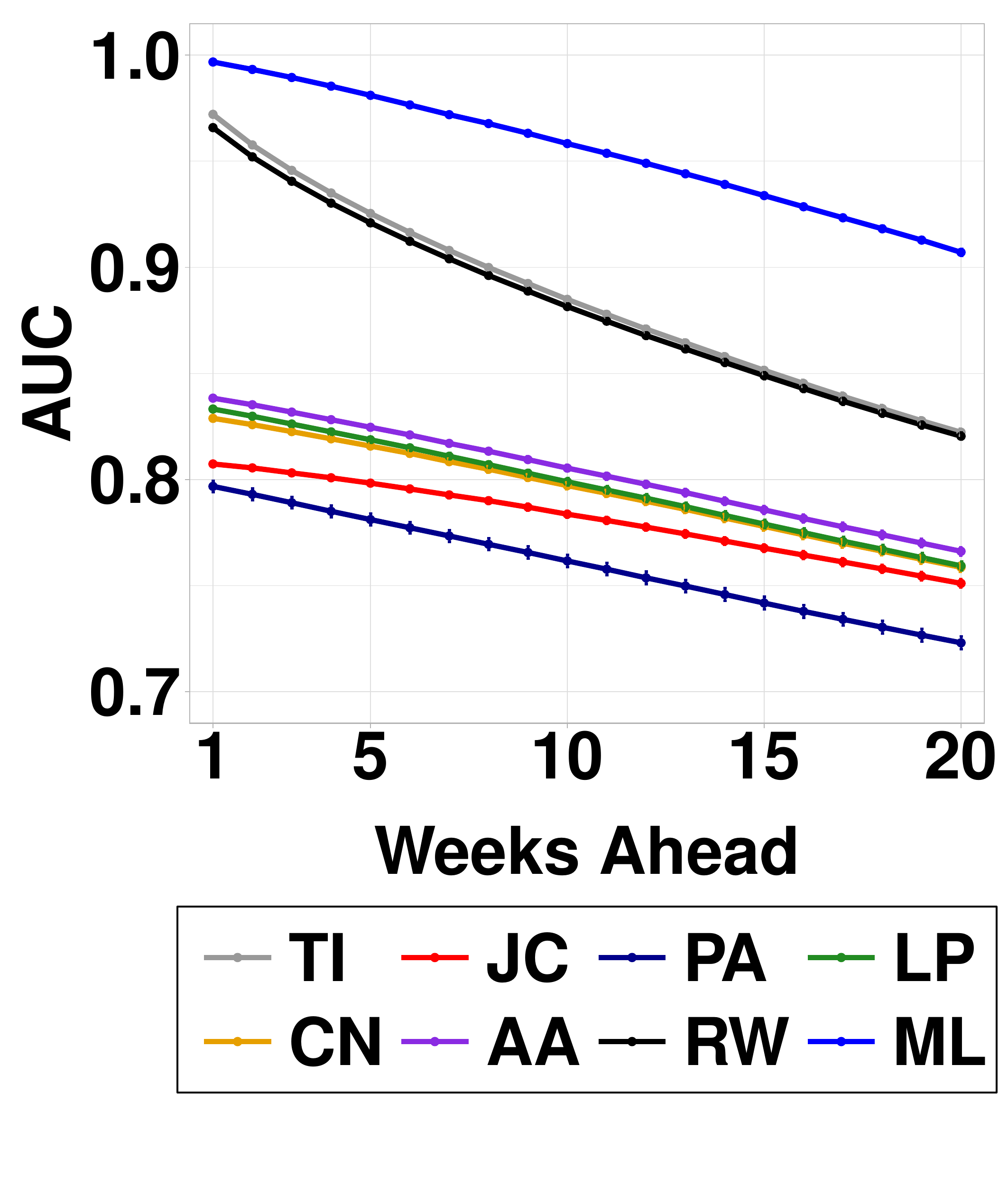}}
	\subfigure[HANGSENG50]{\includegraphics[trim=0.2cm 6.2cm 0.1cm 0.2cm, clip=true, width=0.28\textwidth]{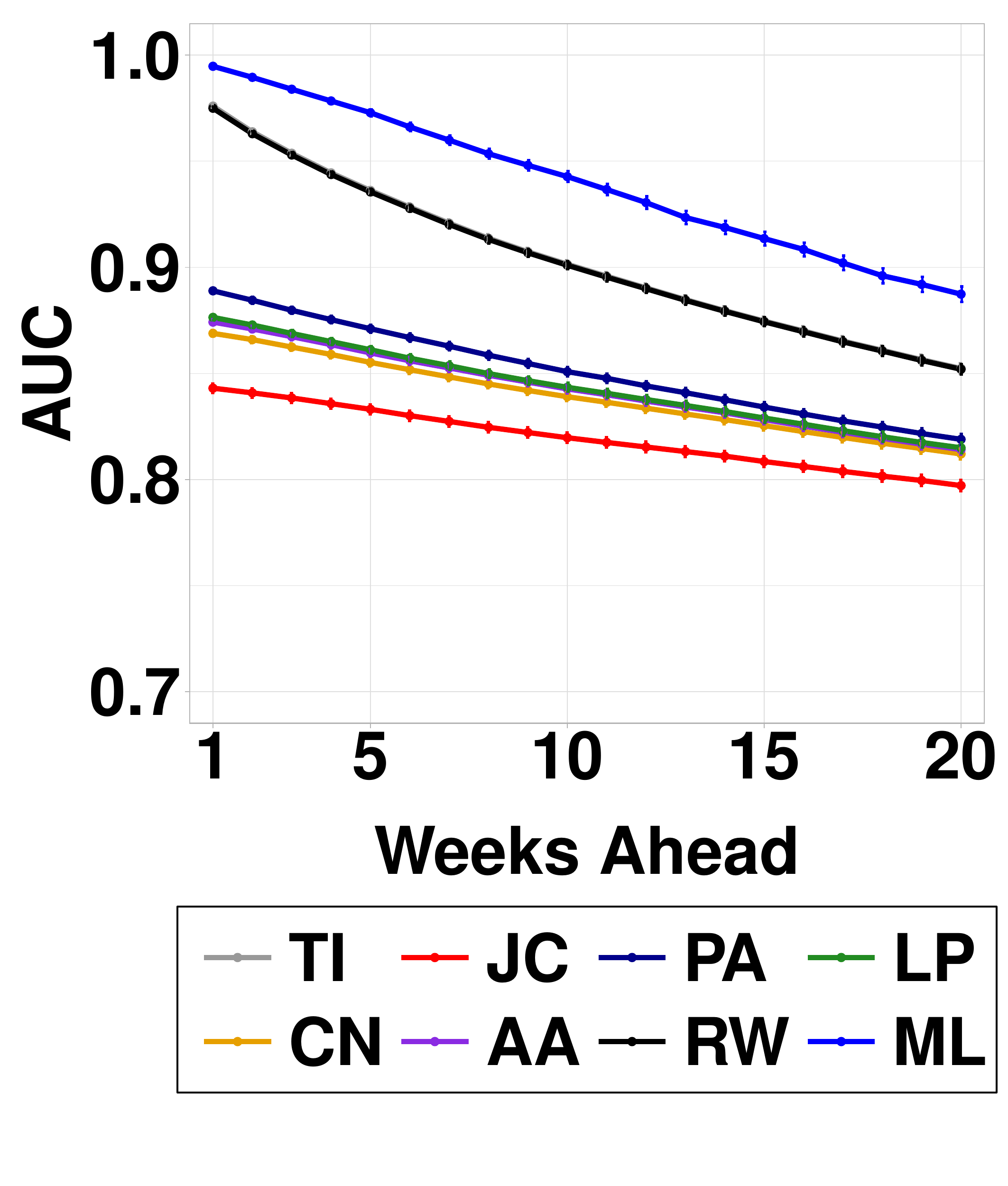}}
	\subfigure[NASDAQ100]{\includegraphics[trim=0.2cm 1.8cm 0.1cm 0.2cm, clip=true, width=0.2804\textwidth]{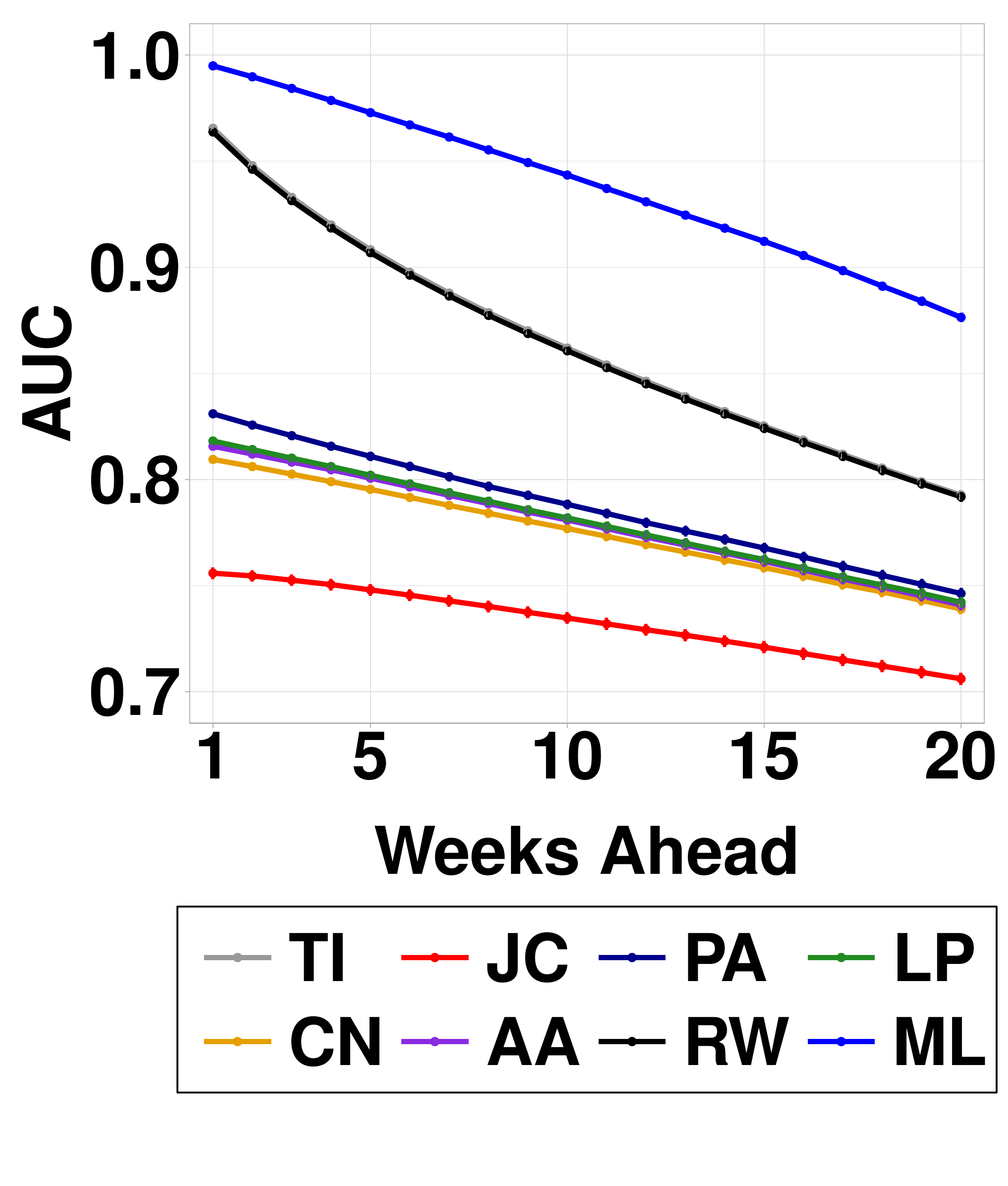}}
	\subfigure[NIFTY50]{\includegraphics[trim=0.2cm 6.2cm 0.1cm 0.2cm, clip=true, width=0.28\textwidth]{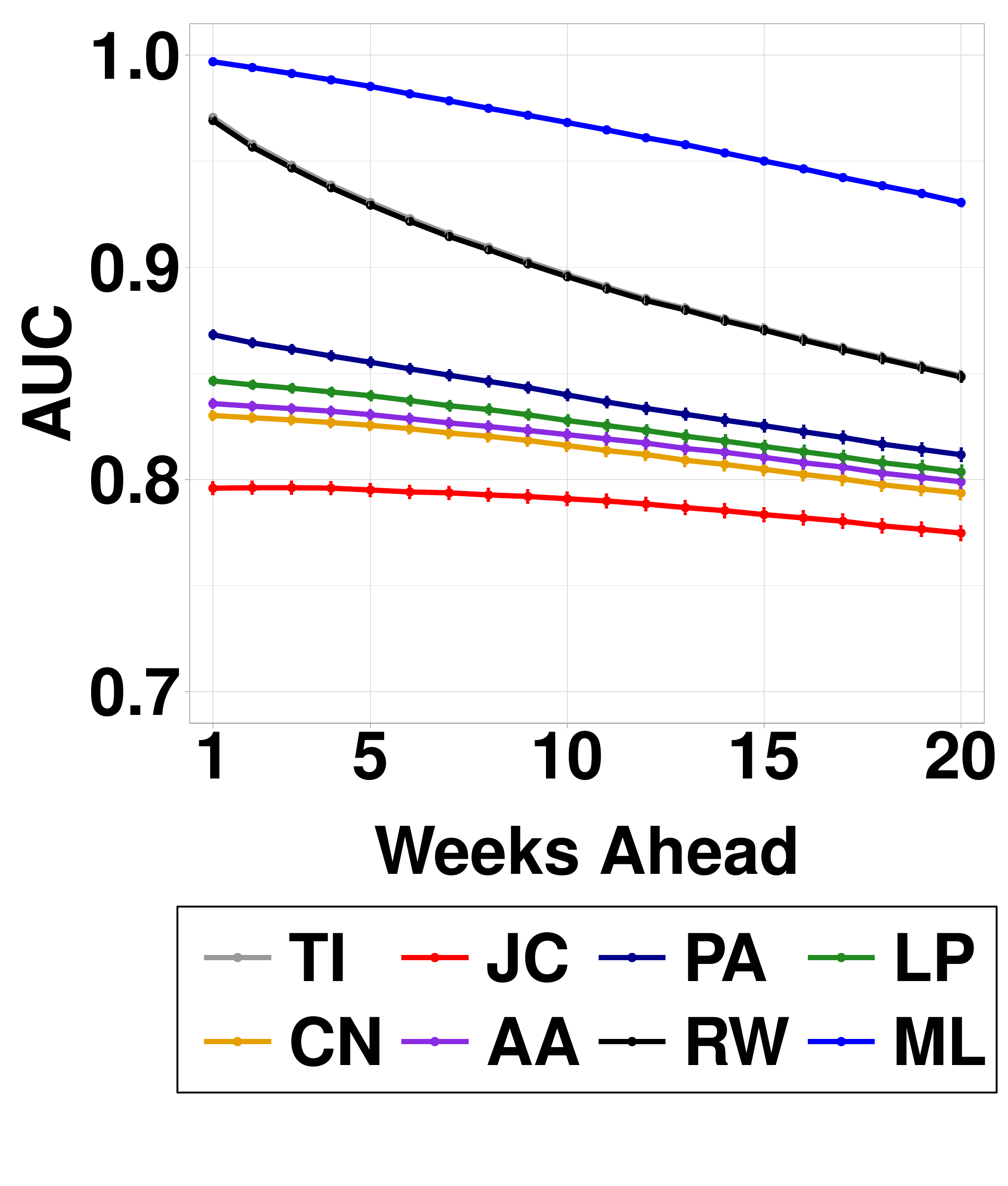}}
	\caption{\textbf{DAG - Predictive performance comparison of all methods.} This figure shows the AUC measure of the machine learning method compared to the baseline methods. For each time step, we calculate the AUC average of each method and its respective standard error over the entire test period. The machine learning method outperforms the baseline methods in all market indices.}
	\label{fig:auc-all-methods-dag}
\end{figure*}

Denoted as ``ML'', the machine learning method outperforms the baseline methods in all market indices and all network filtering methods. In general, predictive performance decreases as the time lag $h$ increases. Despite its simplicity, TI is quite effective and presents good performance across market indices and network filtering methods, similar to RW algorithm. Figure~\ref{fig:auc-all-methods-dag} presents results for the DAG network filtering method, suggesting that market indices with a small number of constituents have a higher AUC than markets with a large number of constituents. Results also suggest that the RW algorithm produces a edge ranking quite similar to TI. The JC method presents the worst predictive performance in all market indices, except for FTSE100 in which PA presents lower AUC values for the DAG network filtering method. 

Figure~\ref{fig:auc-all-methods-dtn} presents results for the DTN network filtering method. ML results are superior in all markets and suggest the proposed method can accurately identify links with high correlation due the main purpose of DTN method. We can observe that baseline algorithms have worst results for HANGSENG50, NASDAQ100 and NIFTY50 indices. As presented in Figure~\ref{fig:degree-cdf}, these market indices have expressive number of nodes without connections. TI algorithm outperforms baseline algorithms in DAX30, EUROSTOXX50 and NASDAQ100. Figure~\ref{fig:auc-all-methods-dmst} presents results related to the DMST network filtering method. Baseline methods have the worst results among the three filtering methods, except for the TI and RW algorithms. ML outperforms the benchmark methods in all markets. 

\begin{figure*}[h!]
	\centering
	\subfigure[DAG (AUC)]{\includegraphics[trim=0.0cm 0.5cm 9.2cm 1.3cm, clip=true, width=0.29\textwidth]{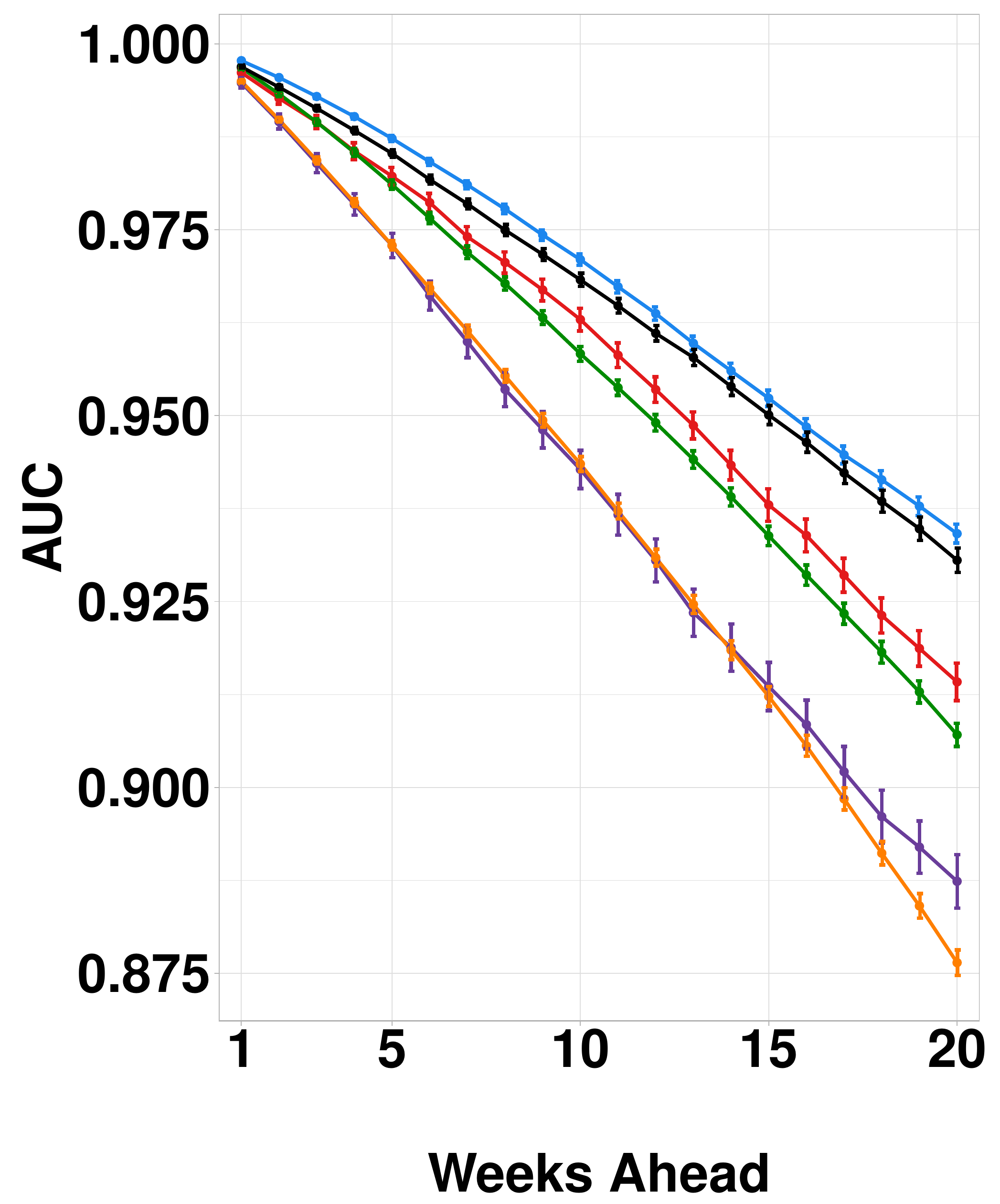}} 
	\subfigure[DAG (AUC$^\ast$)]{\includegraphics[trim=0.0cm 0.5cm 0.0cm 1.3cm, clip=true, width=0.416875\textwidth]{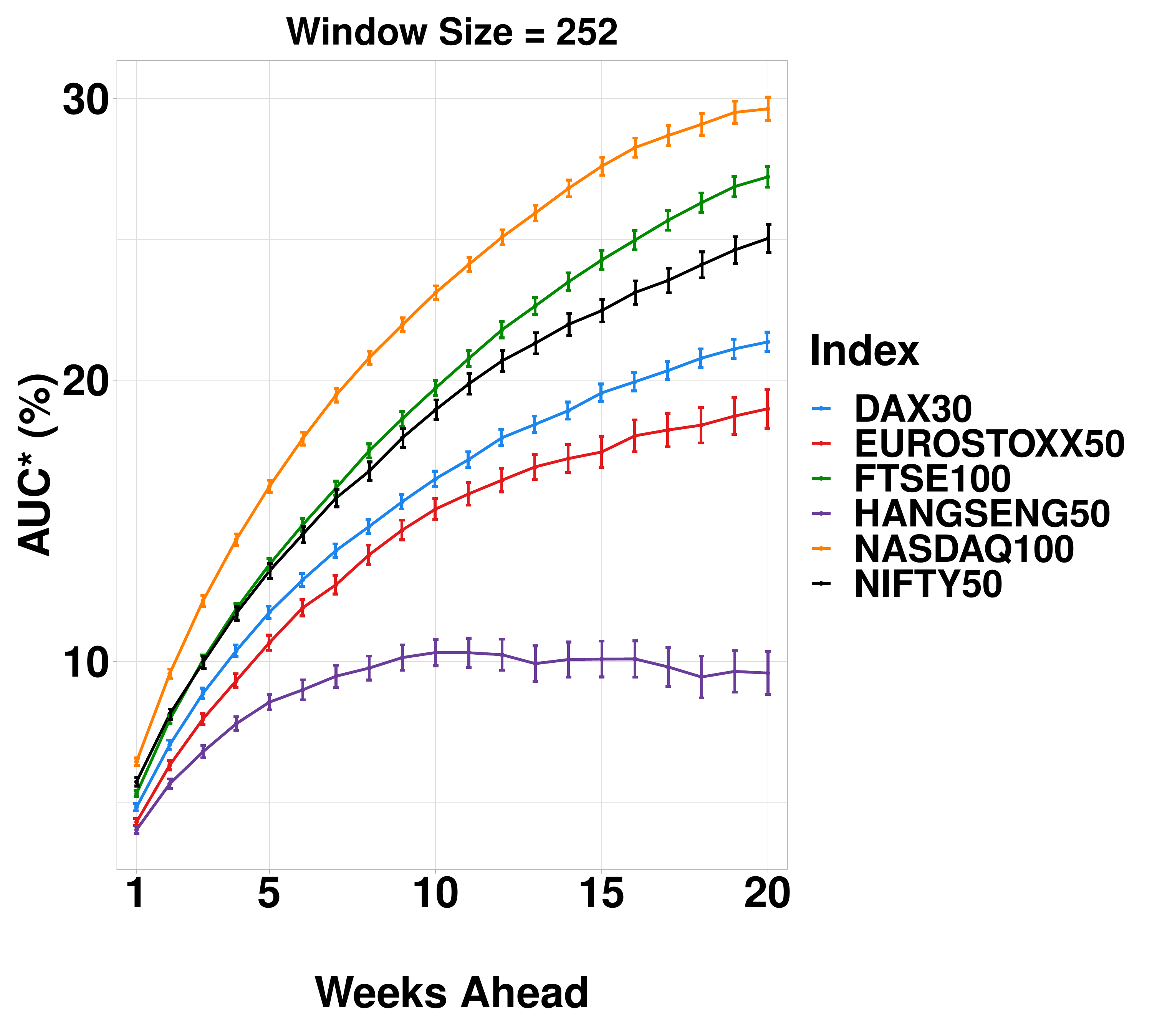}}
	\subfigure[DTN (AUC)]{\includegraphics[trim=0.0cm 0.5cm 9.2cm 1.3cm, clip=true, width=0.29\textwidth]{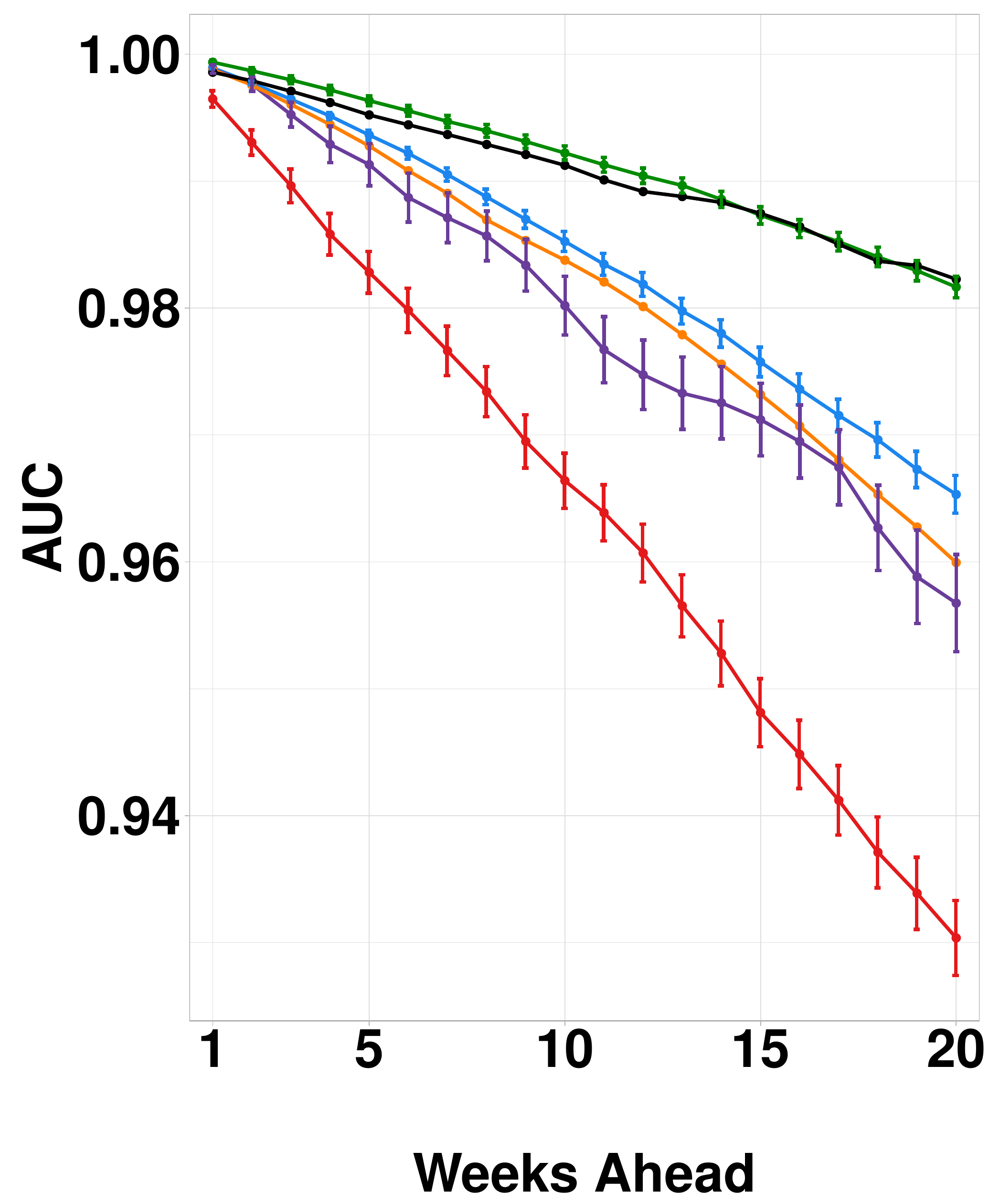}} 
	\subfigure[DTN (AUC$^\ast$)]{\includegraphics[trim=0.0cm 0.5cm 0.0cm 1.3cm, clip=true, width=0.416875\textwidth]{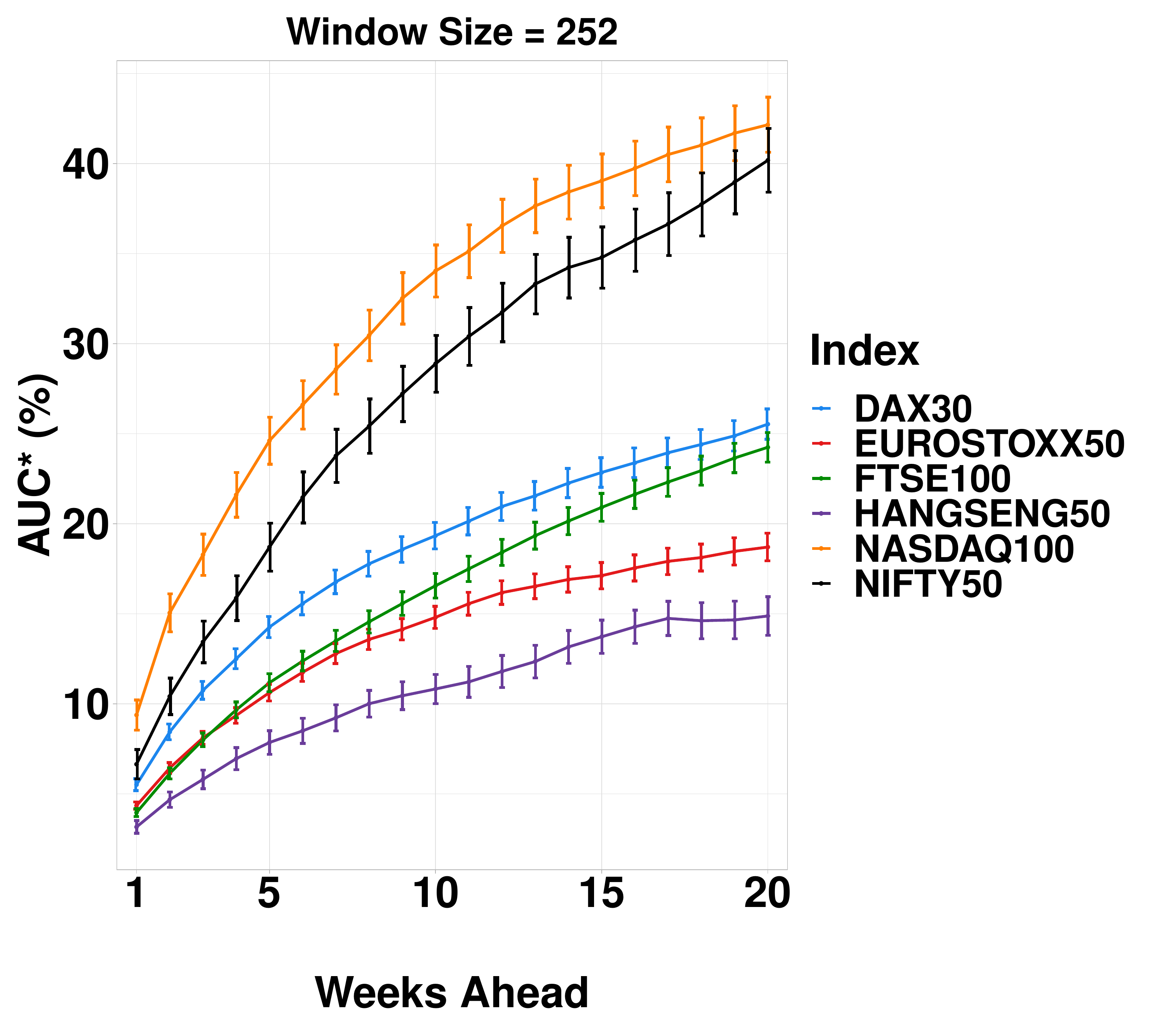}}
	\subfigure[DMST (AUC)]{\includegraphics[trim=0.0cm 0.5cm 9.2cm 1.3cm, clip=true, width=0.29\textwidth]{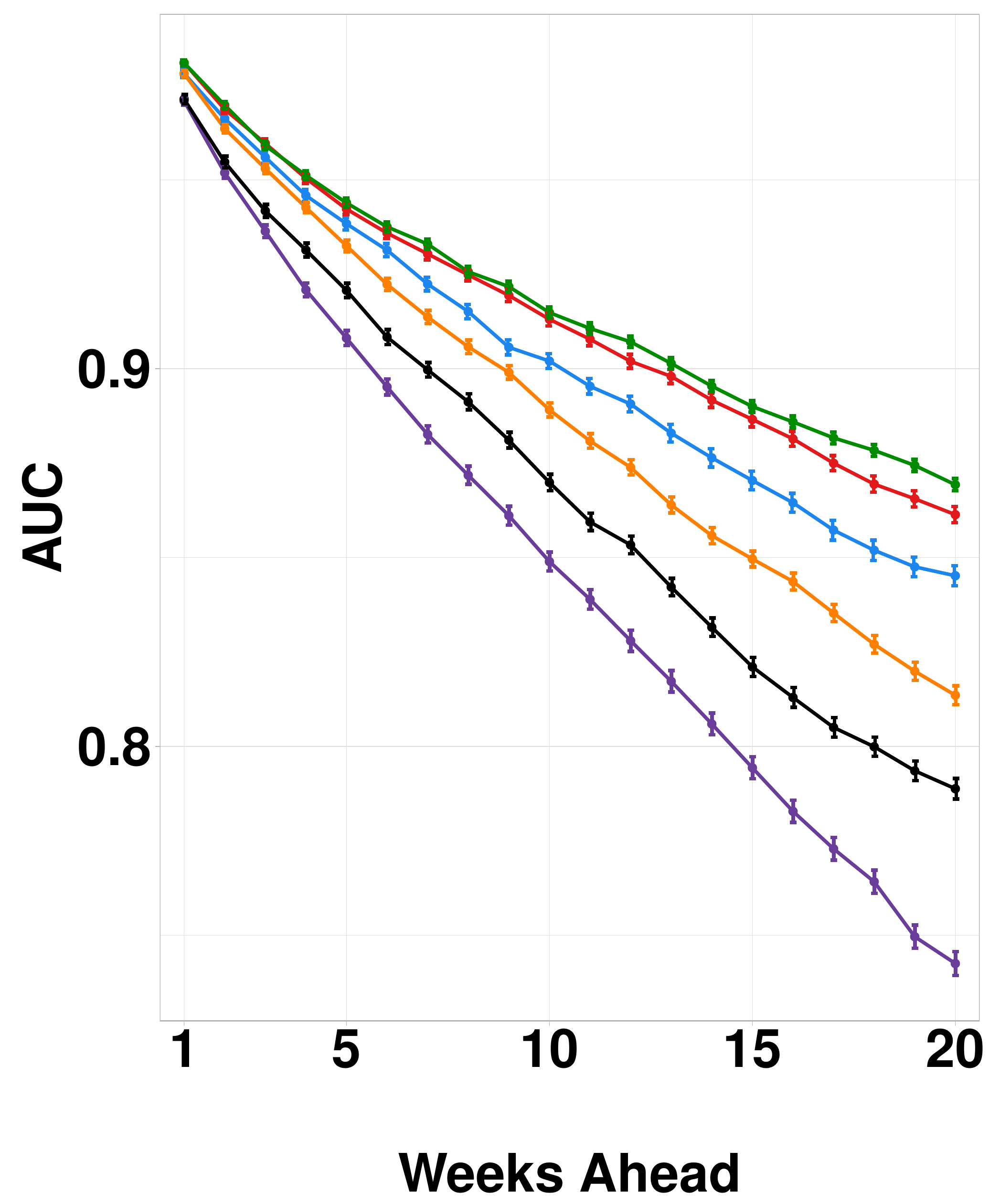}} 
	\subfigure[DMST (AUC$^\ast$)]{\includegraphics[trim=0.0cm 0.5cm 0.0cm 1.3cm, clip=true, width=0.416875\textwidth]{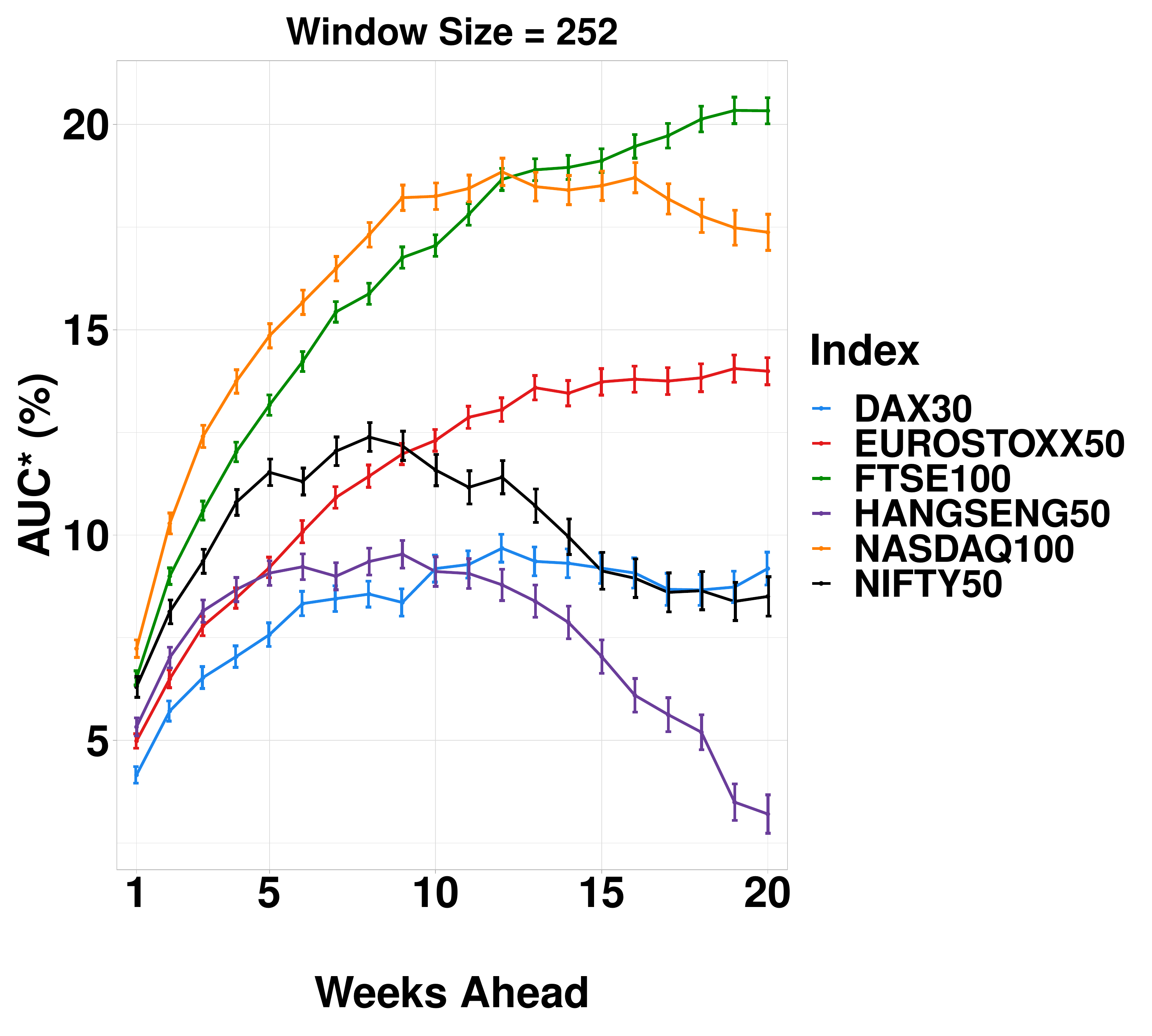}}
	\caption{\textbf{Machine learning AUC and AUC$^\ast$ for DAG, DTN and DMST network filtering methods.} Panels (a), (c) and (e) present the machine learning AUC measure and its standard error for $h$ trading weeks ahead ($1 \leq h \leq 20)$. Panels (b), (d) and (f) present the AUC improvement over the benchmark time-invariant method and its standard error. Results for $L = 252$.}
	\label{fig:auc-all-comparison}
\end{figure*}

Figure~\ref{fig:auc-all-comparison} presents the proposed method AUC performance for $h$ trading weeks ahead ($1 \leq h \leq 20)$ using the DAG, DTN and DMST network filtering methods. The AUC measure decreases as the time lag $h$ increases. We also compared our results against the benchmark time invariant method TI, where the network $G(t)$ is used as the forecast $G(t+h)$. We choose TI to compare our method due to its superior performance over all benchmark methods presented in the previous analysis. Moreover, we selected the TI method because it is derived from information from the pair-wise correlation, as described in Table~\ref{tab:linkfeatures}. The AUC* improvement is calculated as follows: 

\begin{equation}
    AUC^\ast = (AUC_m - 0.5) / (AUC_b - 0.5) - 1,
\end{equation}

\noindent where $AUC_m$ is the machine learning AUC and $AUC_b$ is the benchmark's AUC. Figures~\ref{fig:auc-all-comparison}(b),~\ref{fig:auc-all-comparison}(d) and~\ref{fig:auc-all-comparison}(f) present AUC$^\ast$ improvement results and their standard errors for DAG, DTN and DMST network filtering methods. 

The proposed method presents similar AUC results for all network filtering methods. Results using DAG shown in Figure~\ref{fig:auc-all-comparison}(a) suggest that networks with fewer constituents have better AUC results. Figure~\ref{fig:auc-all-comparison}(b) shows that the highest AUC$^\ast$ improvement is from NASDAQ100, reaching almost $30\%$ for $h = 20$ weeks ahead. On the other hand, for the DTN method shown in Figure~\ref{fig:auc-all-comparison}(c), the best results are FTSE100 and NIFTY50, in which EUROSTOXX50 is the most distinct result. The biggest AUC$^\ast$ improvement related to DTN shown in Figure~\ref{fig:auc-all-comparison}(d) is over NASDAQ100 and NIFTY50, reaching almost $40\%$. Results shown in Figure~\ref{fig:auc-all-comparison}(e) are related to the DMST network filtering method and have a similar decay of AUC for all markets, where DAX30 is the best result. Interestingly, the AUC$^\ast$ improvement shown in Figure~\ref{fig:auc-all-comparison}(e) presents similar curves to NIFTY50 and HANGSENG50 markets. Results show that AUC$^\ast$ improvement for NIFTY50 and HANGSENG50 increases until approximately $h = 9$, achieving almost $12\%$ on NIFTY50. After this max value, the AUC$^\ast$ improvement decreases as $h$ increases. NASDAQ100 presents the best AUC$^\ast$ improvement, reaching almost $19\%$ for $h = 15$ trading weeks ahead.

\subsubsection{Model Interpretability}

\begin{figure*}[t!]
	\centering
	\subfigure[DAG ($L = 126$)]{\includegraphics[trim=0.1cm 6.1cm 0.3cm 0cm, clip=true, clip=true, width=0.30\textwidth]{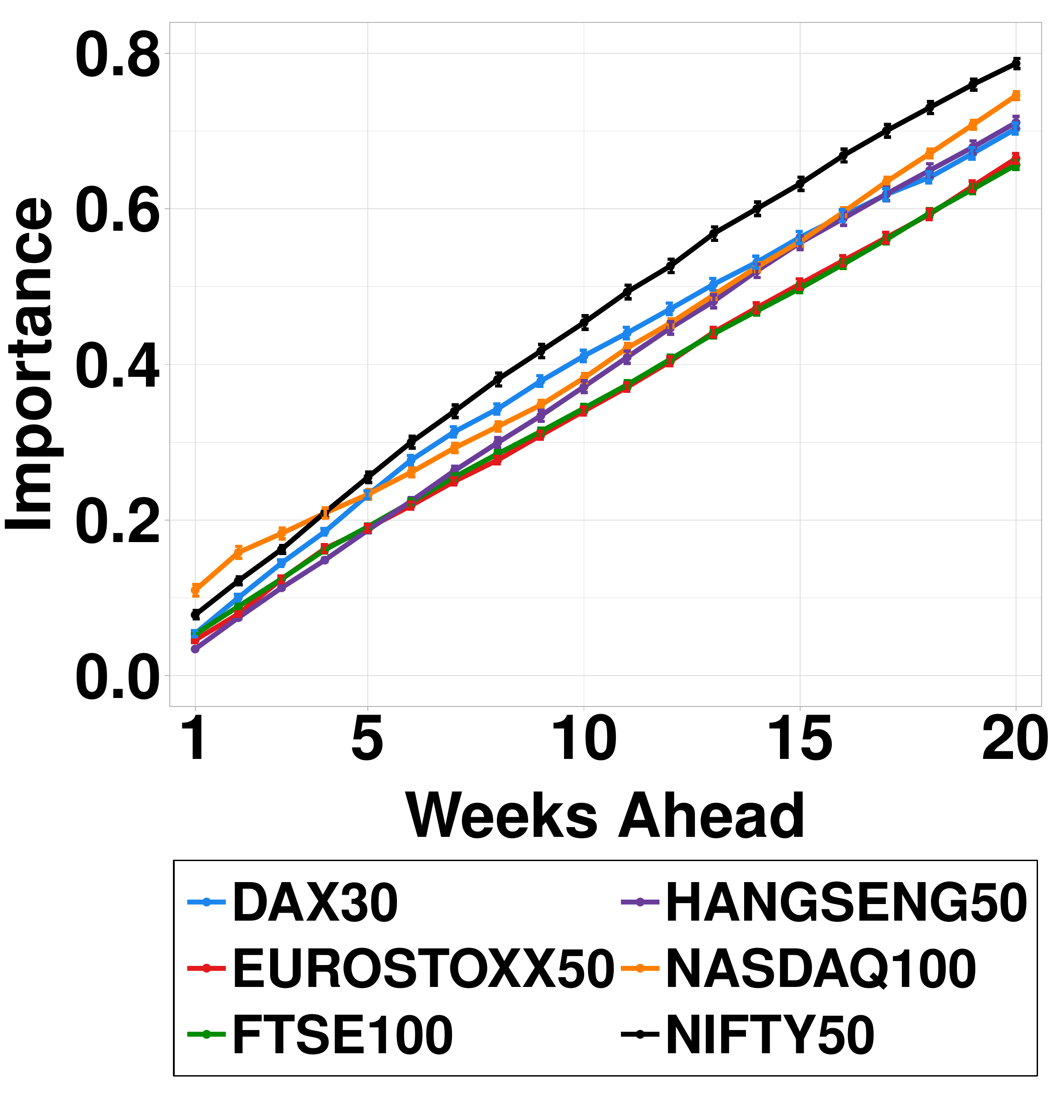}}
	\subfigure[DAG ($L = 252$)]{\includegraphics[trim=0.1cm 6.1cm 0.3cm 0cm, clip=true, clip=true, clip=true, width=0.30\textwidth]{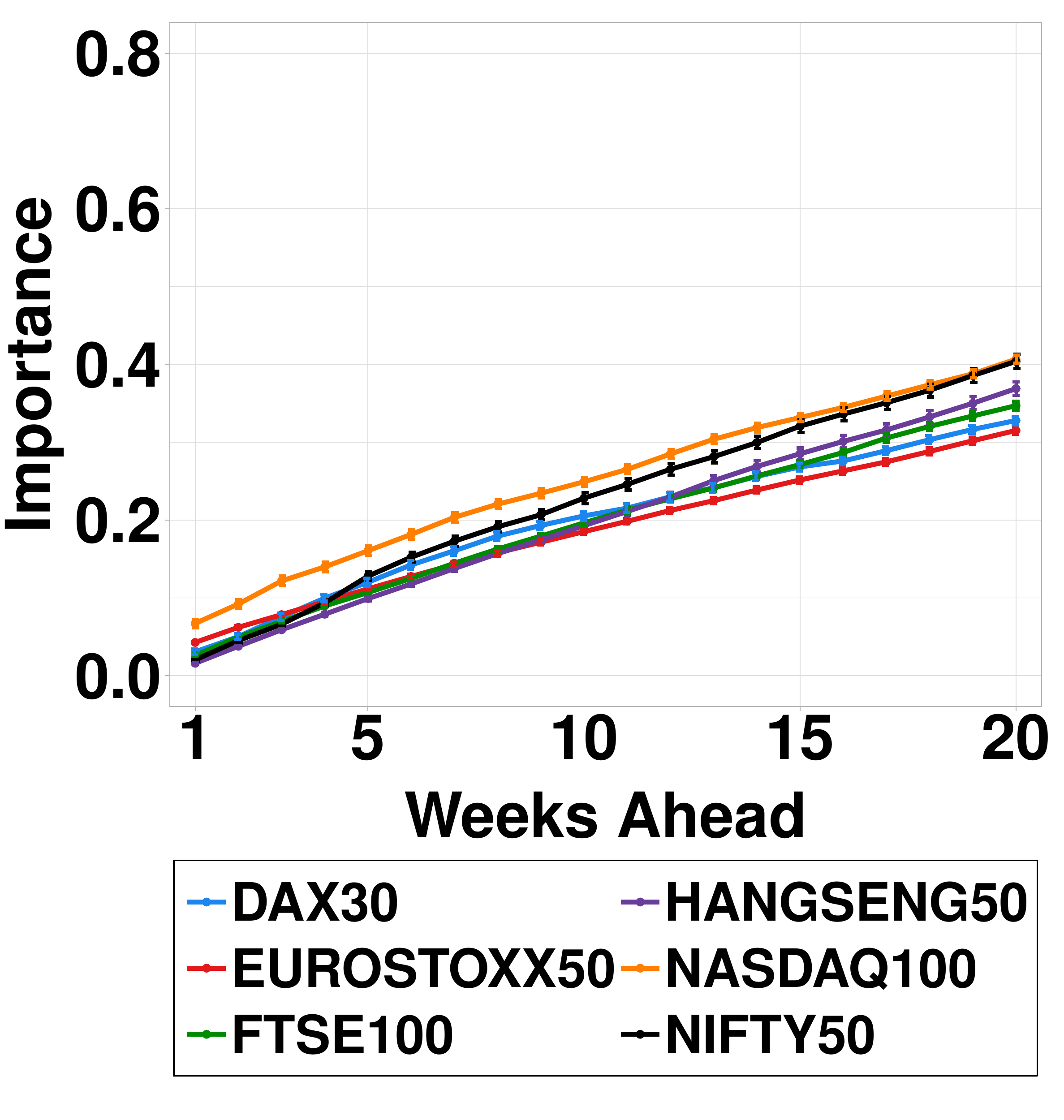}}
	\subfigure[DAG ($L = 504$)]{\includegraphics[trim=0.1cm 6.1cm 0.3cm 0cm, clip=true, clip=true, width=0.30\textwidth]{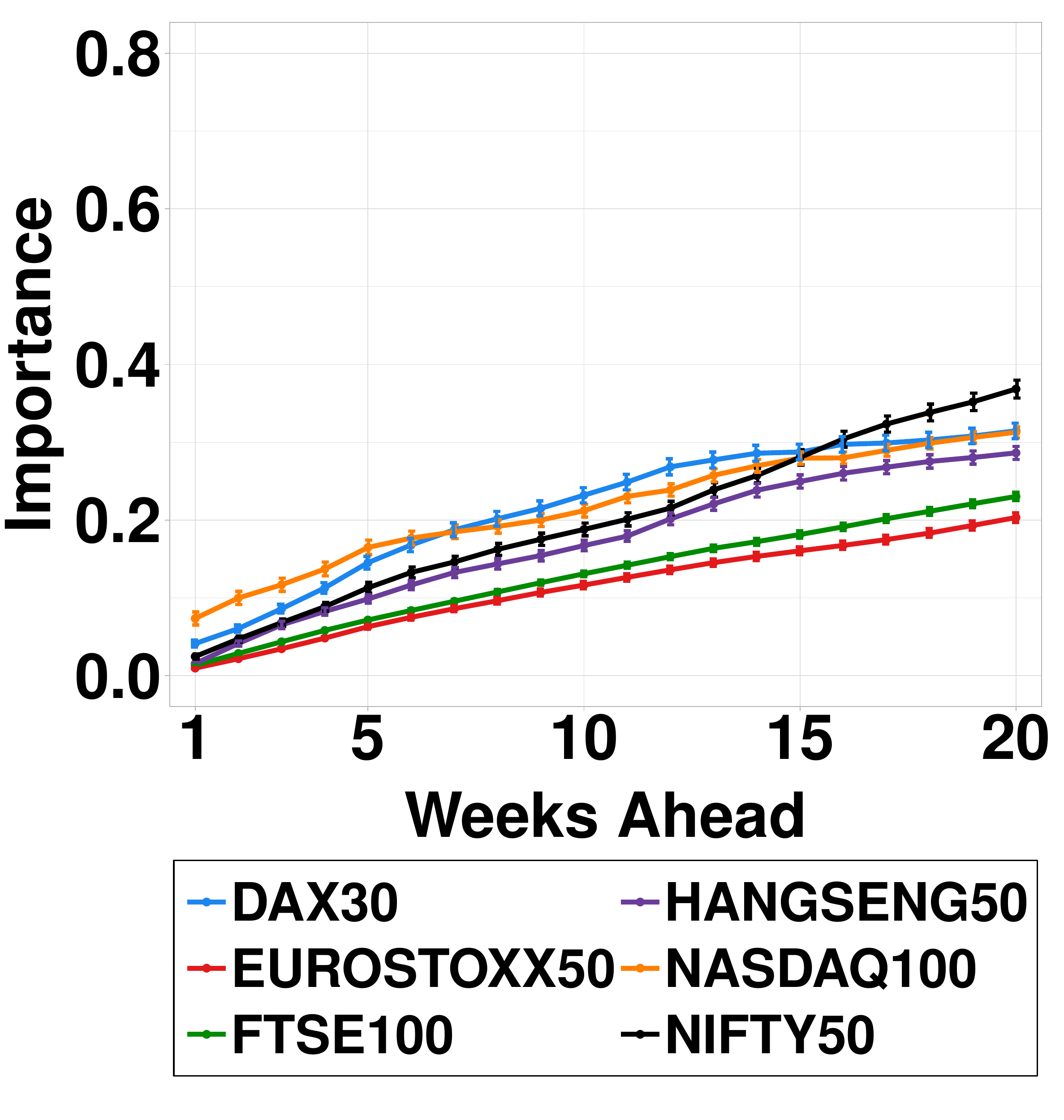}}
	\subfigure[DTN ($L = 126$)]{\includegraphics[trim=0.1cm 6.1cm 0.3cm 0cm, clip=true, clip=true, width=0.30\textwidth]{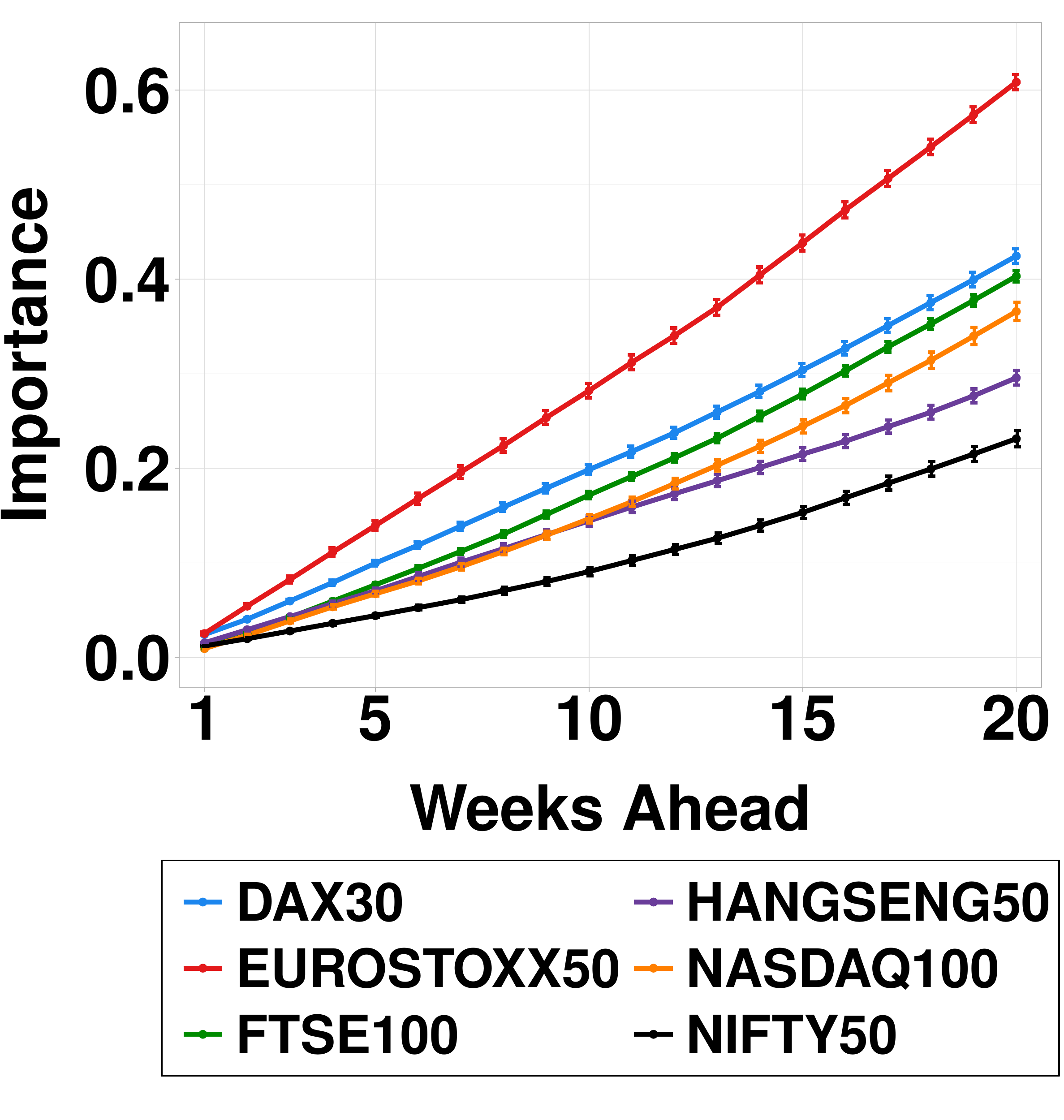}}
	\subfigure[DTN ($L = 252$)]{\includegraphics[trim=0.1cm 6.1cm 0.1cm 0cm, clip=true, clip=true, clip=true, width=0.30\textwidth]{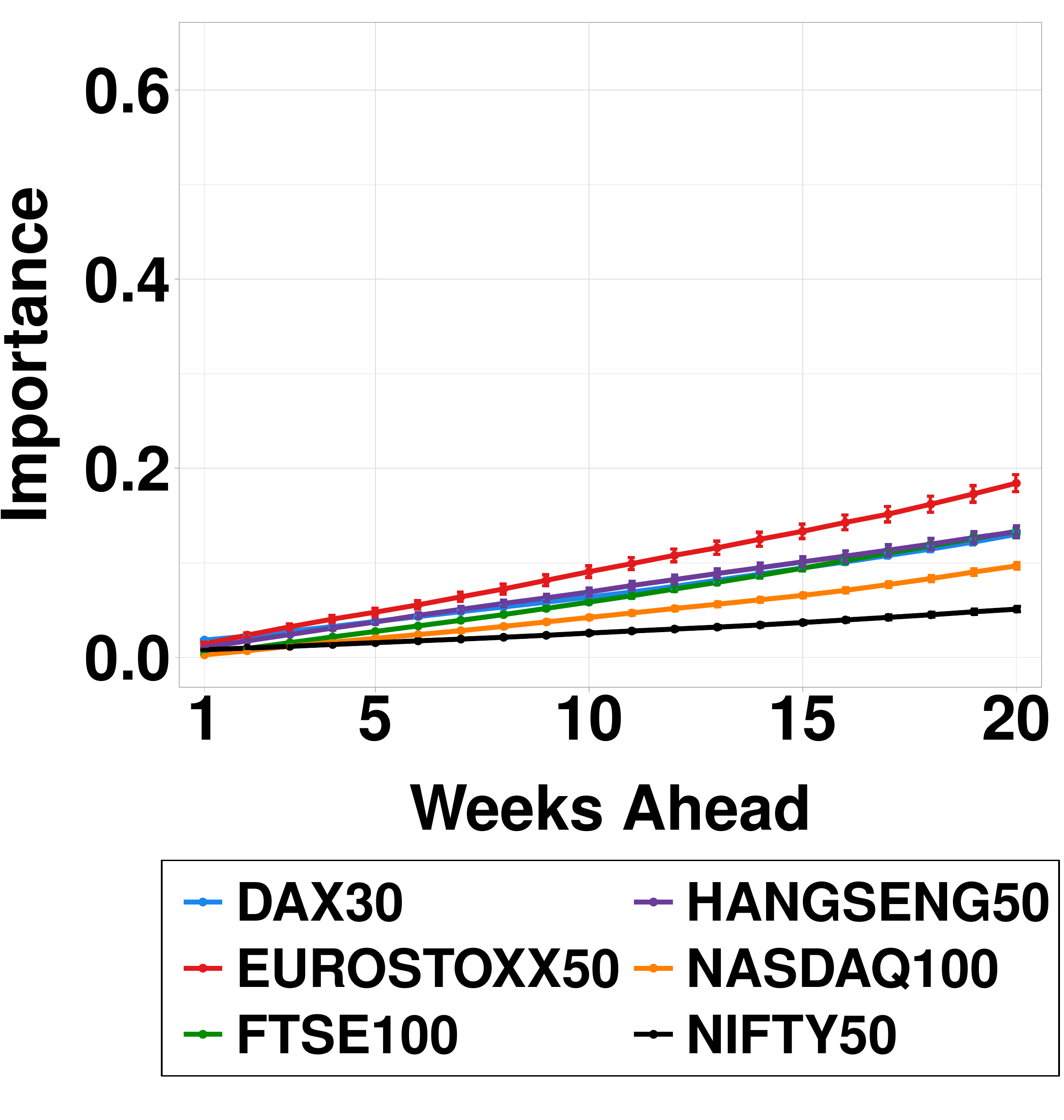}}
	\subfigure[DTN ($L = 504$)]{\includegraphics[trim=0.1cm 6.1cm 0.3cm 0cm, clip=true, clip=true, width=0.30\textwidth]{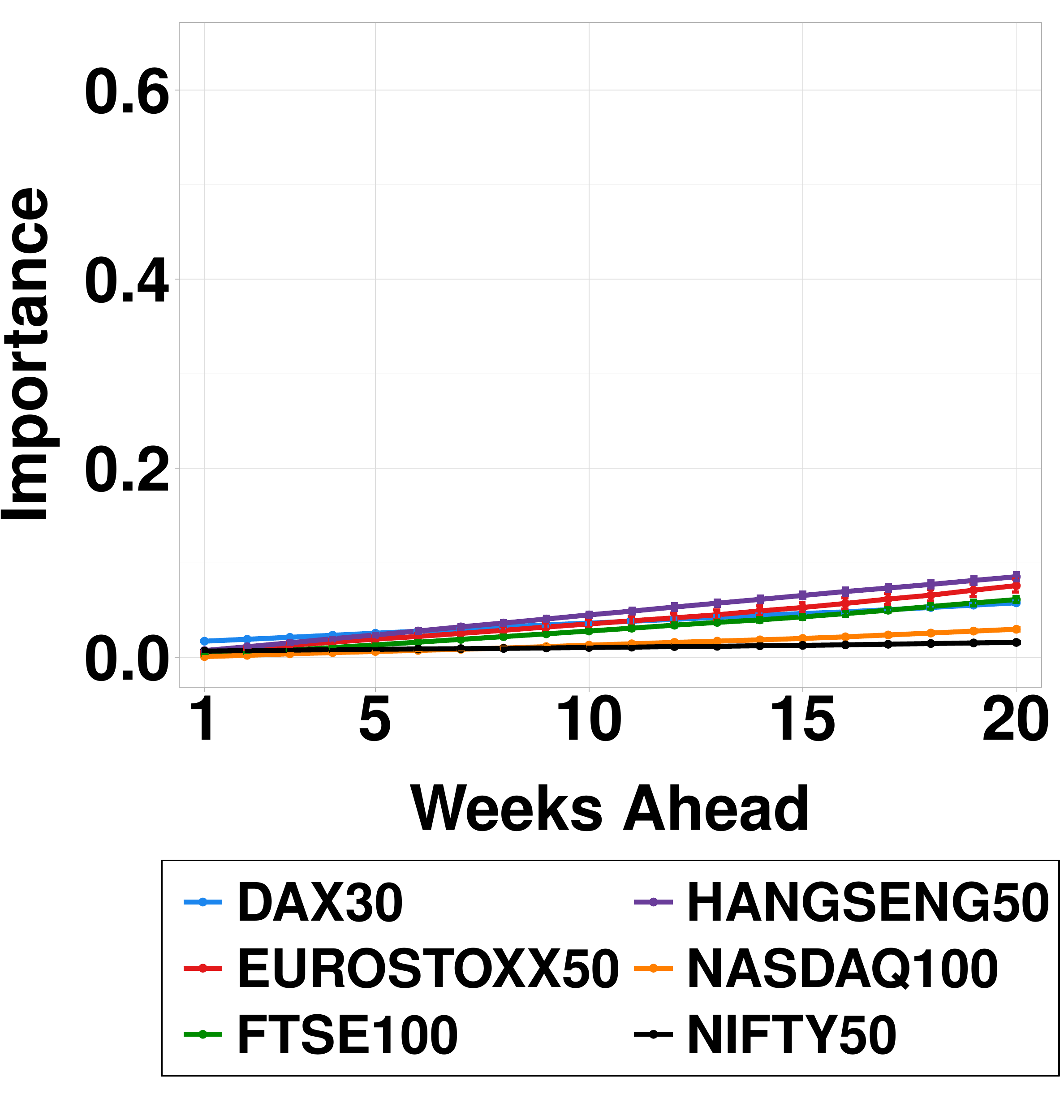}}
	\subfigure[DMST ($L = 126$)]{\includegraphics[trim=0.1cm 6.1cm 0.3cm 0cm, clip=true, clip=true, width=0.30\textwidth]{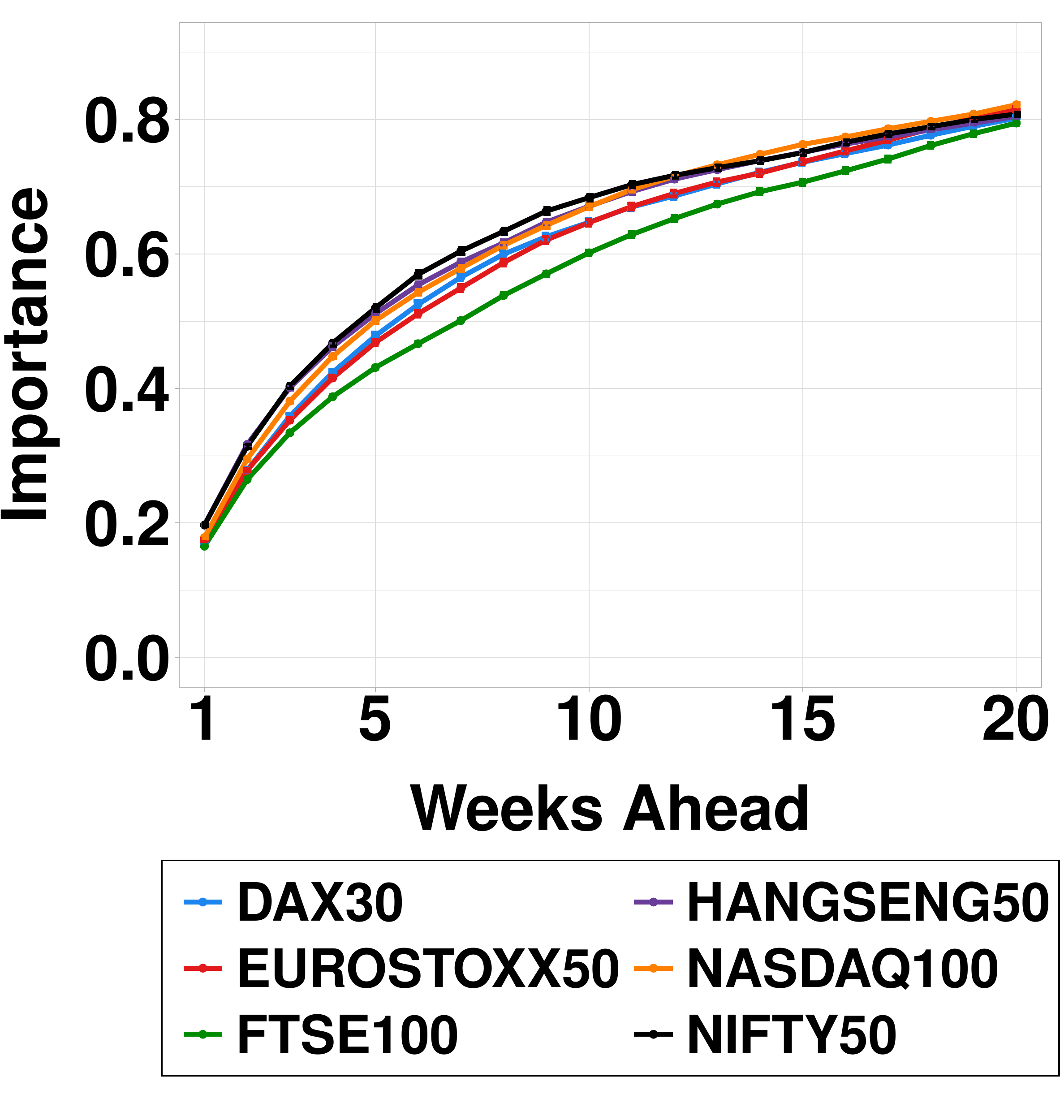}}
	\subfigure[DMST ($L = 252$)]{\includegraphics[trim=0.1cm 0.9cm 0.1cm 0cm, clip=true, clip=true, clip=true, width=0.30\textwidth]{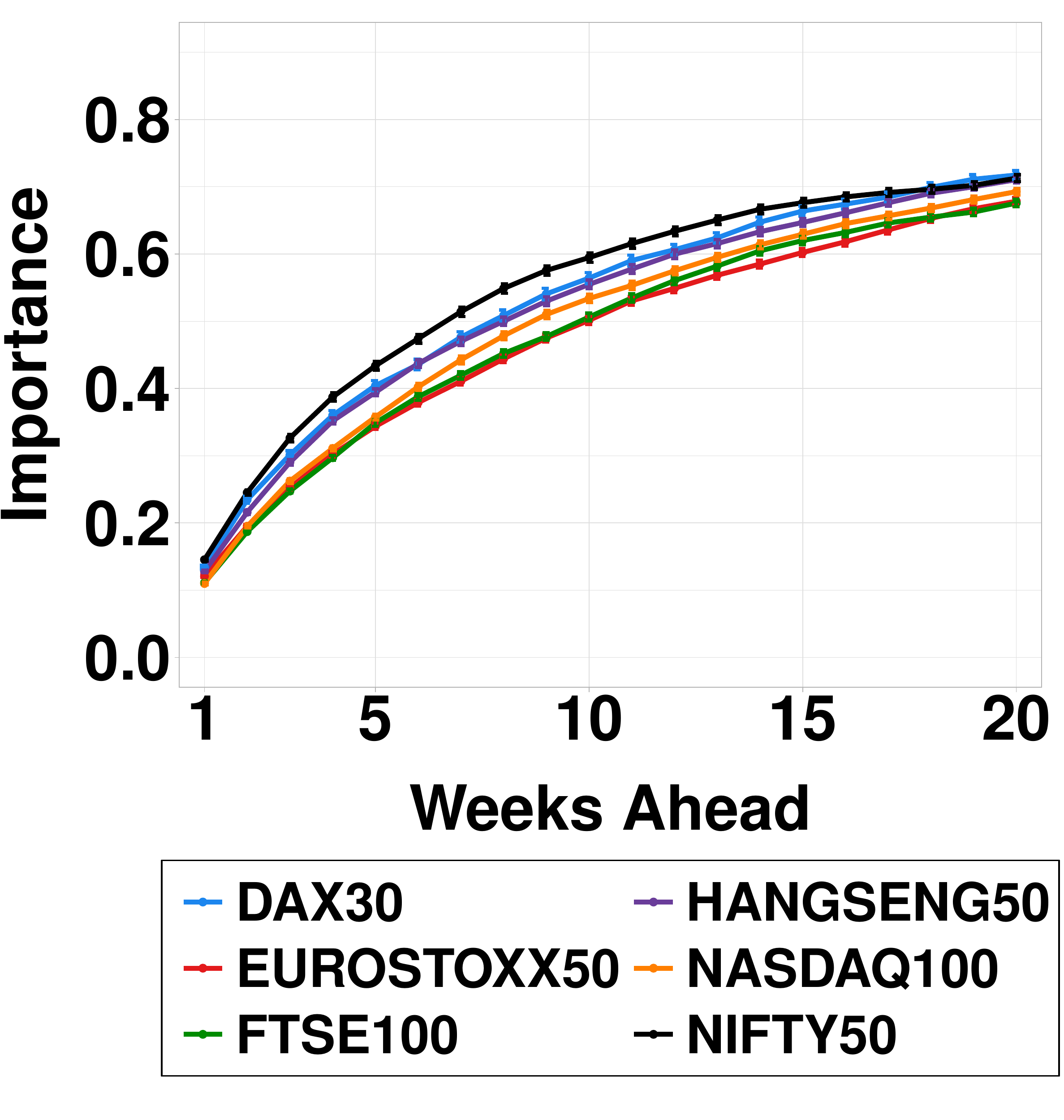}}
	\subfigure[DMST ($L = 504$)]{\includegraphics[trim=0.1cm 6.1cm 0.3cm 0cm, clip=true, clip=true, width=0.30\textwidth]{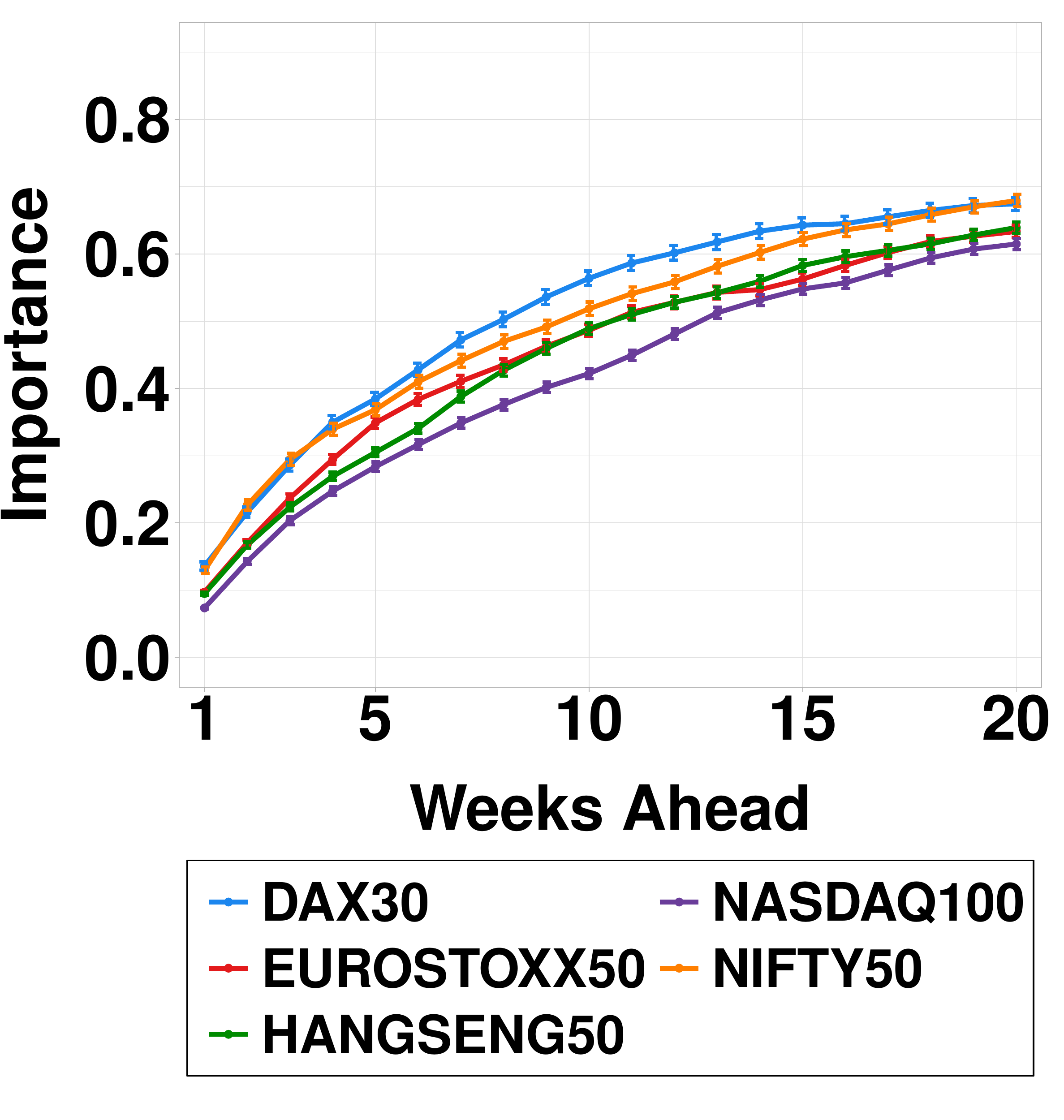}}
	\caption{\textbf{Importance of non pair-wise correlation features for DAG, DTN and DMST.} Figure shows the aggregate importance for non pair-wise correlation features using the size of rolling window $L = \lbrace 126, 252,504 \rbrace $ trading days and DAG, DTN and DMST network filtering methods. Results show the importance of these features increases with the time step $h$. The importance of non pair-wise correlation features for $L = 126$ trading days is higher than $L = 252$ and $L = 504$ for all network filtering methods. The growth of the importance of this subset is consistent across all markets. An interesting result is that the importance of non pair-wise correlation features changes according to the network filtering method.}
	\label{fig:importance}
\end{figure*}

In finance, particularly in portfolio management, the investment risk is calculated using the correlation among portfolio assets. This is the main information used to estimate risk and, given its importance in financial analyses, we also explore them as an input feature for market structure forecasting. However, we want to measure how the topology of the network helps forecast the future network itself. In other words, we are interested in evaluating the importance of non pair-wise correlation features for the forecasting market structure. As described in Section~\ref{sec:network-based-features}, we separated the feature set into two subsets: pair-wise correlation features and non pair-wise correlation features. After constructing the boosted trees in the XGBoost model, we can estimate the importance of each individual attribute. The importance of an attribute is related to the number of times that it is used to create relevant split decisions, i.e., split points that improve the performance metrics~\cite{hastie2009elements}. For each market index, we calculate the average and standard error of aggregate importance of pair-wise correlation and non pair-wise correlation features. Figure~\ref{fig:importance} presents results related to the importance of non pair-wise correlation features, considering the network filtering methods DAG, DTN and DMST and $L \in \lbrace 126, 252,504 \rbrace $ trading days as the rolling window size. It is important to note that the importance of the two feature subsets add up to $1$.

%In addition to the comparison between the proposed machine learning model and other link prediction benchmark algorithms, we are interested in evaluating the importance of non pair-wise correlation features for forecasting market structure.
%: \textit{How does topological network features perform relative to  co-variance data for the market structure prediction task in financial networks?} 
%To address the third research question, we used a feature importance method to asses the importance of topological features. 

%After constructing the boosted trees in the XGBoost model, we can estimate the importance of each individual attribute. The importance of an attribute is related to the number of times that it is used to create relevant split decisions, i.e., split points that improve the performance metrics~\cite{hastie2009elements}. 

%As described in Section~\ref{sec:network-based-features}, we sorted the feature sets into two classes: topological features and pair-wise correlation based, as indicated in Table~\ref{tab:linkfeatures}. 

Results presented in Figure~\ref{fig:importance} show that non pair-wise correlation features help forecast the future market using different network filtering methods. We observe that the importance of non pair-wise correlation features increases with $h$. Moreover, the importance of this subset of features changes according to the network filtering method. Their importance can be observed mainly for smaller $L$, such as $L = 126$, shown in Figures~\ref{fig:importance}(a),~\ref{fig:importance}(d) and~\ref{fig:importance}(g), where their importance for $h =20$ reaches almost $80\%$ for NIFTY50 using the DAG method, $60\%$ for EUROSTOXX50 using DTN and almost $90\%$ for all markets using DMST. For the DMST method, shown in Figures~\ref{fig:importance}(g),~\ref{fig:importance}(h) and~\ref{fig:importance}(i), the importance of non pair-wise correlation features has a similar shape to $L = 126$, $252$ and $504$ rolling window size. DAG results are shown in Figures~\ref{fig:importance}(a),~\ref{fig:importance}(b) and~\ref{fig:importance}(c). For short $h$ values, non pair-wise correlation attributes do not add much information when compared to pair-wise correlation features. However, the importance of these features rapidly increases with the time step $h$, suggesting that these attributes can be more useful than pair-wise correlation attributes for long-horizon forecasting exercises, particularly for short rolling window sizes. For $L = 252$ and $L = 504$, non pair-wise correlation features have less importance in forecasting networks modeled using DAG and DTN network filtering methods. Considering DMST results, the importance of non pair-wise features rapidly increases, even for short $h$ values. This behavior is different from DAG and DTN. A possible explanation for this is the low persistence of trees, as shown in Figure~\ref{fig:cross-similarity-dmst}. Thus, network features are able to add more information to the ML model when compared to pair-wise correlation features.

\section{Conclusion}
\label{sec:conclusion}
In this article, we investigated stock market structure forecasting of multiple financial markets using financial networks modeled using stock returns of major market indices constituents. The stock market structure was modeled as networks, where nodes represent assets and edges represent the relationship among them. Three correlation based filtering methods were used to create stock networks: Dynamic Asset Graphs (DAG), Dynamic Threshold Networks (DTN) and Dynamic Minimal Spanning Tree (DMST). We formulated market structure forecasting as a network link prediction problem, where we aim to accurately predict the edges that will be present in future networks. We proposed and experimentally assessed a machine learning model based on node- and link-based financial network features to forecast future market structure. 

We used data from company constituents of six different stock market indices from the U.S., the U.K., India, Europe, Germany and Hong Kong markets, ranging from $1$ March $2005$ to $18$ December $2019$. To assess the predictive performance of the model, we compared it to seven link prediction benchmark algorithms. Experimental results showed the proposed model was able to forecast the market structure with a performance superior to all benchmark methods and for all market indices, regardless the network filter method. We also measured the improvement against the Time Invariant (TI) algorithm, which assumes that the network does not change over time. Experimental results showed a greater improvement over the TI in networks created using the DTN filtering method, reaching almost $40\%$ improvement for NASDAQ100. Our experimental results also suggested that topological network information is useful in forecasting stock market structure compared to pair-wise correlation measures, particularly for long-horizon predictions. 

As work limitations, we should emphasize that we only used assets that stayed in the market index throughout the whole period, which limits the insertion and removal of nodes in the networks. In addition, for networks with large number of nodes, the execution time increased significantly, both for generating derived features and for training ML models.

%We also concluded that forecasting market structure using different network filter methods present different levels of aggregated importance for topological features, such as DMST in which the importance of topological features is greater than in DAG and DTN methods.

%Additionally, we observed that network structure and its information content were important in determining network predictive power by measuring the relationship between network metrics such as modularity, degree entropy and clustering and model's performance (AUC) when forecasting market structure.

%Our results can be useful to improve some financial strategies based on financial correlation structure, such as portfolio selection, arbitrage strategies and pair-trading algorithms. 

Our results can be useful in the study of stock market dynamics and to improve portfolio selection and risk management on a forward-looking basis and market structure estimation. As future work, we plan to use the predicted stock market structure as input in portfolio and risk management tools to evaluate its usefulness in risk management scenarios. Future work also includes market structure forecasting using order book data for high frequency trading analysis and the study of different asset classes beyond equities.

%Future work also includes the use of tensor, as proposed in~\cite{dunlavy2011temporal}, and the identification of the market structure using order book data for high frequency trading analysis.

\section*{Acknowledgements}

%Acknowledgements should be brief, and should not include thanks to anonymous referees and editors, or effusive comments. Grant or contribution numbers may be acknowledged.

D.C. and A.C.P.L.F.C would like thank to CAPES, Intel and CNPq (grant 202006/2018-2) for their support.

\section*{Author contributions statement}

%Must include all authors, identified by initials, for example:
%A.A. conceived the experiment(s),  A.A. and B.A. conducted the experiment(s), C.A. and D.A. analysed the results.  All authors reviewed the manuscript. 

D.C. and T.T.P.S. developed the proposed model. D.C. and T.T.P.S. conceived and designed the experiments. D.C. and T.T.P.S. prepared figures and tables, implemented and carried out the experiments. All authors analyzed the results and wrote the manuscript. All authors reviewed the article. 

\section*{Additional information}

\noindent \textbf{Competing Interests:} the authors declare no competing financial interests.
\\
\noindent \textbf{Supplementary Information:} provided with this document as Supplementary Material. 

%To include, in this order: \textbf{Accession codes} (where applicable); \textbf{Competing interests} (mandatory statement). 

%The corresponding author is responsible for submitting a \href{http://www.nature.com/srep/policies/index.html#competing}{competing interests statement} on behalf of all authors of the paper. This statement must be included in the submitted article file.

% ----------------------------------------------------------
% Bibliography
% ----------------------------------------------------------

\bibliography{mybib}

% ----------------------------------------------------------
% Appendix
% ----------------------------------------------------------
%\input{appendix}

\end{document}